\documentclass[reprint, superscriptaddress, secnumarabic, amssymb, nobibnotes, prl]{revtex4-2}

\usepackage{times,xspace}
\usepackage{graphicx,graphics,color,epsfig}
\usepackage{amsmath}
\usepackage{amssymb}
\usepackage{amsfonts}
\usepackage{amsfonts}
\usepackage{epstopdf}
\usepackage[T1]{fontenc}
\usepackage{amsmath}
\usepackage{amssymb}
\usepackage{xcolor}
\usepackage{siunitx}
\usepackage{braket}
\usepackage{xcolor}
\usepackage{bbm}
\usepackage{color}
\usepackage{float}
\usepackage[colorlinks=true]{hyperref} 
\usepackage{tikz}

\hypersetup{
    unicode=false,              
    pdftoolbar=true,            
    pdfmenubar=true,            
    pdffitwindow=false,         
    pdfstartview={FitH},        
    pdftitle={Precise Fermi level engineering in a topological Weyl semimetal via fast ion implantation},                
    pdfauthor={Manasi Mandal},  
    pdfsubject={},              
    pdfcreator={},              
    pdfproducer={},             
    pdfkeywords={} {} {},       
    pdfnewwindow=true,          
    colorlinks=true,            
    linkcolor=blue,             
    citecolor=blue,             
    filecolor=magenta,          
    urlcolor=blue               
} 
\usepackage[normalem]{ulem}

\newcommand{\figref}[1]{Fig.~\ref{#1}}

\newcommand{\tableref}[1]{Table~\ref{#1}}

\renewcommand{\approx}{\simeq}

\renewcommand{\vec}[1]{\boldsymbol{#1}}

\def\be{\begin{equation}}
\def\ee{\end{equation}}
\def\bea{\begin{eqnarray}}
\def\eea{\end{eqnarray}}

\makeatletter
\renewcommand*{\@fnsymbol}[1]{\ensuremath{\ifcase#1\or \dagger\or *\or \ddagger\or
   \mathsection\or \mathparagraph\or \|\or **\or \dagger\dagger
   \or \ddagger\ddagger \else\@ctrerr\fi}}
\makeatother

\begin{document}
\title{Precise Fermi level engineering in a topological Weyl semimetal via fast ion implantation}

\author{Manasi Mandal}
\thanks{These authors contribute to this work equally.}
\affiliation{Quantum Measurement Group, MIT, Cambridge, MA 02139, USA}
\affiliation{Department of Nuclear Science and Engineering, MIT, Cambridge, MA 02139, USA}

\author{Abhijatmedhi Chotrattanapituk}
\thanks{These authors contribute to this work equally.}
\affiliation{Quantum Measurement Group, MIT, Cambridge, MA 02139, USA}
\affiliation{Department of Electrical Engineering and Computer Science, MIT, Cambridge, MA 02139, USA}

\author{Kevin Woller}
\affiliation{Department of Nuclear Science and Engineering, MIT, Cambridge, MA 02139, USA}

\author{Haowei Xu}
\affiliation{Department of Nuclear Science and Engineering, MIT, Cambridge, MA 02139, USA}

\author{Nannan Mao}
\affiliation{Department of Electrical Engineering and Computer Science, MIT, Cambridge, MA 02139, USA}

\author{Ryotaro Okabe}
\affiliation{Quantum Measurement Group, MIT, Cambridge, MA 02139, USA}
\affiliation{Department of Chemistry, MIT, Cambridge, MA 02139, USA}

\author{Artittaya Boonkird}
\affiliation{Quantum Measurement Group, MIT, Cambridge, MA 02139, USA}
\affiliation{Department of Nuclear Science and Engineering, MIT, Cambridge, MA 02139, USA}
    
\author{Thanh Nguyen}
\affiliation{Quantum Measurement Group, MIT, Cambridge, MA 02139, USA} 
\affiliation{Department of Nuclear Science and Engineering, MIT, Cambridge, MA 02139, USA}
    
\author{Nathan C. Drucker}
\affiliation{Quantum Measurement Group, MIT, Cambridge, MA 02139, USA}
\affiliation{School of Engineering and Applied Sciences, Harvard University, Cambridge, MA 02138, USA}

\author{Takashi Momiki}
\affiliation{Aisin Technical Center of America, Inc, CA 95110, USA}

\author{Ju Li}
\affiliation{Department of Nuclear Science and Engineering, MIT, Cambridge, MA 02139, USA}
\affiliation{Department of Materials Science and Engineering, MIT, Cambridge, MA 02139, USA}

\author{Jing Kong}
\affiliation{Department of Electrical Engineering and Computer Science, MIT, Cambridge, MA 02139, USA}

\renewcommand{\thefootnote}{\fnsymbol{1}}
    
\author{Mingda Li\footnote{}}
\email[]{manasim@mit.edu}
\email[]{mingda@mit.edu}
\affiliation{Quantum Measurement Group, MIT, Cambridge, MA 02139, USA}
\affiliation{Department of Nuclear Science and Engineering, MIT, Cambridge, MA 02139, USA}

\date{\today}

\begin{abstract}
The precise controllability of the Fermi level is a critical aspect of quantum materials. For topological Weyl semimetals, there is a pressing need to fine-tune the Fermi level to the Weyl nodes and unlock exotic electronic and optoelectronic effects associated with the divergent Berry curvature. However, in contrast to 2D materials, where the Fermi level can be controlled through various techniques, the situation for bulk crystals beyond laborious chemical doping poses significant challenges. Here, we report the meV-level ultra-fine-tuning of the Fermi level of bulk topological Weyl semimetal TaP using accelerator-based high-energy hydrogen implantation and theory-driven planning. By calculating the desired carrier density and controlling the accelerator profiles, the Fermi level can be fine-tuned from 5 meV to only $\sim$0.5 meV (DFT calculations) away from the Weyl nodes. The Weyl nodes are preserved, while the carrier mobility is largely retained. Our work demonstrates the viability of this generic approach to tune the Fermi level in semimetal systems and could serve to achieve property fine-tuning for other bulk quantum materials with ultrahigh precision. 
\end{abstract}

\maketitle

The precise controllability of the Fermi level and carrier doping in a solid-state material holds paramount significance in condensed matter physics and materials science. In microelectronics such as high-electron-mobility transistors, the ability to fine-tune the Fermi level is crucial to control the operation thresholds and high-frequency responses \cite{hemaja2022, Lee2016, hemt1}. In strongly correlated systems such as high-temperature cuprate superconductors, achieving optimal hole doping is essential for enhancing the superconducting critical temperature \cite{sc1, sc2, sc3}. As for energy harvesting applications such as thermoelectrics, a well-adjusted Fermi level strikes a delicate balance between electrical conductivity and thermopower, which leads to an optimized power factor \cite{thermo1, thermo2, thermo3, thermo4}. A particularly noteworthy utilization for Fermi level fine-tuning lies in topological materials, where only a precisely positioned Fermi level can unveil the topological properties. For instance, in topological insulators, dissipationless electronic states such as quantum anomalous Hall states can only manifest when the Fermi level is located precisely within the small surface bandgap. This can only be achieved through a strategic blend of chemical doping \cite{chang2013, qah1, qah2} and electrostatic gating in thin heterostructures \cite{qahe1, qahe2, deng2020, lin2022}. For topological Weyl semimetals (WSM), exotic phenomena like large charge-to-spin interconversion \cite{zhao2020, mendes2022}, strong higher-order photoresponse \cite{osterhoudt2019, ma2019}, anomalous Nernst effect \cite{wsm1, roychowdhury2023}, and quantized thermoelectric Hall effect \cite{TaP, wsm4} emerge due to a divergent Berry curvature, which can only happen when the Fermi level is tuned to be exactly at the Weyl nodes.

However, unlike in lower dimensional materials, in which the Fermi level can be continuously fine-tuned through a variety of accessible techniques like electrostatic and ionic gating, achieving this level of control in three-dimensional bulk materials has presented a longstanding challenge. Chemical doping can only be performed during the synthesis and offers limited precision, which does not meet the stringent requirement for topological semimetals. The electronic states in topological semimetals, such as from the Weyl and Dirac bands, have a linear dispersion with zero density-of-states at the topological singularities, making the Fermi level location extremely sensitive to the carrier concentration \cite{dirac1, dirac2}. Therefore, there is an urgent need to attain enhanced fine-tunability of topological semimetals for more practical energy and information applications.
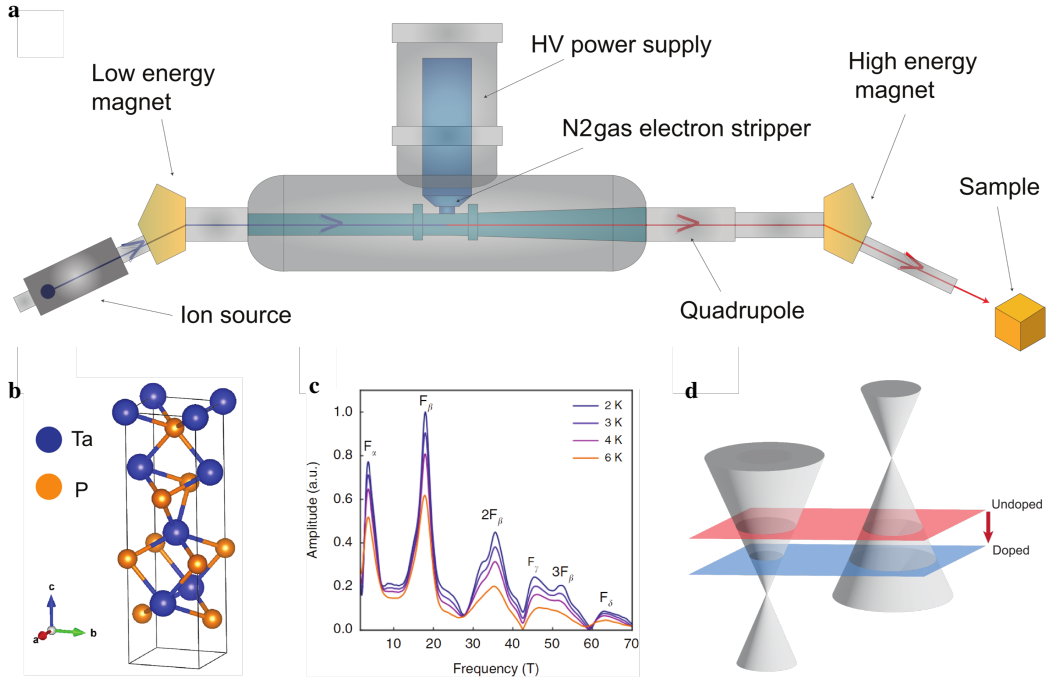
\begin{figure*}[ht!]
    \centering
    \begin{tikzpicture}
        \node[anchor=north west] (image) at (0,0) 
        {\includegraphics[width=1.6\columnwidth]{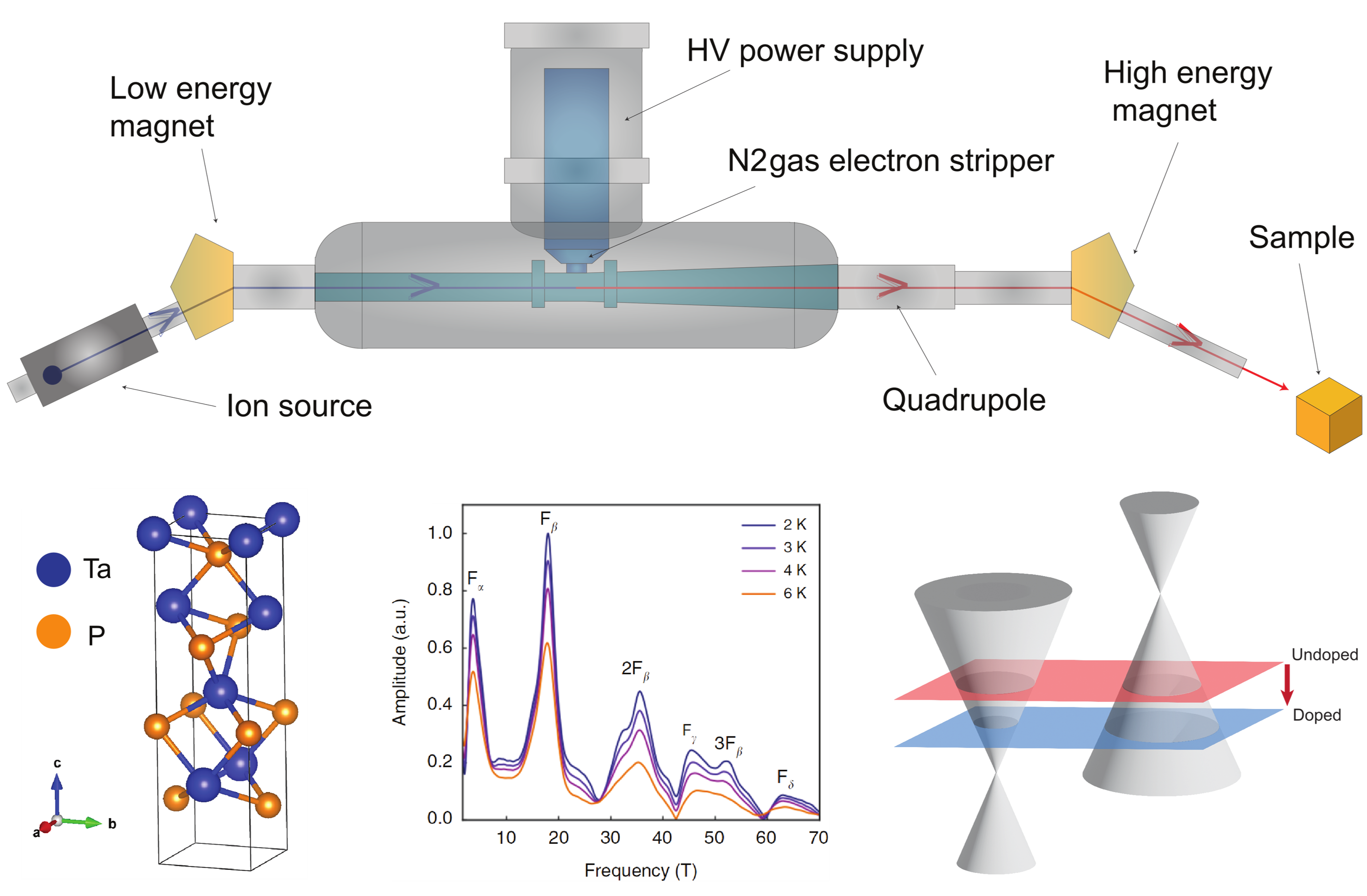}};
        \node[anchor=north west] (a) at (0,0) {\textbf{a}};
        \node[anchor=north west] (b) at (0,-5) {\textbf{b}};
        \node[anchor=north west] (c) at (4,-5) {\textbf{c}};
        \node[anchor=north west] (d) at (9,-5) {\textbf{d}};
    \end{tikzpicture}
    \caption{\textbf{a)} Tandem accelerator schematic. It includes an ion source to produce negative ions. Selected ion species are injected toward the terminal using a low-energy magnet. The ion beam is focused by a magnetic quadrupole lens and the desired ion species are directed into the beamline for the experiment. \textbf{b)} Crystal structure of the Weyl semimetal TaP. \textbf{c)} Fast Fourier transform of the longitudinal magnetoresistance of TaP reveals quantum oscillations with specific frequencies (reproduced from \cite{TaP}). \textbf{d)} Illustration of the irradiation effect on the Fermi level of TaP. The red arrow indicates the shift of the Fermi level towards a Weyl node.}
    \label{fig1}
\end{figure*}

In this work, we use high-energy, accelerator-based hydrogen implantation to achieve ultra-fine carrier doping in WSM of tantalum phosphide (TaP) \cite{TaP}. Although hydrogenation and low-energy ion implantation have been used to tune materials properties \cite{hydrogen1, hydrogen2, bulk1, Ferry, nano1, nano2}, there are a few key distinct features presented in the current setup. Foremost, the fine controllability of the advanced accelerator technique enables an ultra-fine tuning of the Fermi level down to the meV regime, far beyond the previous approaches. In addition, the high energy beam also enables the doping of a bulk crystal beyond a thin film with tens of nanometers \cite{bulk1, Ferry, k_thesis}. Moreover, thanks to the rapid development of the accelerator technology with high precision control of the accelerator beam energy and flux of ions, we were able to achieve \textit{a priori} doping planning through a combination of DFT and Monte Carlo calculations, which eventually shows quantitative agreement with actual experiments. This sets up an example for ultrafine doping planning experiments. The Weyl nodes are preserved after doping, as shown from the quantum oscillation measurements, while the temperature to reach the charge neutral point is nearly doubled, without lowering much on the carrier mobility. 
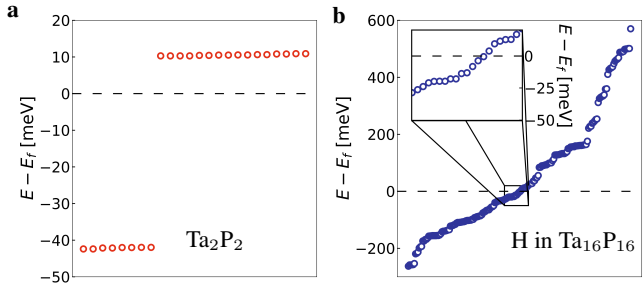
\begin{figure}[ht!]
    \centering
    \begin{tikzpicture}
        \node[anchor=north west] (image) at (0,0) 
        {\includegraphics[width=0.5\columnwidth]{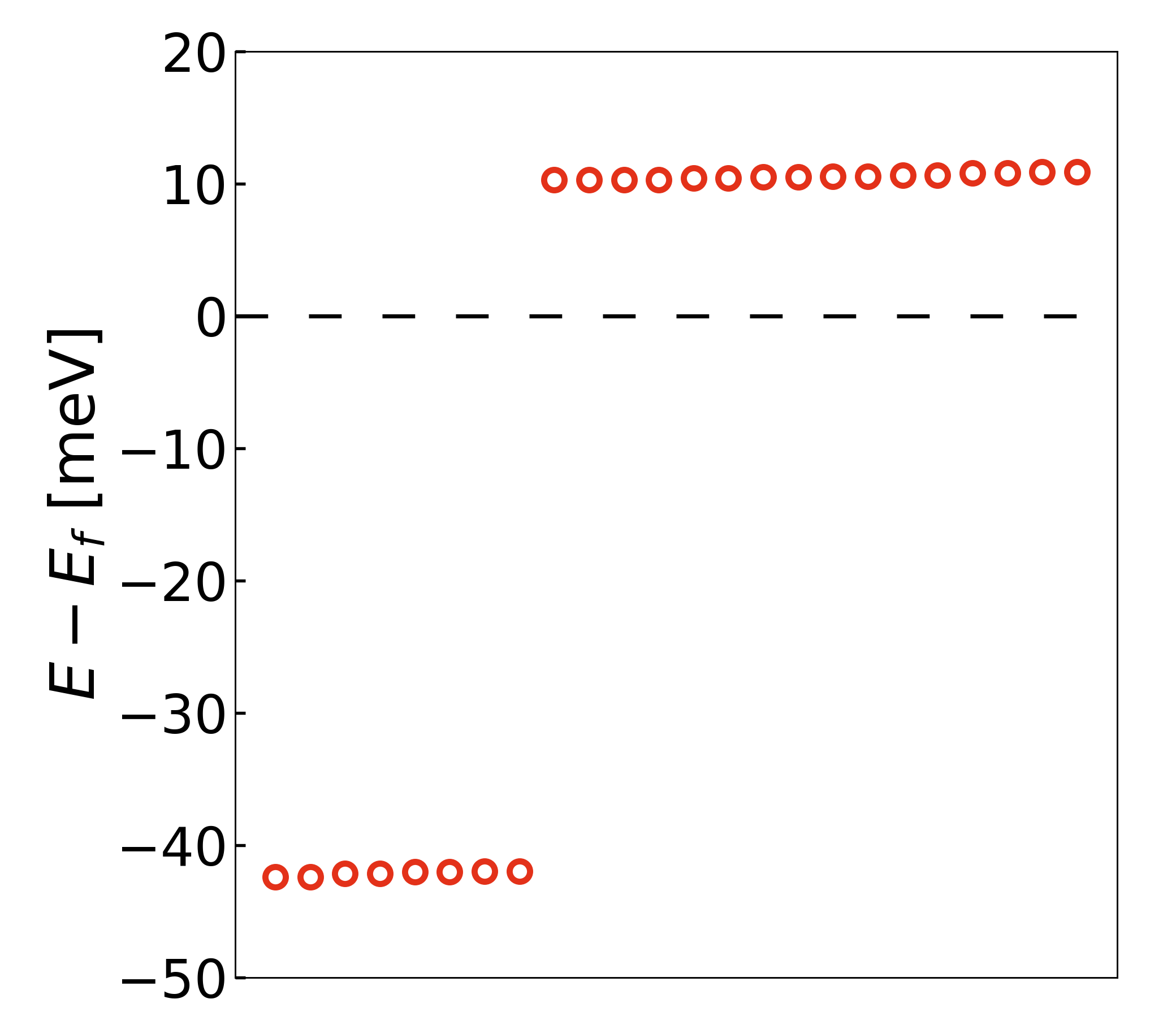}};
        \node[anchor=north west] (image) at (0.50\columnwidth,0) 
        {\includegraphics[width=0.5\columnwidth]{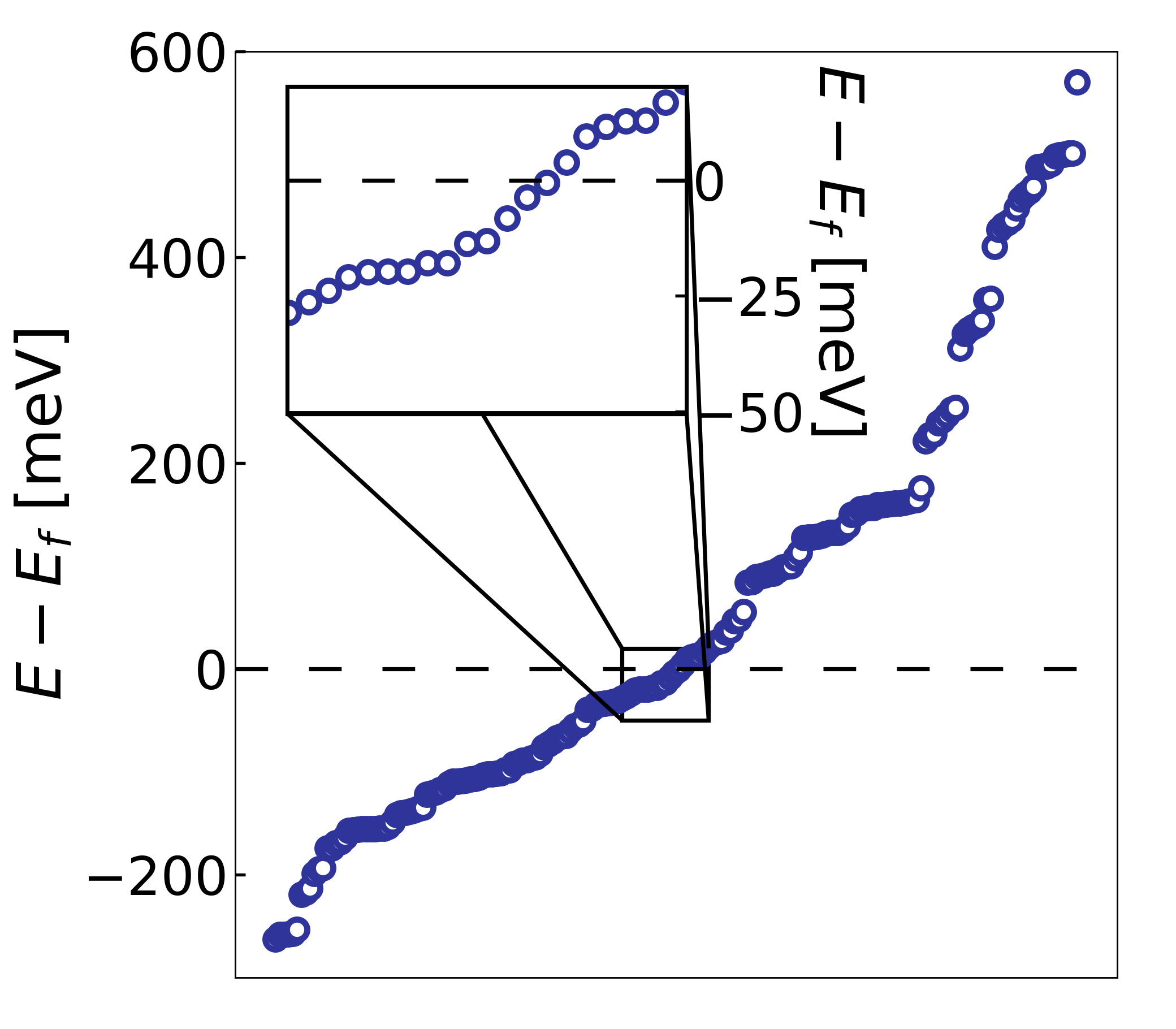}};
        \node[anchor=north west] (a) at (0,0) {\textbf{a}};
        \node[anchor=north west] (b) at (0.5\columnwidth,0) {\textbf{b}};
        \node[anchor=north west] at (0.275\columnwidth,-3) {Ta$_2$P$_2$};
        \node[anchor=north west] at (0.775\columnwidth,-3) {H in Ta$_{16}$P$_{16}$};
    \end{tikzpicture}
    \caption{Ab initio calculation of Weyl nodes' energy relative to Fermi level ($E_f$) of \textbf{a)} pristine TaP, and \textbf{b)} H doped TaP ($2\times2\times2$ TaP supercell with one H insertion). Each plot point represents a Weyl node placed in its energy ascending order from left to right. The inset in \textbf{b} shows the zoom-in view of the Weyl nodes with energy near Fermi level}
    \label{fig2}
\end{figure}

The doping experiment uses the Tandem accelerator located at the Center for Science and Technology with Accelerators and Radiation (CSTAR). The schematic used in our experiment is shown in \figref{fig1}a. This type of accelerator consists of two successive linear accelerators with the same power supply. A cesium sputtering source is used as the ion source to generate negative ions from most of the elements with low vapor pressure (excluding noble gases). This source works by cesium sputtering a solid target, which produces negative ions that are then subjected to electron exchange with neutral cesium. Negative ions are initially introduced into the first stage (blue arrow), where they are accelerated toward the positive high-voltage (HV) terminal. Subsequently, depending on the experimental preferences, nitrogen gas can be injected into the HV terminal to strip some of the electrons from the ions, which transforms them into positive ions. As a result, the Tandem accelerator enables a remarkable capability to dope tens of different elements in the period table of elements. We have chosen negative hydrogen as the dopant, which offers several advantages in materials science due to its small atomic size, i.e., it does not significantly alter the structure of pristine materials \cite{H1}. Electrochemical hydrogenation can also be used for electron doping of semiconductors with good doping regime controllability, but not high-precision doping level control \cite{yajima}. In addition, hydrogen can exhibit a range of exotic properties as a dopant that makes it particularly interesting for research and applications, including metal-to-insulator transitions \cite{H_phase}, magnetostriction \cite{H_Sr}, enhanced superconductivity \cite{H_sc}, and paraelectric-to-ferroelectric phase transitions \cite{H_ferro}.

\begin{figure*}[ht!]
    \centering
    \begin{tikzpicture}
        \node[anchor=north west] (image) at (0,0) 
        {\includegraphics[width=0.66\columnwidth]{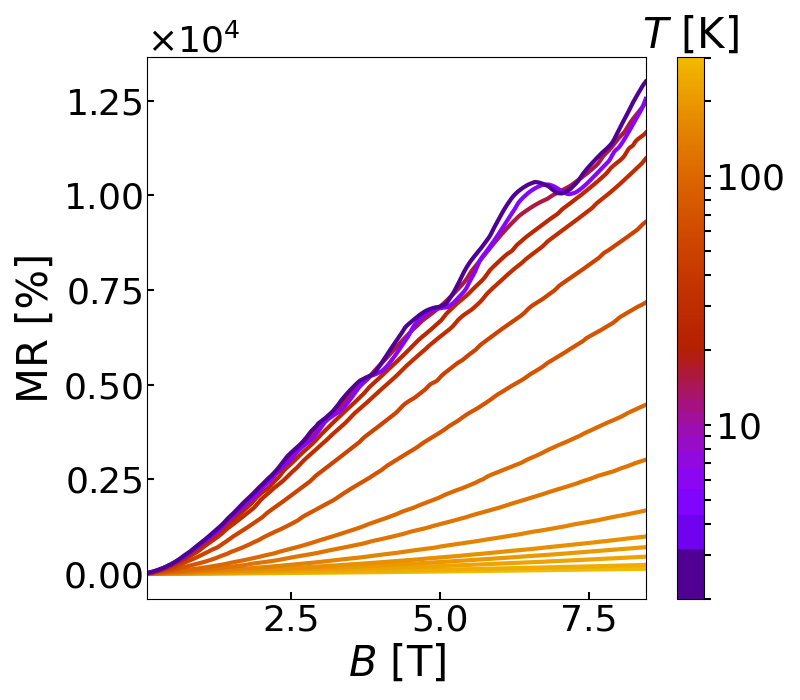}};
        \node[anchor=north west] (image) at (0.66\columnwidth,0) 
        {\includegraphics[width=0.66\columnwidth]{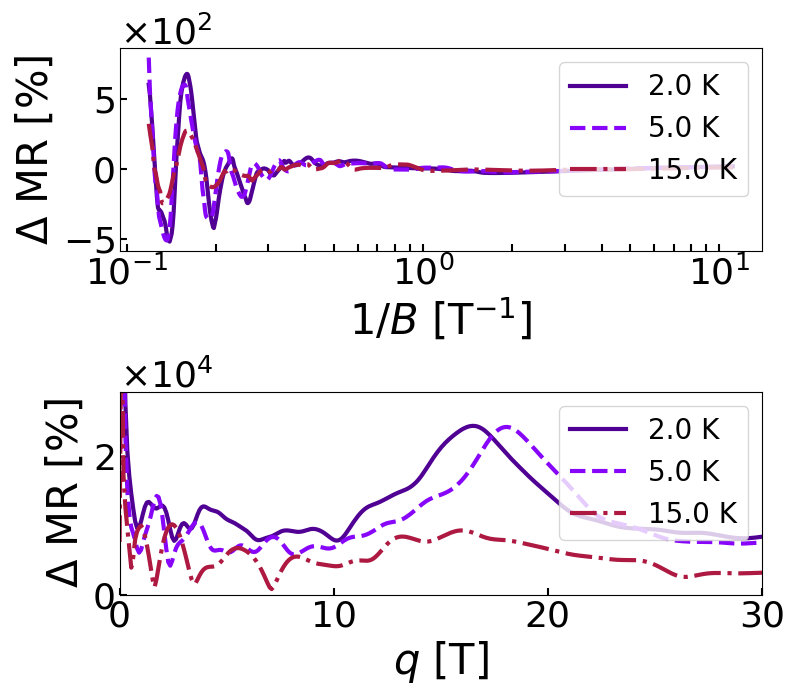}};
        \node[anchor=north west] (image) at (1.32\columnwidth,0) 
        {\includegraphics[width=0.66\columnwidth]{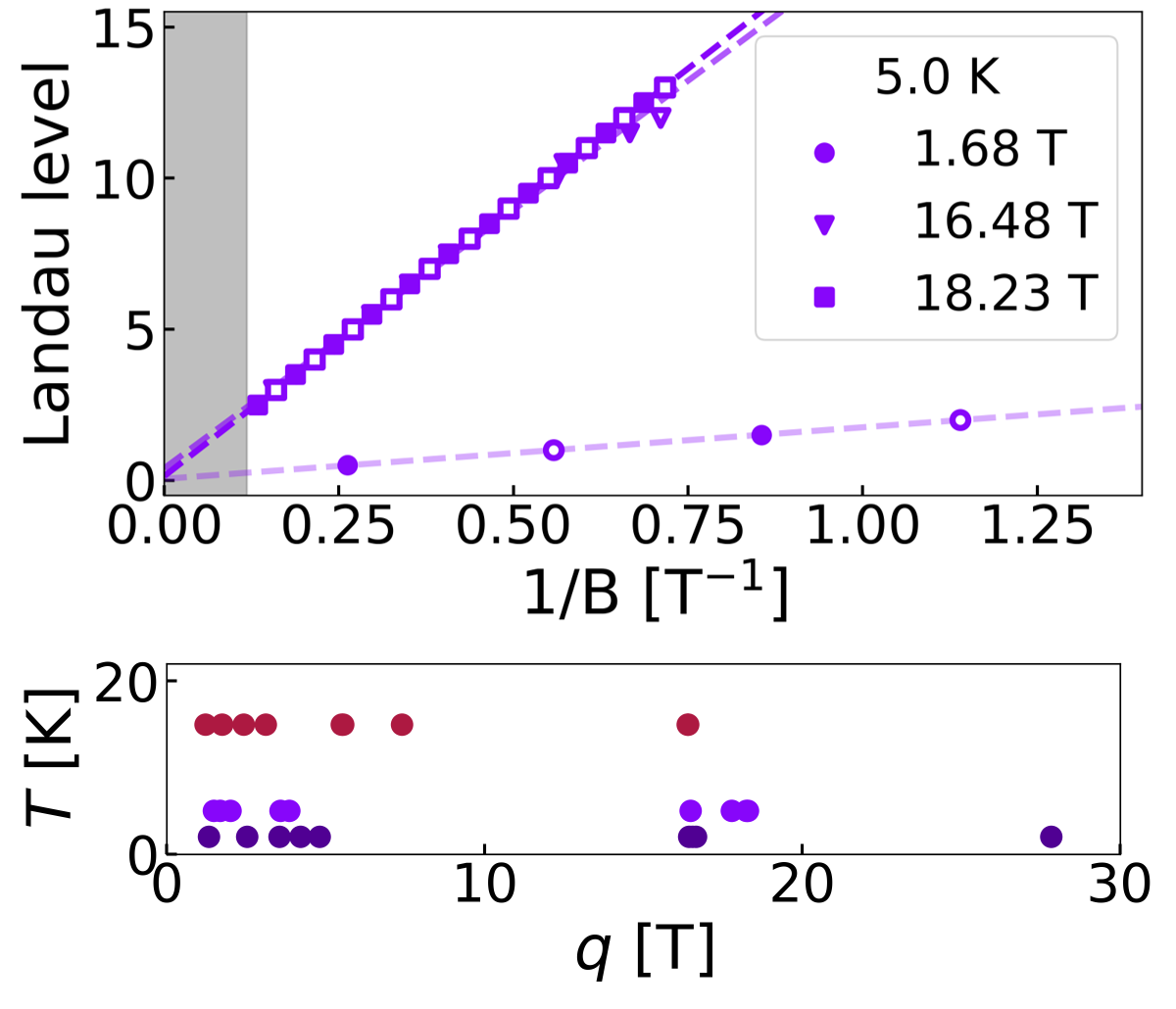}};
        \node[anchor=north west] (a) at (0,0) {\textbf{a}};
        \node[anchor=north west] (b) at (0.66\columnwidth, 0) {\textbf{b}};
        \node[anchor=north west] (c) at (0.66\columnwidth,-2.5) {\textbf{c}};
        \node[anchor=north west] (d) at (1.32\columnwidth, 0) {\textbf{d}};
        \node[anchor=north west] (e) at (1.32\columnwidth,-3) {\textbf{e}};

        \node[anchor=north west] at (0.2\columnwidth,-0.5) {S$_{3m}$};
        \node[anchor=north west] at (0.86\columnwidth,-0.5) {S$_{3m}$};
        \node[anchor=north west] at (0.86\columnwidth,-3) {S$_{3m}$};
        \node[anchor=north west] at (1.52\columnwidth,-0.5) {S$_{3m}$};
        \node[anchor=north west] at (1.82\columnwidth,-3.3) {S$_{3m}$};
    \end{tikzpicture}
    \begin{tikzpicture}
        \node[anchor=north west] (image) at (0,0) 
        {\includegraphics[width=0.66\columnwidth]{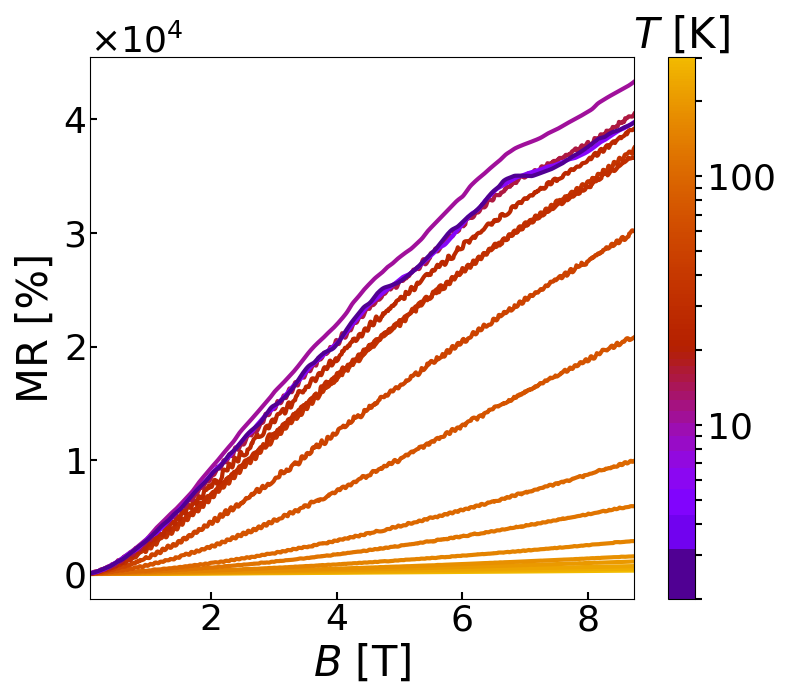}};
        \node[anchor=north west] (image) at (0.66\columnwidth,0) 
        {\includegraphics[width=0.66\columnwidth]{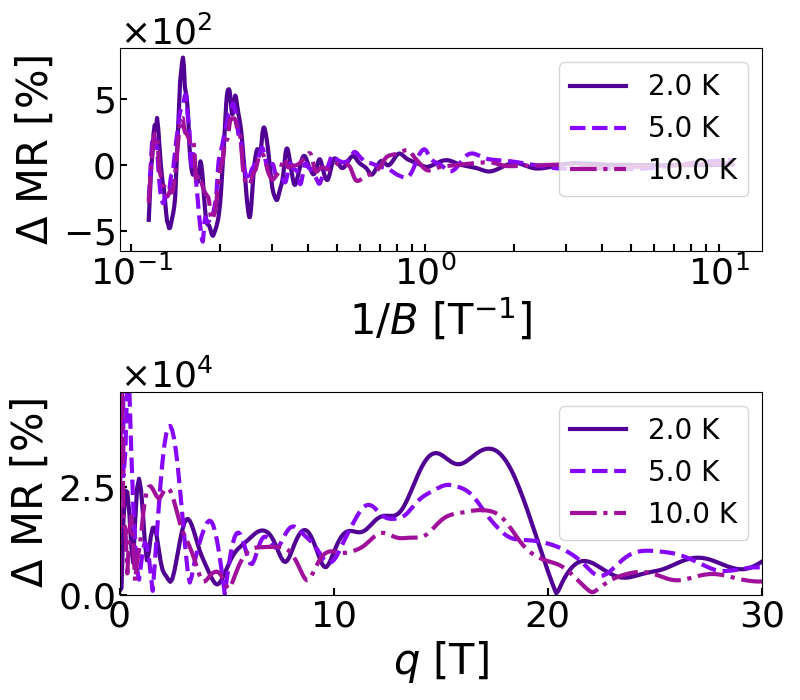}};
        \node[anchor=north west] (image) at (1.32\columnwidth,0) 
        {\includegraphics[width=0.66\columnwidth]{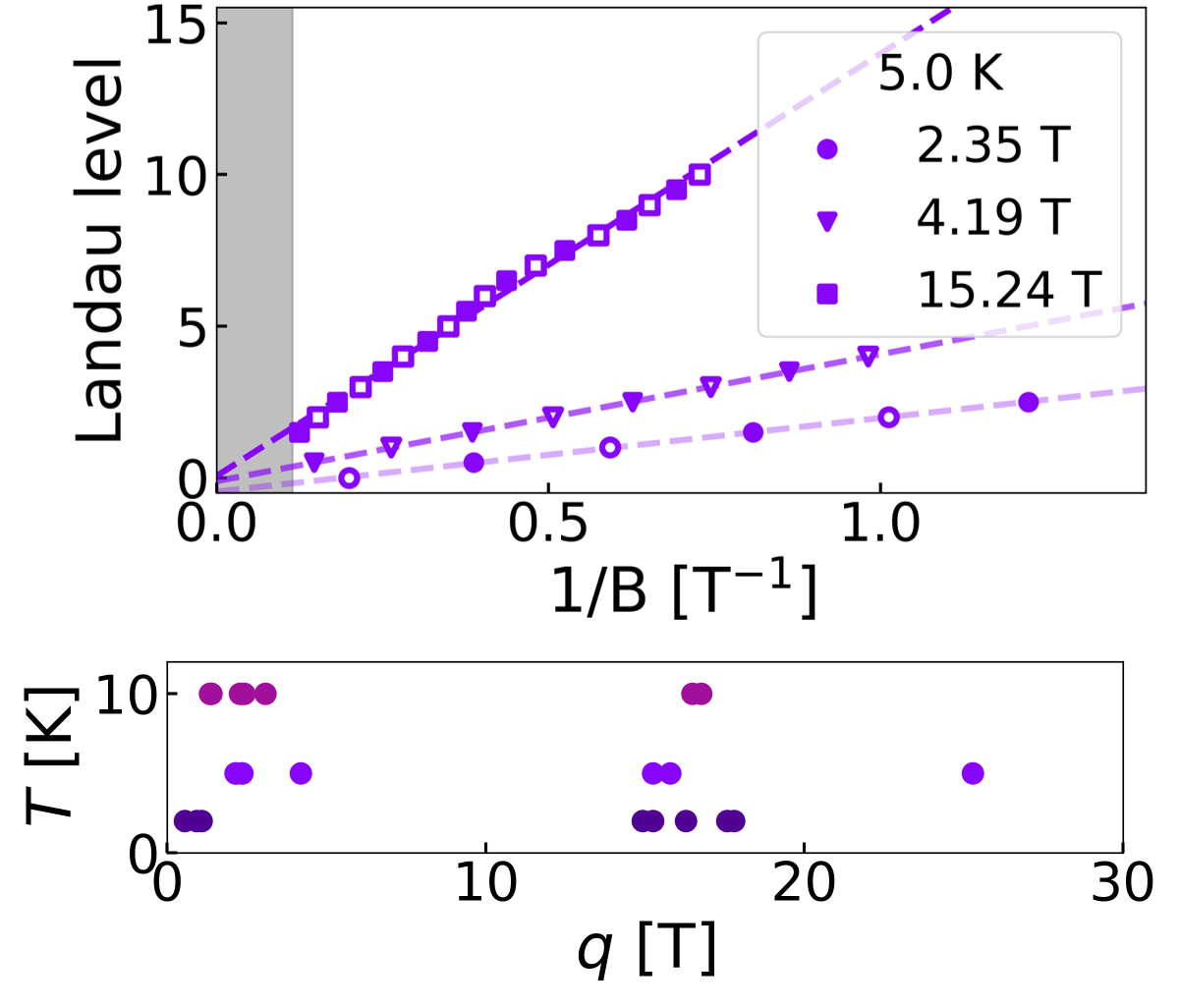}};
        \node[anchor=north west] (f) at (0,0) {\textbf{f}};
        \node[anchor=north west] (g) at (0.66\columnwidth, 0) {\textbf{g}};
        \node[anchor=north west] (h) at (0.66\columnwidth,-2.5) {\textbf{h}};
        \node[anchor=north west] (i) at (1.32\columnwidth, 0) {\textbf{i}};
        \node[anchor=north west] (j) at (1.32\columnwidth,-3) {\textbf{j}};

        \node[anchor=north west] at (0.2\columnwidth,-0.5) {S$_{20m}$};
        \node[anchor=north west] at (0.86\columnwidth,-0.5) {S$_{20m}$};
        \node[anchor=north west] at (0.86\columnwidth,-3) {S$_{20m}$};
        \node[anchor=north west] at (1.52\columnwidth,-0.5) {S$_{20m}$};
        \node[anchor=north west] at (1.82\columnwidth,-3.3) {S$_{20m}$};
    \end{tikzpicture}
    \caption{\textbf{a)} Variation of magnetoresistance (MR) with magnetic field ($B$) at different temperatures for sample S$_{3m}$. \textbf{b)} Background-subtracted MR ($\Delta$MR) of \textbf{a} versus 1/B for a few temperatures that exhibit Shubnikov-de Haas (SdH) oscillations. This $\Delta$MR is calculated using q polynomial-fitted background subtraction. \textbf{c)} Fast Fourier transform (FFT) of \textbf{b}. \textbf{d)} Example of a Landau level fan diagram analysis which indicates the presence of lower frequencies in \textbf{b} at 5 K than TaP \cite{TaP}. \textbf{e)} Scatter plot of the oscillation frequencies extracted from \textbf{b} using various background subtraction methods at 5 K. \textbf{f)}-\textbf{j)} are the same as \textbf{a)}-\textbf{e)}, but for the S$_{20m}$ sample.}
    \label{fig3}
\end{figure*}
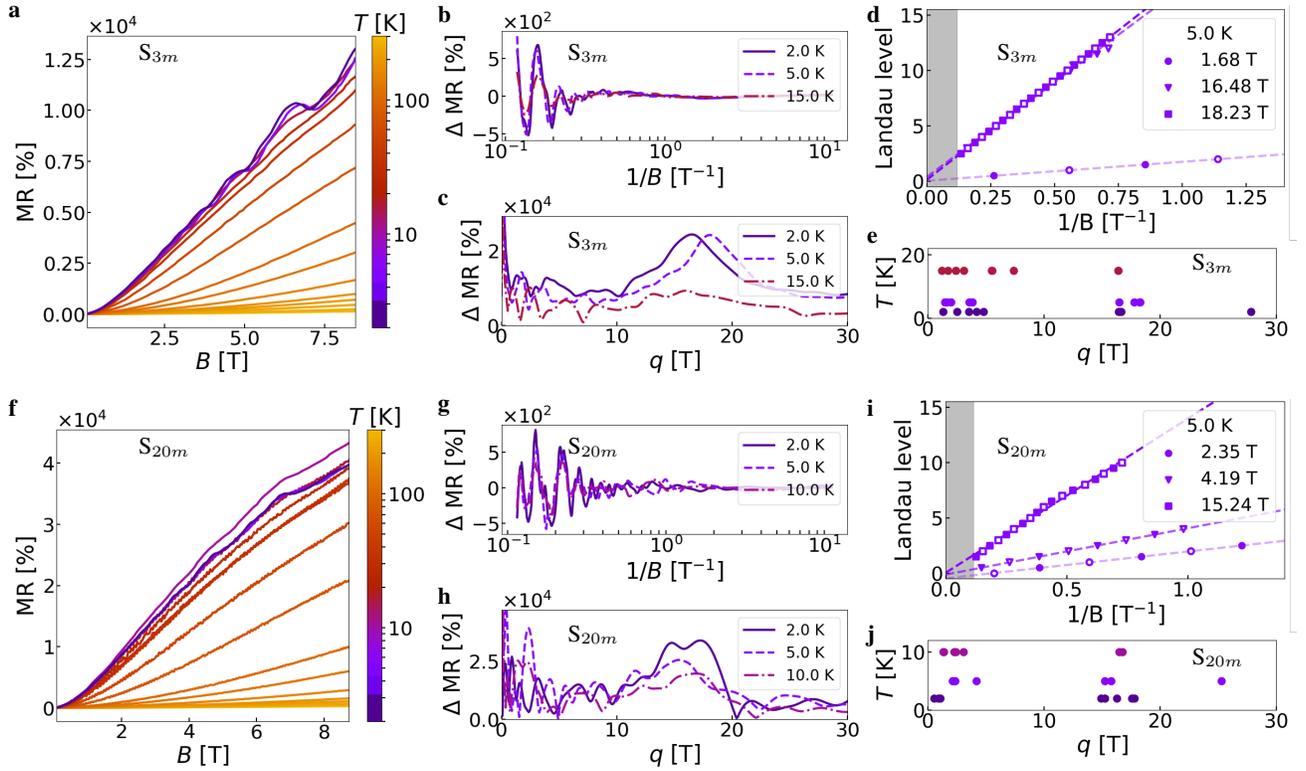

The earlier study \cite{TaP} highlights TaP as a remarkable topological Weyl semimetal with an inversion symmetry-breaking crystal structure (\figref{fig1}b). It exhibits a non-saturating thermopower, a giant magnetoresistance, and quantized thermoelectric Hall effect. Shubnikov-de Haas oscillations (SdH) are observed in the background-subtracted MR ($\Delta$MR) data at temperatures below 25 K. The Fourier transform of $\Delta$MR showed four pockets, with two small carrier pockets, one with a low-frequency F$_{\alpha}$ = 4 T and the other with F$_{\beta}$ = 18 T indicating the presence of the W2 Weyl node, in addition to the electron pocket from the W1 Weyl node contributing to F$_{\beta}$ (\figref{fig1}c) \cite{TaP}. Introducing H$^{-}$ ions through irradiation is expected to shift the Fermi level closer to the Weyl node by carrier doping, as illustrated in the schematic diagram in \figref{fig1}d. 

By employing the Vienna ab initio simulation package (VASP), we conducted density functional theory (DFT) calculations. These calculations provide a qualitative insight into the energy redistribution of Weyl node energy levels upon H doping. The energy of the Weyl nodes with respect to the Fermi level ($E_f$) is shown in \figref{fig2}a for TaP and \figref{fig2}b for the doped TaP system (one H in Ta$_{16}$P$_{16}$). Here, the horizontal axis is the index of the Weyl nodes. The Weyl nodes are ranked according to their energy, so the index itself does not carry additional information. Notably, the doped TaP system exhibits Weyl nodes that are distributed somewhat continuously. Comparing the doped TaP to the pristine TaP, our calculations indicate that certain Weyl nodes in the doped system lie merely 0.5 meV away from the Fermi level, whereas, in the pristine system, they are much further at 10 meV from the Fermi level. A theoretical model and ab initio calculations estimating the correlation between hydrogen doping concentration in TaP and the shift in the Fermi level are discussed in Supplementary Information (SI).

To calculate the implanted ion energy, penetration depth within the crystal, and damage to the target material, the stopping and range of ions in matter software (SRIM/TRIM) \cite{SRIM} was employed. The detailed calculations are presented in the SI. Single crystals of TaP were irradiated with 20 keV H$^{-}$ ions with varying durations to achieve the desired doping concentration. The summary of different samples with their respective dose of irradiation and time duration is presented in \tableref{tab1}. 

The irradiated samples named S$_{3m}$ and S$_{20m}$ exhibit a significantly high magnetoresistance (MR) on the order of 10$^{4}$\% and accompanied by SdH oscillations at temperatures below 15 K (see \figref{fig3}a, and \figref{fig3}f, respectively). In contrast, the S$_{2h}$ sample does not display such oscillations even up to 9 T (details in SI). It is important to note that the analysis of carrier pockets based on quantum oscillations can be influenced by the choice of the background for MR. Therefore, we calculated the background-subtracted MR ($\Delta$MR) using different background subtraction methods (details in SI).
\begin{table}[ht!]
    \centering 
    \renewcommand{\arraystretch}{1.3}
    \begin{tabular}{|c|c|c|}   
    \hline
     Sample & Dose of irradiation  & Irradiation time \\
     & (cm$^{-2}$)  & (minutes) \\
     \hline
       S$_{3m}$ &  1.25$\times$10$^{15}$  & 3 \\
       S$_{20m}$ &  1.23$\times$10$^{16}$  & 20 \\
       S$_{2h}$ &  1.23$\times$10$^{17}$  & 120 \\
    \hline     
    \end{tabular}
    \caption{Summary of TaP samples irradiated with H$^{-}$ ions for different durations.}
    \label{tab1}
\end{table}
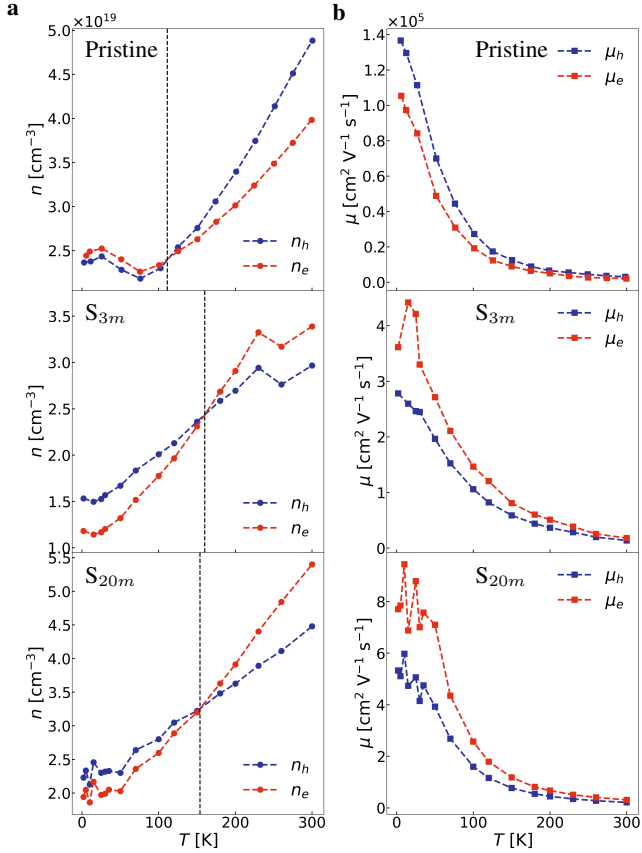
\begin{figure}[ht!]
    \centering
    \begin{tikzpicture}
        \node[anchor=north west] (image) at (0,0) 
        {\includegraphics[width=\columnwidth]{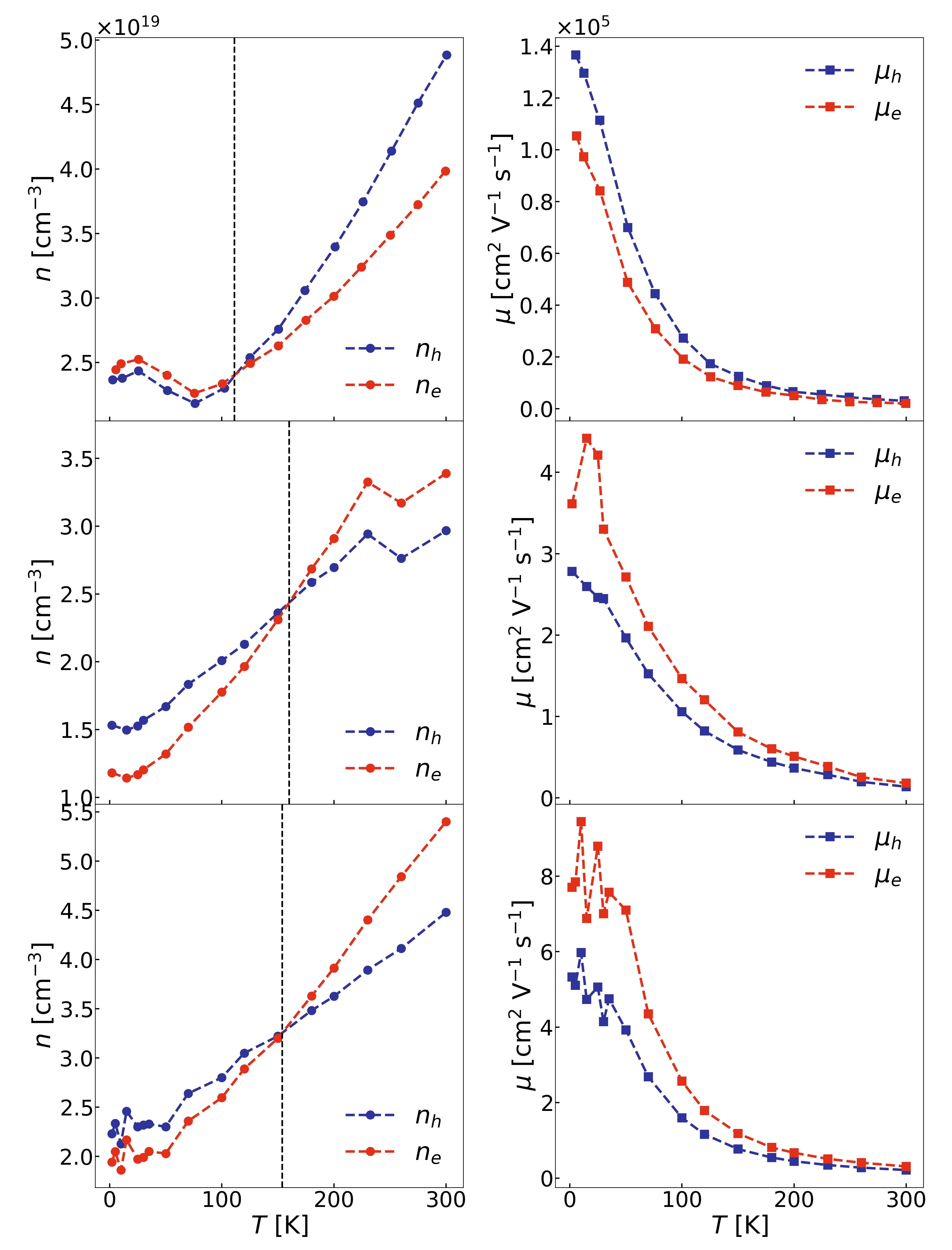}};
        \node[anchor=north west] (a) at (0,0) {\textbf{a}};
        \node[anchor=north west] (b) at (0.5\columnwidth,0) {\textbf{b}};

        \node[anchor=north west] at (0.12\columnwidth,-0.5) {Pristine};
        \node[anchor=north west] at (0.12\columnwidth,-4) {S$_{3m}$};
        \node[anchor=north west] at (0.12\columnwidth,-7.5) {S$_{20m}$};

        \node[anchor=north west] at (0.72\columnwidth,-0.5) {Pristine};
        \node[anchor=north west] at (0.72\columnwidth,-4) {S$_{3m}$};
        \node[anchor=north west] at (0.72\columnwidth,-7.5) {S$_{20m}$};
    \end{tikzpicture}
    \caption{\textbf{a)} Temperature-dependent profiles of electron ($n_{e}$) and hole concentrations ($n_{h}$). \textbf{b)} Temperature-dependent trends of electron ($\mu_{e}$) and hole mobilities ($\mu_{h}$). The top profiles serve as references for TaP. Our analysis involves the extraction of equivalent profiles from the two-band model fitting, applied to the $S_{3m}$ and $S_{20m}$ samples as annotated in the plots.}
    \label{fig4}
\end{figure}
\figref{fig3}b, and \figref{fig3}g illustrate the $\Delta$MR obtained with a polynomial-fitted background subtraction method, while \figref{fig3}c, and \figref{fig3}h show the corresponding fast Fourier transform (FFT) results for S$_{3m}$ and S$_{20m}$, respectively. These FFT plots reveal a noticeably lower oscillation frequency compared to the pristine TaP \cite{TaP}. Furthermore, \figref{fig3}d, and \figref{fig3}i present the results of the Landau level (LL) fan diagram analysis which indicates the presence of more oscillations with lower frequencies. The oscillation frequencies have been repeatedly extracted using various background subtraction methods, and scatter plotted together to show agreement between methods (see \figref{fig3}e, and \figref{fig3}j). This finding supports and validates our DFT calculations concerning the continuous energy distribution of Weyl nodes in proximity to the Fermi surface compared to pristine TaP.

\begin{figure}[ht!]
    \centering
    \begin{tikzpicture}
        \node[anchor=north west] (image) at (0,0) 
        {\includegraphics[width=0.5\columnwidth]{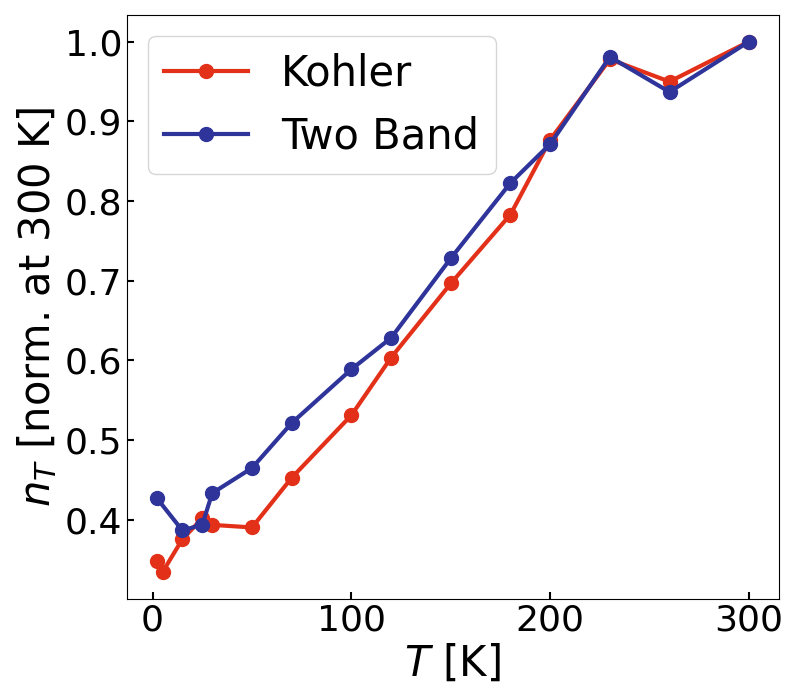}};
        \node[anchor=north west] (image) at (0.50\columnwidth,0) 
        {\includegraphics[width=0.5\columnwidth]{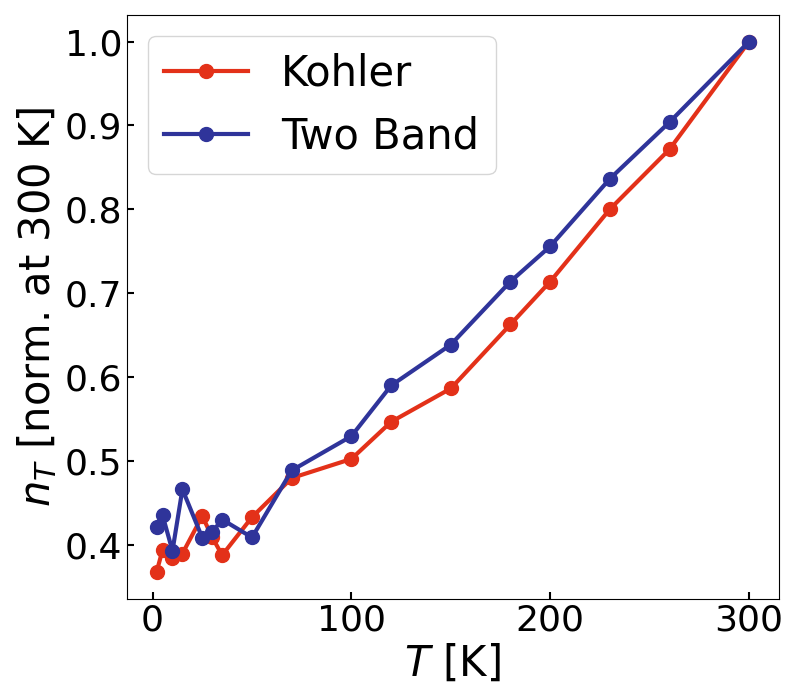}};
        \node[anchor=north west] (a) at (0,0) {\textbf{a}};
        \node[anchor=north west] (b) at (0.5\columnwidth,0) {\textbf{b}};

        \node[anchor=north west] at (0.4\columnwidth,-2.5) {S$_{3m}$};
        \node[anchor=north west] at (0.9\columnwidth,-2.5) {S$_{20m}$};
    \end{tikzpicture}
    \caption{Direct comparison of the temperature dependence of the normalized adjusted parameter $n_{T}$, as extracted from both the extended Kohler’s rule and the two-band model for \textbf{a)} S$_{3m}$, and \textbf{b)} S$_{20m}$. The plots show good agreement between simple extended Kohler's rule and fitted two-band model.}
    \label{fig5}
\end{figure}
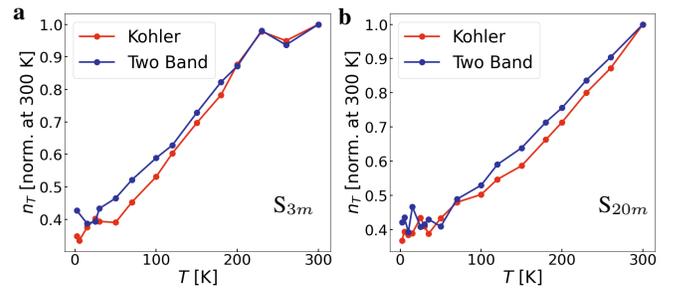

To find the effect of irradiation on the carrier concentration and the mobility \figref{fig4}, the longitudinal and transverse conductivity data were fitted with a two-band model. The details of the data extraction from these measurements are discussed in the SI. The compensation temperature increased to 160 K and 155 K for S$_{3m}$ and S$_{20m}$, respectively, whereas it is 110 K for pristine TaP \cite{TaP}. \figref{fig5} shows the normalized carrier concentration $n_T$ extracted from the simple extended Kohler's rule. The consistent temperature-dependent trend observed in our analysis, in accordance with the fitted two-band model, serves as compelling evidence for the accuracy and validity of our findings (see SI for detailed calculation).

In summary, we have addressed a novel method for ultra-fine Fermi level tuning in the prototype bulk WSM TaP using high-energy hydrogen implantation facilitated by accelerator-based techniques. Our transport measurements demonstrate a successful increment in the charge neutral point temperature through precise tuning of the Fermi level in proximity to the Weyl nodes. These experimental results are achievable by the guidance from DFT calculations. The approach presented in this study represents a highly controllable and universally applicable technique for fine-tuning the Fermi level and quantum orderings in bulk quantum materials. By enabling precise control over the Fermi level in bulk crystals, the work could open up new possibilities for exploring and manipulating the unique properties of topological WSM and other quantum materials. This advancement holds significant promise for the development of future quantum technologies and applications.

\begin{acknowledgments}
\noindent \textbf{Acknowledgements}\\
The authors Manasi Mandal and Abhijatmedhi Chotrattanapituk contributed equally to this work. MM and AC acknowledge support from the US Department of Energy (DOE), Office of Science (SC), Basic Energy Sciences (BES), Award No. DE-SC0020148. TN is supported by NSF Designing Materials to Revolutionize and Engineer our Future (DMREF) Program with Award No. DMR-2118448. TB is supported by the NSF Convergence Accelerator Award No. 2235945. JL acknowledges support by NSF, DMR-2132647. ML acknowledges the support from Aisin Inc, the Class of 1947 Career Development Professor Chair, and the support from R. Wachnik.  
\end{acknowledgments}

\bibliographystyle{apsrev4-2}
\bibliography{references} 

\onecolumngrid
\appendix

\clearpage
\renewcommand{\thefigure}{S\arabic{figure}}
\setcounter{figure}{0}
\section{Supplementary Information}

\subsection{DFT}

The density functional theory (DFT) \cite{1,2} calculations are performed using the Vienna ab initio simulation package (VASP) \cite{3,4}. Generalized gradient approximation (GGA) in the form of Perdew-Burke-Ernzerhof (PBE) \cite{5} is used to treat the exchange-correlation interactions. Core and valence electrons are treated by the projector augmented wave (PAW) method \cite{6} and a plane wave basis, respectively. A tight-binding Hamiltonian is built from the DFT results using the Wannier90 package \cite{7}. The tight-binding Hamiltonian is used to find Weyl nodes with the help of the WannierTools package \cite{8}. In doped TaP, the concentration of H is around 2$\times$10$^{21}$ cm$^{-3}$.

\subsection{Theoretical Model of H-doped TaP}

\begin{figure}[ht!]
    \centering
    \begin{tikzpicture}
        \node[anchor=north west] (image) at (0,0) 
        {\includegraphics[width=0.8\columnwidth]{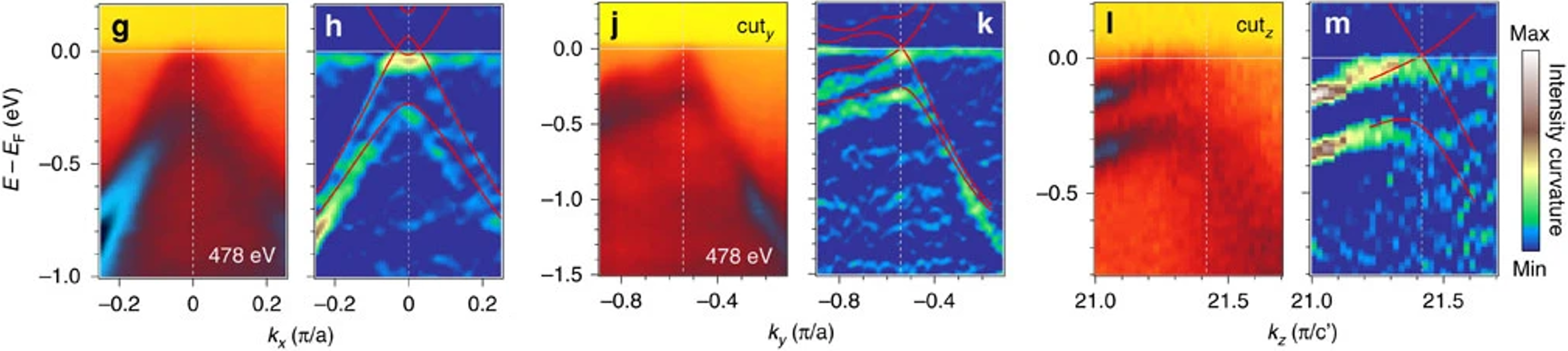}};
        \node[anchor=north west, scale = 1.2] (a) at (0,0) {\textbf{a}};
    \end{tikzpicture}
    \begin{tikzpicture}
        \node[anchor=north west] (image) at (0,0) 
        {\includegraphics[width=0.8\columnwidth]{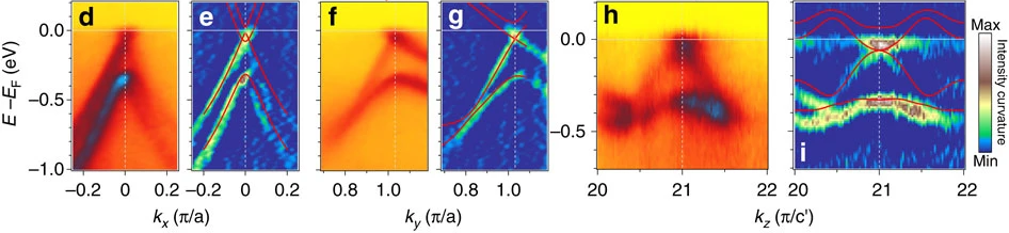}};
        \node[anchor=north west, scale = 1.2] (b) at (0,0) {\textbf{b}};
    \end{tikzpicture}
    \begin{tikzpicture}
        \node[anchor=north west] (image) at (0,0) 
        {\includegraphics[width=0.3\columnwidth]{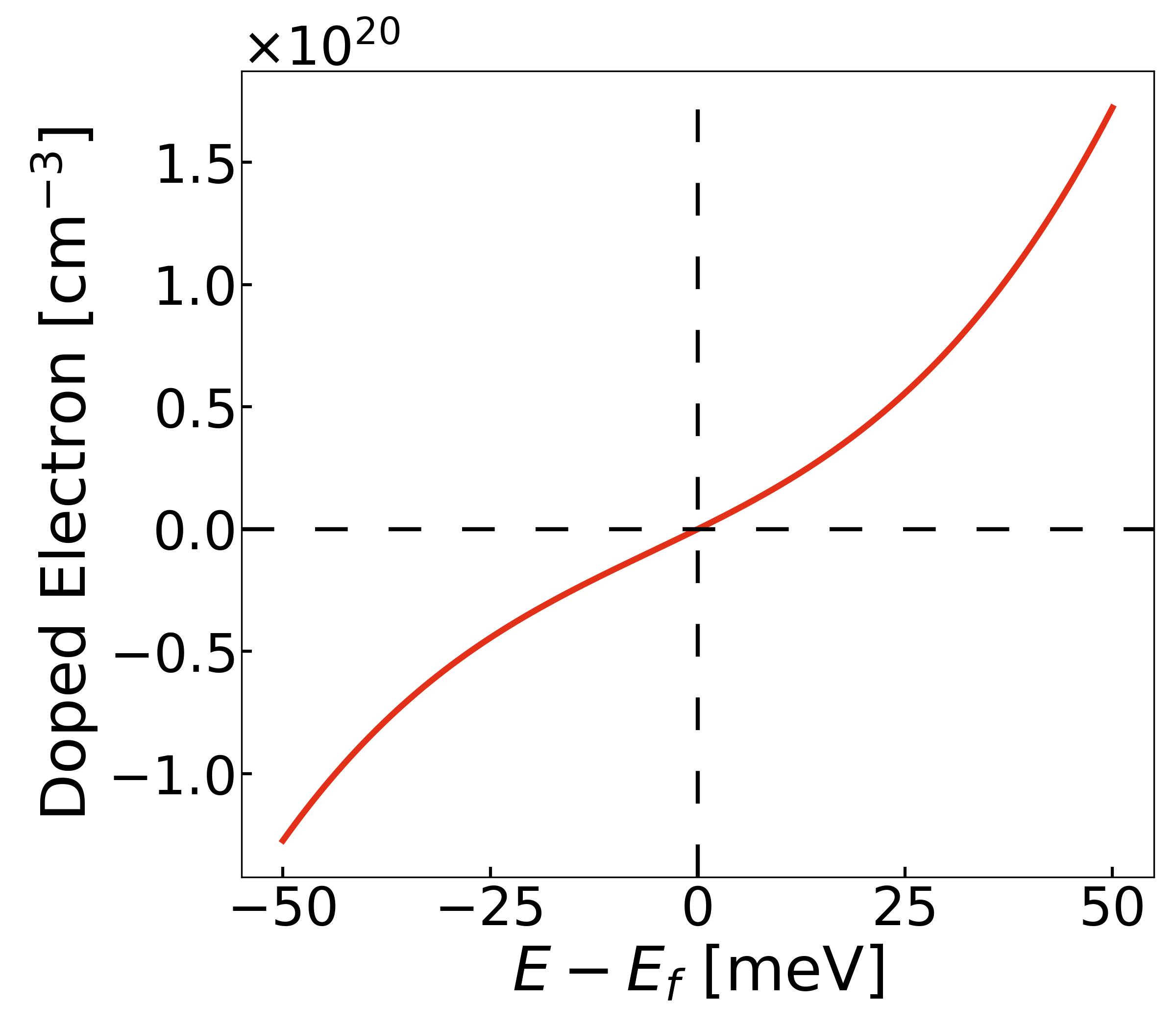}};
        \node[anchor=north west, scale = 1.2] (c) at (0,0) {\textbf{c}};
    \end{tikzpicture}
    \caption{Angle-resolved photoemission spectroscopy (ARPES) intensity plot and its curvature plot with calculated bands for TaP passing through \textbf{a} W1, and \textbf{b} W2 Weyl nodes (recreated from \cite{TaP_model}). \textbf{c} The plot of the equation \eqref{eq_model7} for the range $E \in [-50, 50]$ meV.}
    \label{figS1}
\end{figure}

The analysis was initiated with some key aspects of TaP to set up appropriate assumptions for the model. In TaP conventional unit cell, there are, in total, 24 Weyl nodes, of which 16 are W1 type with energy 24 meV above Fermi level while the remaining 8 are W2 type with energy 40 meV below Fermi level. The dispersion is approximately linear near the nodes, especially the regions between the two node types’ energies. Furthermore, since the dispersion generally does not change abruptly over a small region in momentum space, we would approximate all Weyl nodes’ dispersions with tilted 4-dimensional cones. This means the constant energy surface near nodes’ levels is approximately ellipsoids. To further simplify the model, we considered the experimental results of \cite{TaP_model}. We adapted \figref{figS1}a for W1 and \figref{figS1}b for W2 from figures 2 and 3 of the article \cite{TaP_model}. The cones are slightly tilted for both W1 and W2. Furthermore, the dispersion slopes of principal directions are different as well. However, because of the large uncertainty of slopes one could extract from the angle-resolved photoemission spectroscopy (ARPES) experimental results, and the order of doping magnitude is what we are concerned about (We ignored the red lines in these figures since they are from the ab initio calculation). The nodes were approximated with upright but distorted cones, and their dispersion slopes were extracted from the available ARPES figure of each node type.

With this approximated model, the constant energy (E) surfaces are ellipsoids with semi-principal axes
\begin{equation}\label{eq_model1}   
    r_{x,y,z} = \left| \vec{k} - \vec{k}_0 \right| = \frac{\left| \mathbf{E} - \mathbf{E}_0 \right|}{d_{x,y,z}}.
\end{equation}
Here, $E_0$ is the node’s energy level, $\vec{k}_0$ is the node’s location, and $d$ is the dispersion slope. This implies that, for each node, the energy increment from the Fermi level ($E_f$) to $E$ would lead to an expansion in the volume of momentum space located below the Fermi level by
\begin{equation}\label{eq_model2}
\begin{aligned}
    \left( \frac{4}{3} \right) \pi \dfrac{\left( E - E_0 \right)^3}{d_xd_yd_z} - \left( \frac{4}{3} \right) \pi \dfrac{\left( E_f - E_0 \right)^3}{d_xd_yd_z} = \frac{4 \pi}{3 d_x d_y d_z} \left( \left( E - E_0 \right)^3 - \left( E_f - E_0 \right)^3 \right).
\end{aligned}
\end{equation}
The absolute value is omitted from energy differences with $E_0$, resulting in negative volumes for energies below $E_0$. This adjustment ensures the accurate representation of momentum space volume increments. Given the standard practice of expressing energy relative to the Fermi level, the formulas were subsequently simplified using this notation as follows:
\begin{equation}\label{eq_model3}
\begin{aligned}
        \left( \frac{4\pi}{3d_x d_y d_z} \right) \left( (E - E_0)^3 + E_0^3 \right).
\end{aligned}
\end{equation}

Consider energy levels $E_1$ and $E_2$ corresponding to the nodes of W1 and W2, respectively, where $E_1 = 24$ meV and $E_2 = -40$ meV. Assuming that $E$ is close to 0 ($E_f$) enough such that only the nodes contribute to the total momentum space volume increment. The total increment in momentum space volume (sixteen from W1 and eight from W2) would be
\begin{equation}\label{eq_model4}
\begin{aligned}
       \left( \frac{32\pi}{3} \right) \left(\frac{2}{{d_{1,x} d_{1,y} d_{1,z}}} \left( (E - E_1)^3 + E_1^3 \right) + \frac{1}{{d_{2,x} d_{2,y} d_{2,z}}} \left( (E - E_2)^3 + E_2^3 \right) \right),
\end{aligned}
\end{equation}
where $d_1$ and $d_2$ are average dispersion slopes of W1 and W2 nodes respectively. This signifies that the number of electronic states increases by
\begin{equation}\label{eq_model5}
\begin{aligned}
        2\left( \frac{V}{{8\pi^3}} \cdot \frac{32\pi}{3} \right) \left( \frac{2}{{d_{1,x} d_{1,y} d_{1,z}}} \left( (E - E_1)^3 + E_1^3 \right) + \frac{1}{{d_{2,x} d_{2,y} d_{2,z}}} \left( (E - E_2)^3 + E_2^3 \right) \right).
\end{aligned}    
\end{equation}
In an alternative perspective, the requirement arises for TaP to undergo doping with an additional electron concentration of
\begin{equation}
\begin{aligned}
    & \left( \frac{8}{{3\pi^2}} \right) \left( \frac{2}{{d_{1,x} d_{1,y} d_{1,z}}} \left( (E - E_1)^3 + E_1^3 \right) + \frac{1}{{d_{2,x} d_{2,y} d_{2,z}}} \left( (E - E_2)^3 + E_2^3 \right) \right).
\end{aligned}
\label{eq_model6}
\end{equation}

It can be observed from the extracted figures that
\begin{align*}
d_{1,x} &= 3.3 \text{ \AA eV}, d_{1,y} = 1.9 \text{ \AA eV}, d_{1,z} = 0.3 \text{ \AA eV}, \\
d_{2,x} &= 3.0 \text{ \AA eV}, d_{2,y} = 1.9 \text{ \AA eV}, d_{2,z} = 0.2 \text{ \AA eV}.
\end{align*}
Therefore, the doping electron concentration in cm$^{-3}$ of TaP for the desired Fermi level shift, E, in meV is
\begin{equation}
\begin{aligned}    
        \left( \frac{8 \times 10^{15}}{3\pi^2} \right) \left( \frac{2}{1.9} \left( (E - 24)^3 + 24^3 \right) + \frac{1}{1.1} \left( (E + 40)^3 - 40^3 \right) \right).
\end{aligned}
\label{eq_model7}
\end{equation}
The plot depicting the equation \eqref{eq_model7} for $E \in [-50, 50]$ meV is shown in \figref{figS1}c. By substituting $E$ with $E_1$, and $E_2$, the model predicts the doping electron concentration of 5.3$\times$10$^{19}$ cm$^{-3}$ for W1, and doping hole concentration of 8.6$\times$10$^{19}$ cm$^{-3}$ for W2, respectively.

\subsection{Ab initio Calculation of H-doped TaP}

\begin{figure}[ht!]
    \centering
    \begin{tikzpicture}
        \node[anchor=north west] (image) at (0,0) 
        {\includegraphics[width=0.3\columnwidth]{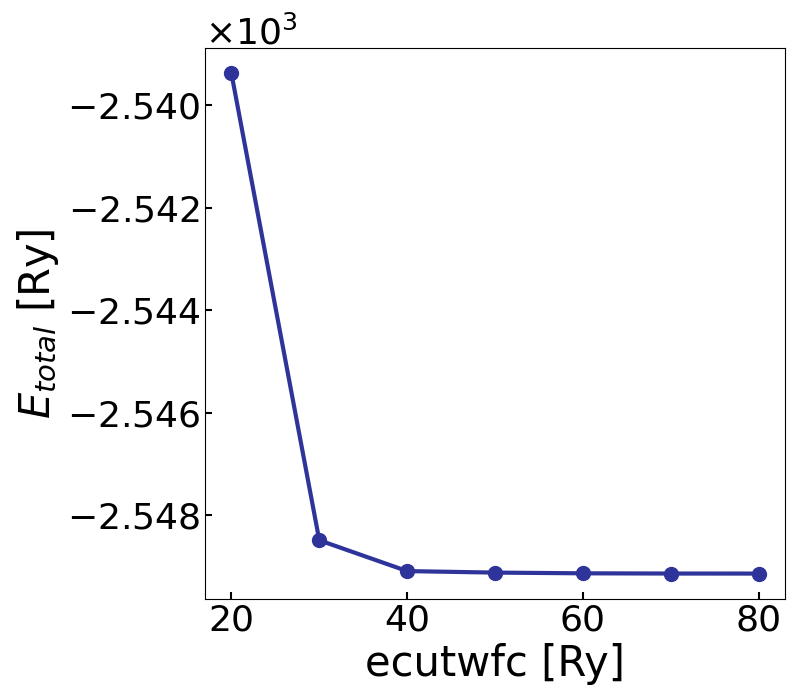}};
        \node[anchor=north west] (image) at (0.3\columnwidth,0) 
        {\includegraphics[width=0.3\columnwidth]{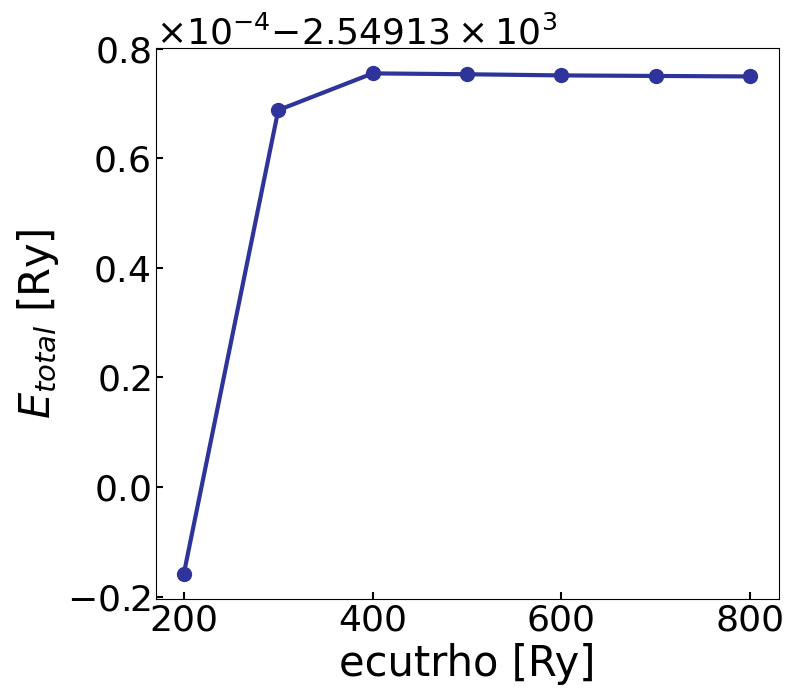}};
        \node[anchor=north west, scale = 1.2] (a) at (0,0) {\textbf{a}};
        \node[anchor=north west, scale = 1.2] (b) at (0.3\columnwidth,0) {\textbf{b}};
    \end{tikzpicture}
    \begin{tikzpicture}
        \node[anchor=north west] (image) at (0,0) 
        {\includegraphics[width=0.3\columnwidth]{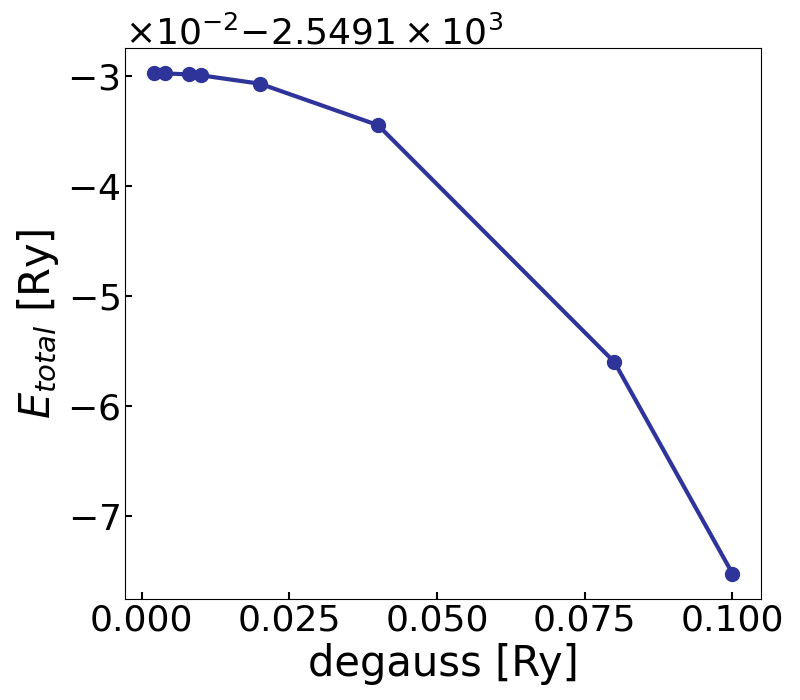}};
        \node[anchor=north west] (image) at (0.3\columnwidth,0) 
        {\includegraphics[width=0.3\columnwidth]{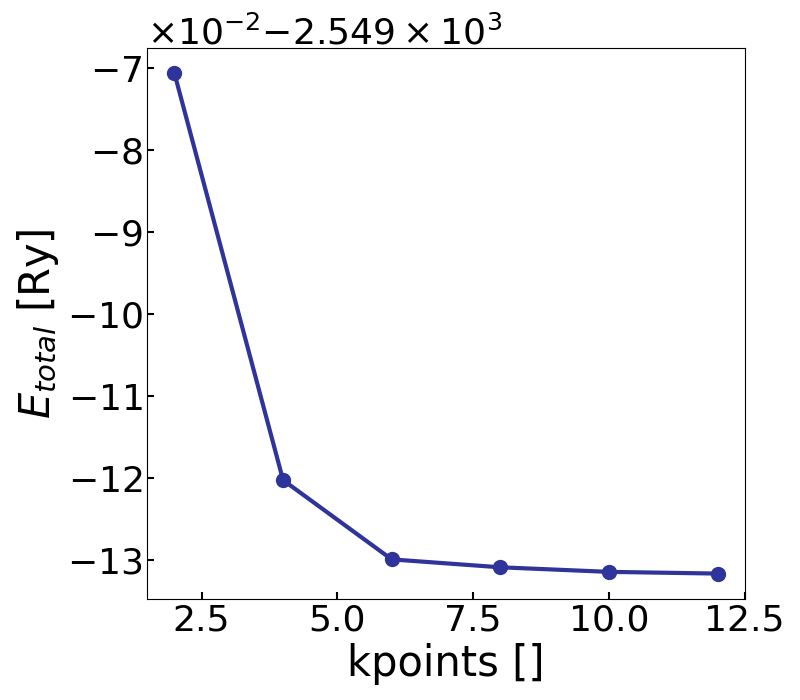}};
        \node[anchor=north west, scale = 1.2] (c) at (0,0) {\textbf{c}};
        \node[anchor=north west, scale = 1.2] (d) at (0.3\columnwidth,0) {\textbf{d}};
    \end{tikzpicture}
    \caption{Convergence test on simulation's hyperparameters: \textbf{a} kinetic energy cutoff for wavefunction (ecutwfc), \textbf{b} kinetic energy cutoff for charge density (ecutrho), \textbf{c} Gaussian spreading for Brillouin-zone integration in metals (degauss), and \textbf{d} number of k-point grids on each axis.}
    \label{figS2}
\end{figure}

To estimate the correlation between hydrogen doping concentration in TaP and the shift in the Fermi level, ab initio calculations were done. It involves determining the count of unoccupied states situated between the initial and adjusted Fermi levels. In this work, Quantum Espresso version 7.2 was used as the ab initio calculation program. The input material structure is obtained from the MaterialsProject in the primitive cell format \cite{ab1}. The pseudopotential files for Ta and P (both with scalar and full relativistic) are from the Quantum Espresso database. All pseudo potentials are Projector-Augmented Wave (PAW) types with the Perdew-Burke-Ernzerhof approximation method for solid (PBEsol). It was initiated by performing convergence test on simulation hyperparameters: kinetic energy cutoff for wavefunction (ecutwfc), and charge density (ecutrho), Gaussian spreading for Brillouin-zone integration in metals (degauss), and number of k-point grid on each axis (nk1, nk2, nk3). \figref{figS2} show the convergence test results.

\begin{figure}[ht!]
    \centering
    \begin{tikzpicture}
        \node[anchor=north west] (image) at (0,0) 
        {\includegraphics[width=0.3\columnwidth]{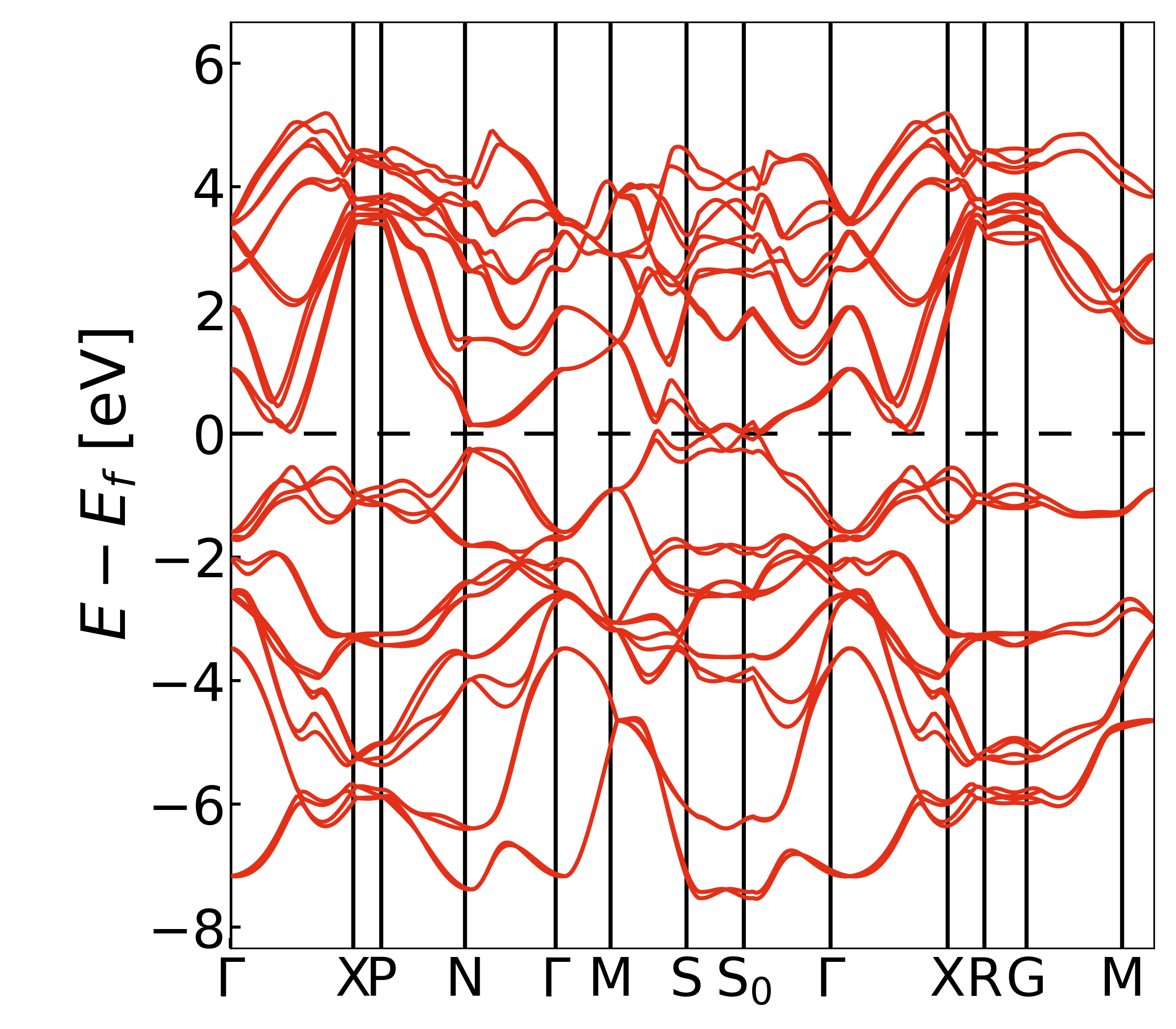}};
        \node[anchor=north west] (image) at (0.3\columnwidth,0) 
        {\includegraphics[width=0.3\columnwidth]{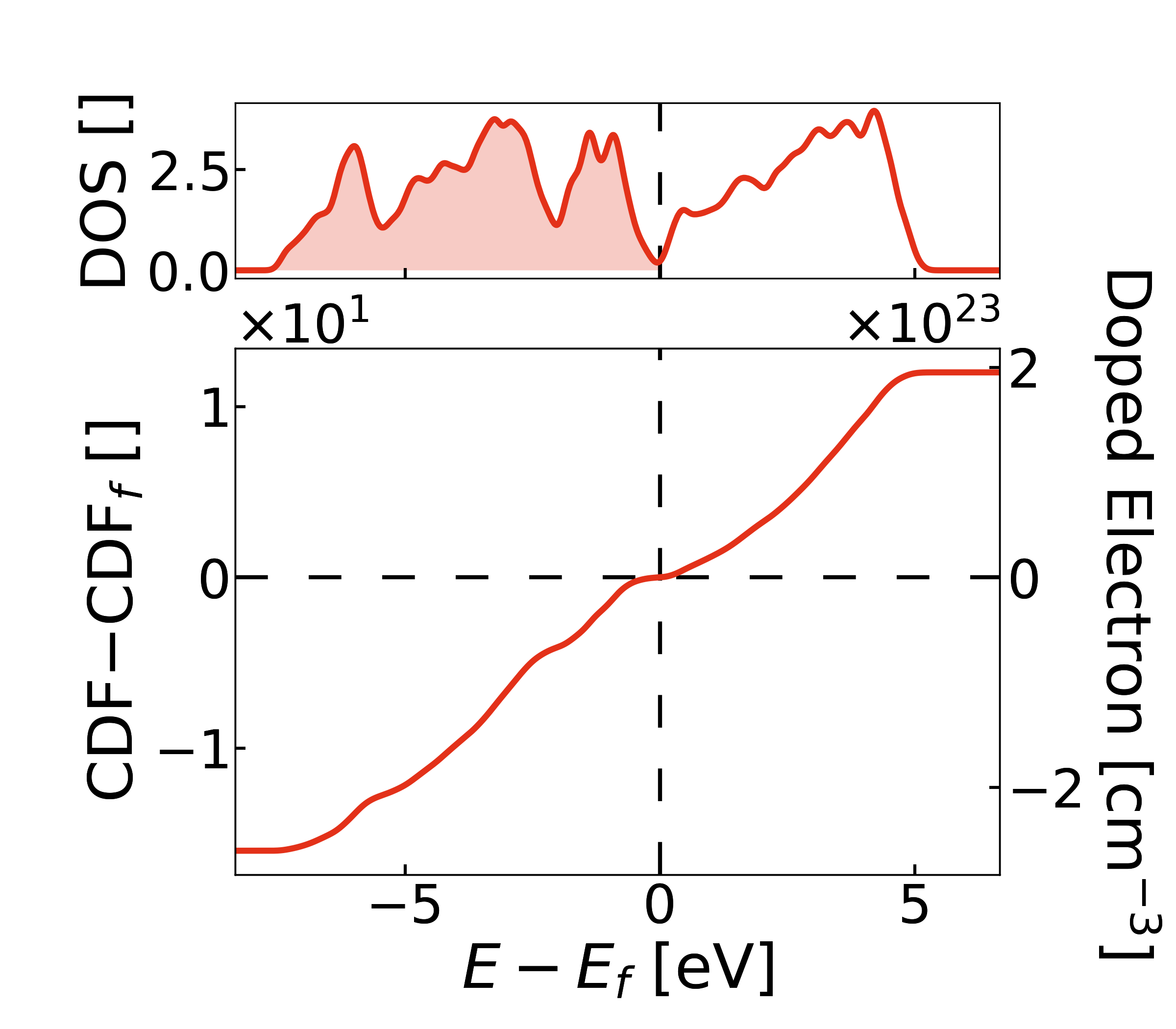}};
        \node[anchor=north west] (image) at (0.6\columnwidth,0) 
        {\includegraphics[width=0.3\columnwidth]{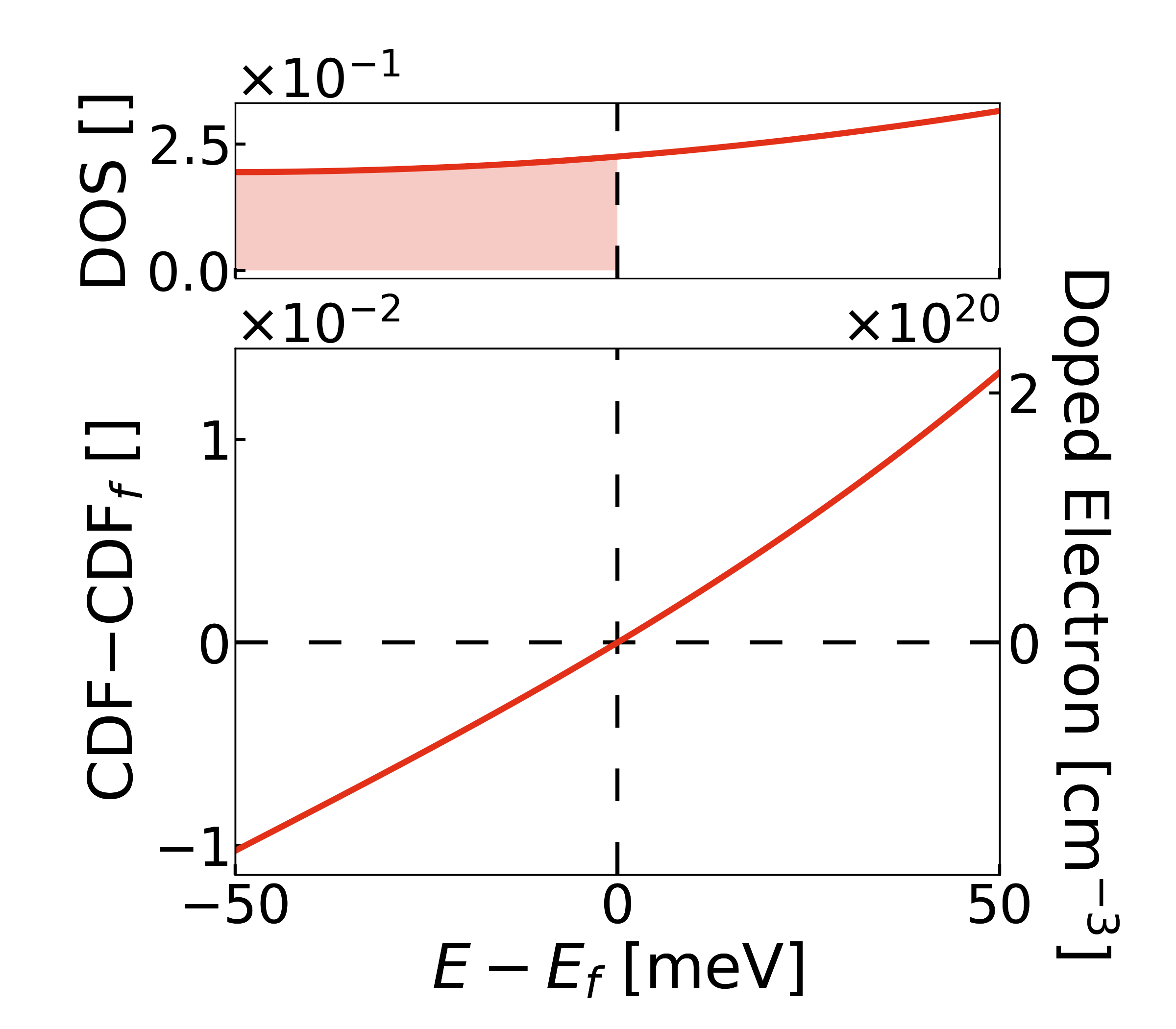}};
        \node[anchor=north west, scale = 1.2] (a) at (0,0) {\textbf{a}};
        \node[anchor=north west, scale = 1.2] (b) at (0.3\columnwidth,0) {\textbf{b}};
        \node[anchor=north west, scale = 1.2] (c) at (0.6\columnwidth,0) {\textbf{c}};
    \end{tikzpicture}
    \caption{\textbf{a} The electronic band structure of TaP. \textbf{b} The energy dependence of the density of states (\textbf{Top}), and number of states per unit cell (\textbf{Bottom}, left-hand axis) that needed to be filled with the doped electron to effectuate the shift of the Fermi level. The variation of doping electron concentration is shown by right-hand axis. \textbf{c} A magnified view of \textbf{b} for the range $E\in [-50, 50]$ meV with respect to Fermi level.}
    \label{figS3}
\end{figure}

Subsequently, structure relaxation was performed on the structure from the MaterialsProject to get a stable structure according to the pseudopotentials used. This stable structure was used in Quantum Espresso to calculate the self-consistent function, and non-self-consistent  calculation with high k-point density grids \cite{ab2}. Then, we perform density of states and band structure calculation pathway. The band structure and density of the states plot are shown in \figref{figS3}a and \figref{figS3}b, respectively. The doping relation was achieved by integrating the density of states starting from the Fermi level. This allowed the determination of the number of states per unit cell that needed to be filled with the doped electron to effectuate the shift of the Fermi level (shown in the bottom of \figref{figS3}b). Finally, the doping electron concentration was calculated by dividing the unit cell volume. Our calculation predicts the doping electron concentration of 9.7$\times$10$^{19}$ cm$^{-3}$ for W1, and doping hole concentration of 1.4$\times$10$^{20}$ cm$^{-3}$ for W2.

\subsection{SRIM and TRIM Calculation}
We employed the SRIM (stopping and range of ions in matter) and TRIM (transport of ions in matter) software \cite{SRIM} to simulate realistic irradiation conditions and investigate the defect profile. TRIM simulations provided us with the flexibility to adjust parameters such as target type, thickness, and incident ion beam energy. Through the TRIM simulations, we obtained valuable results that revealed the depth profile generated by both the incident ions and recoils. For our study, we deliberately chose H$^{-}$ ions with an energy of 20 keV to irradiate the TaP crystal. This selection was driven by our specific interest in doping the ions near the surface of the sample ($\sim$ 5000 \AA). By considering the beam current and aperture size, we calculated the incident ion flux. Furthermore, we accounted for the subtraction of secondary electrons and the total duration of irradiation, enabling us to determine the number of implanted ions in each sample. In our experimental setup, we utilized a 4 mm diameter beam to irradiate samples of size 2$\times$2 mm$^{2}$. The beam current for S$_{3m}$ was 140 nA, and for S$_{20m}$ and S$_{2h}$ was 170 nA. We varied the irradiation time accordingly to achieve the targeted doping concentration of $\approx$ 10$^{19}$ cm$^{-3}$, 10$^{20}$ cm$^{-3}$ and 10$^{21}$ cm$^{-3}$ for S$_{3m}$, S$_{20m}$ and S$_{2h}$ samples as per our mathematical and computational calculations. The total doping concentration in the sample is determined by multiplying the irradiation dose for each sample by the penetration depth. The irradiation dose is summarized in the main text (\tableref{tab1}).

\subsection{Carrier concentration and mobility}

\begin{figure}[ht!]
    \centering
    \begin{tikzpicture}
        \node[anchor=north west] (image) at (0,0) 
        {\includegraphics[width=0.3\columnwidth]{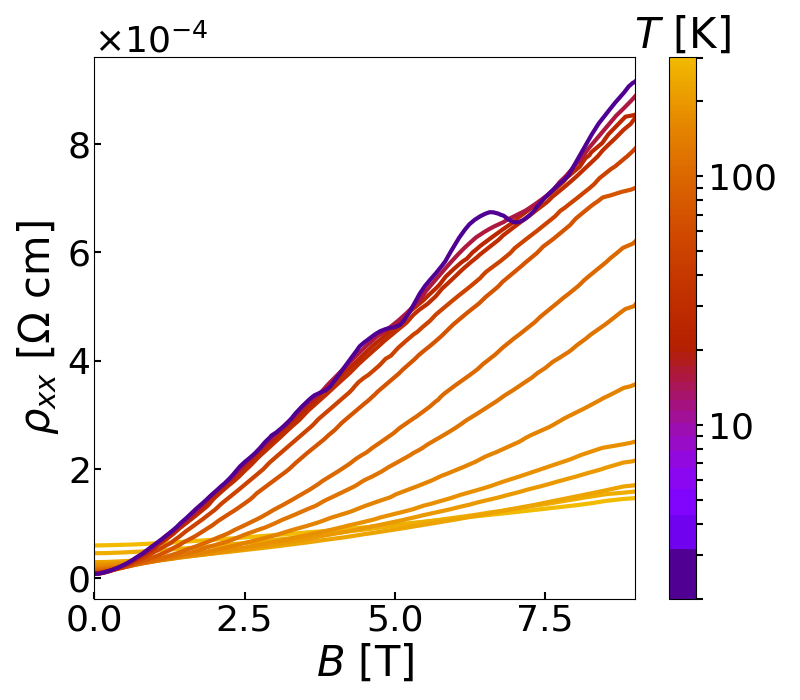}};
        \node[anchor=north west] (image) at (0.3\columnwidth,0) 
        {\includegraphics[width=0.3\columnwidth]{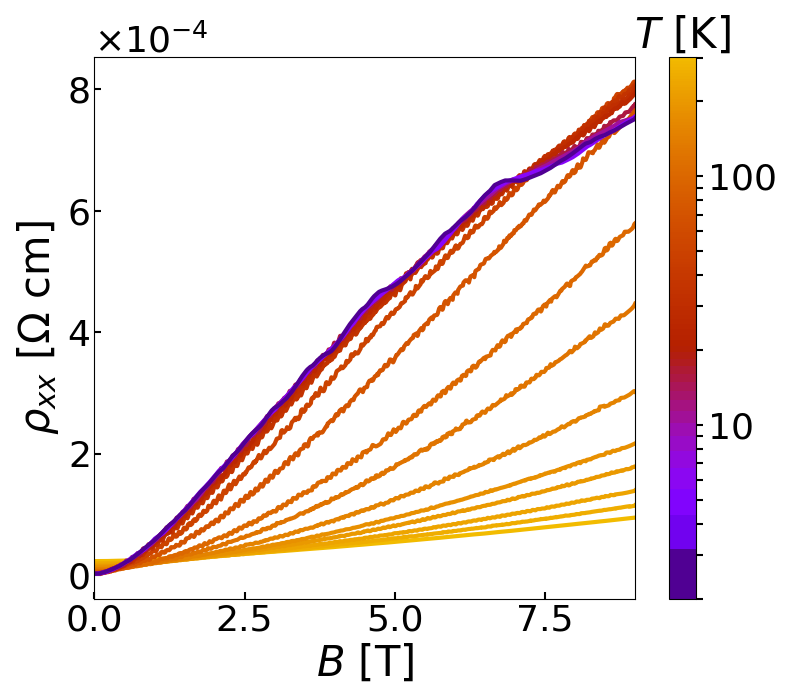}};
        \node[anchor=north west] (image) at (0.6\columnwidth,0) 
        {\includegraphics[width=0.3\columnwidth]{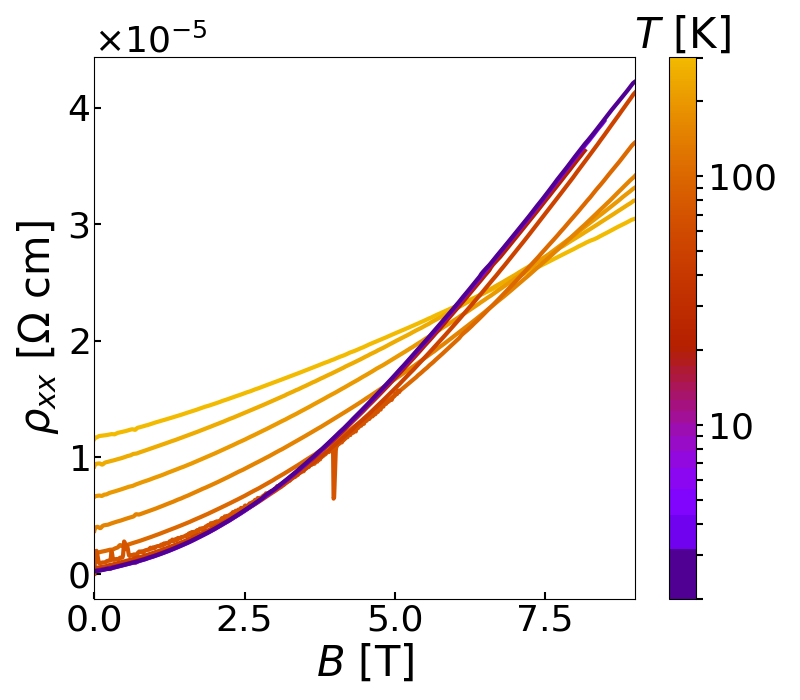}};
        \node[anchor=north west, scale = 1.2] (a) at (0,0) {\textbf{a}};
        \node[anchor=north west, scale = 1.2] (b) at (0.3\columnwidth,0) {\textbf{b}};
        \node[anchor=north west, scale = 1.2] (c) at (0.6\columnwidth,0) {\textbf{c}};

        \node[anchor=north west, scale = 1.2] at (0.05\columnwidth,-0.75) {S$_{3m}$};
        \node[anchor=north west, scale = 1.2] at (0.35\columnwidth,-0.75) {S$_{20m}$};
        \node[anchor=north west, scale = 1.2] at (0.65\columnwidth,-0.75) {S$_{2h}$};
    \end{tikzpicture}
    \begin{tikzpicture}
        \node[anchor=north west] (image) at (0,0) 
        {\includegraphics[width=0.3\columnwidth]{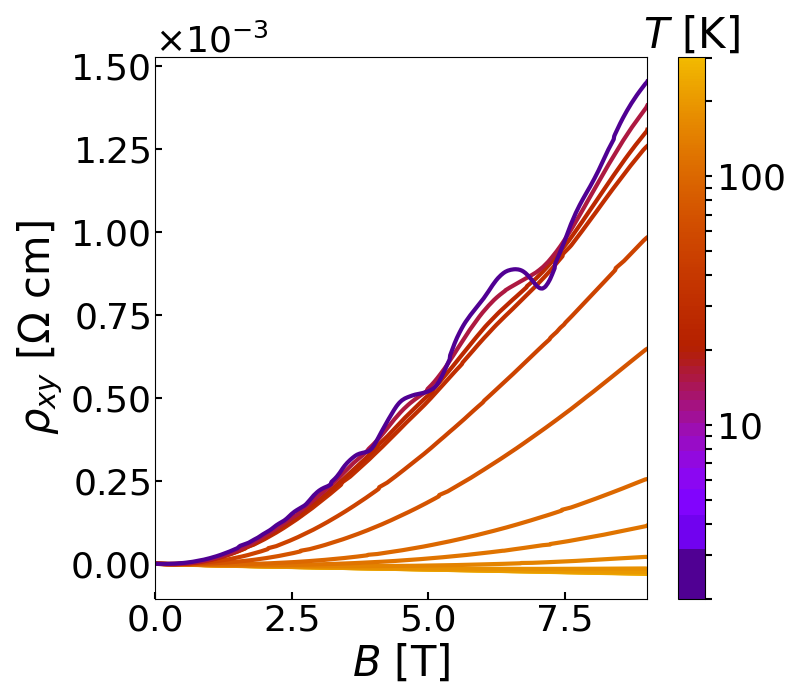}};
        \node[anchor=north west] (image) at (0.3\columnwidth,0) 
        {\includegraphics[width=0.3\columnwidth]{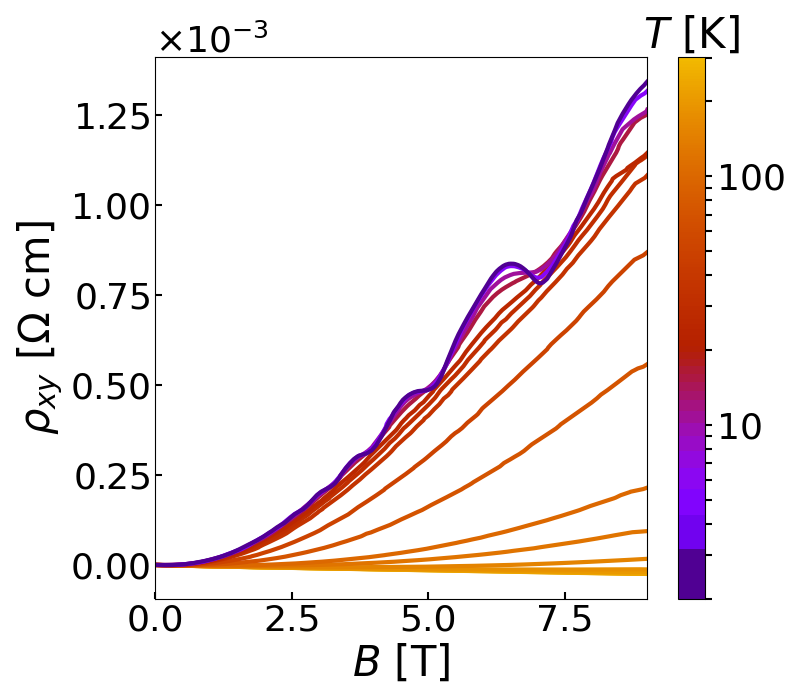}};
        \node[anchor=north west] (image) at (0.6\columnwidth,0) 
        {\includegraphics[width=0.3\columnwidth]{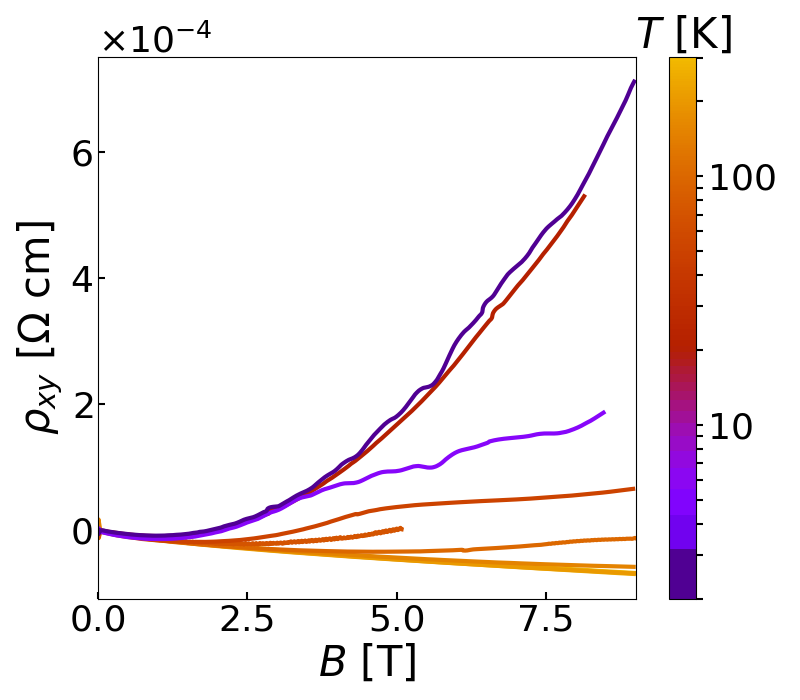}};
        \node[anchor=north west, scale = 1.2] (d) at (0,0) {\textbf{d}};
        \node[anchor=north west, scale = 1.2] (e) at (0.3\columnwidth,0) {\textbf{e}};
        \node[anchor=north west, scale = 1.2] (f) at (0.6\columnwidth,0) {\textbf{f}};

        \node[anchor=north west, scale = 1.2] at (0.075\columnwidth,-0.75) {S$_{3m}$};
        \node[anchor=north west, scale = 1.2] at (0.375\columnwidth,-0.75) {S$_{20m}$};
        \node[anchor=north west, scale = 1.2] at (0.65\columnwidth,-0.75) {S$_{2h}$};
    \end{tikzpicture}
    \caption{Magnetic field dependent of longitudinal resistivity, $\rho_{xx}$, at different temperatures up to 300 K for \textbf{a} S$_{3m}$, \textbf{b} S$_{20m}$, and \textbf{c} S$_{2h}$. \textbf{d}-\textbf{f} The same as \textbf{a}-\textbf{c}, respectively, but for $\rho_{xy}$.}
    \label{figS4}
\end{figure}

The electrical measurements of the samples were carried out using the electrical transport option (ETO) of the physical property measurement system (PPMS). We utilized the ETO with a six-probe configuration to obtain the required data, enabling simultaneous measurements of both $\rho_{xx}$ and $\rho_{xy}$. We employed a symmetric probe configuration to minimize the influence of manually made contact misalignment. In this setup, the longitudinal resistivity $\rho_{xx}$ exhibited symmetry with respect to the applied magnetic field, while the transverse resistivity $\rho_{xy}$ displayed antisymmetry. We employed the following equations to achieve the symmetric probe:
\begin{equation}
    \rho_{xx}(B) = \frac{\rho_{xx}(+B) + \rho_{xx}(-B)}{2},\, 
    \rho_{xy}(B) = \frac{\rho_{xy}(+B) - \rho_{xy}(-B)}{2}.
\end{equation}

The magnetic field dependence of $\rho_{xx}$ and $\rho_{xy}$ at different temperatures is illustrated in \figref{figS4}. Shubnikov-de Haas (SdH) oscillations can be observed in both $\rho_{xx}$ and $\rho_{xy}$ at low temperatures, up to 15 K for S$_{3m}$, and 10 K for S$_{20m}$. However, no such oscillations are observed for S$_{2h}$ sample, indicating that the presence of high disorder suppresses the quantum oscillations. Thus, our further experimental analysis primarily focuses on the S$_{3m}$, and S$_{20m}$ samples as our main interest lies in precise Fermi level engineering in a topological Weyl semimetal.

\begin{figure}[ht!]
    \centering
    \begin{tikzpicture}
        \node[anchor=north west] (image) at (0,0) 
        {\includegraphics[width=0.6\columnwidth]{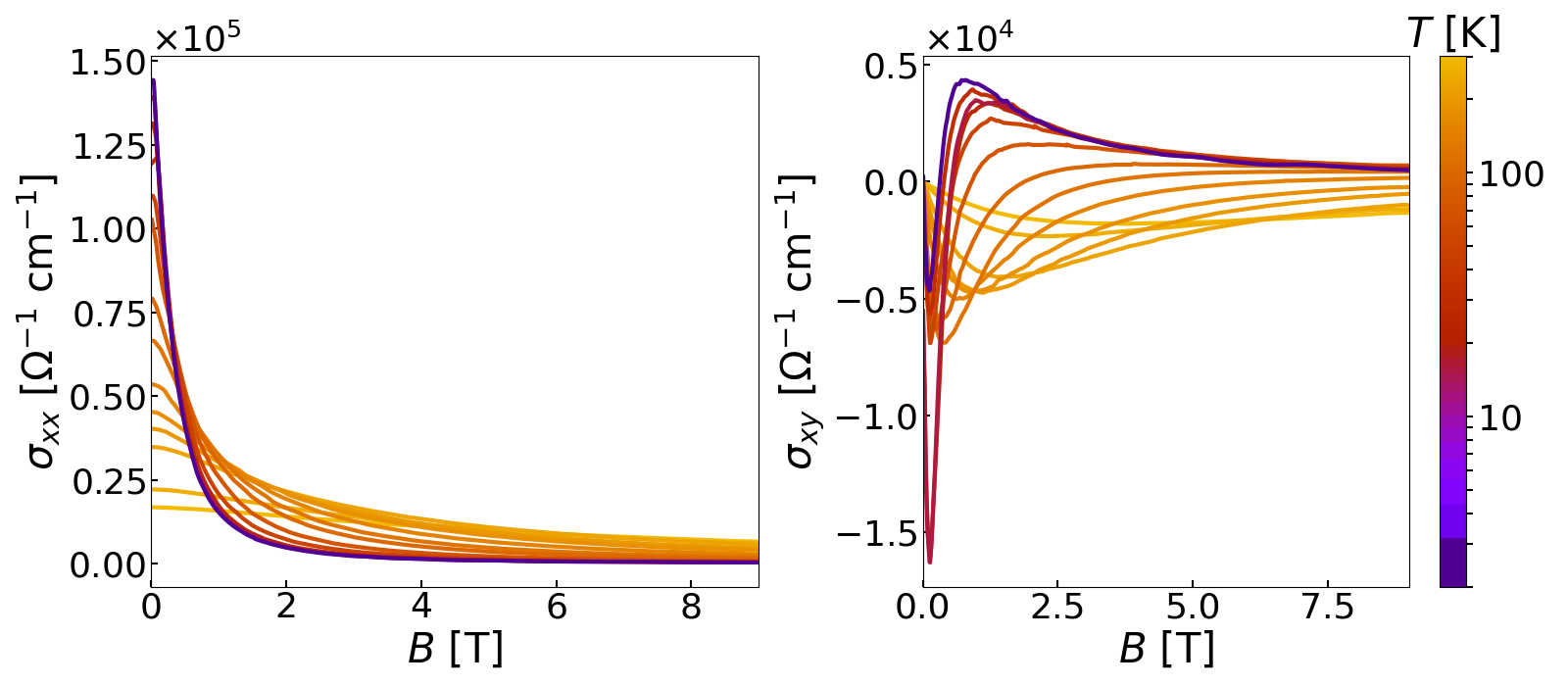}};
        \node[anchor=north west, scale = 1.2] (a) at (0,0) {\textbf{a}};
        \node[anchor=north west, scale = 1.2] (b) at (0.3\columnwidth,0) {\textbf{b}};

        \node[anchor=north west, scale = 1.2] at (0.225\columnwidth,-1) {S$_{3m}$};
    \end{tikzpicture}
    \begin{tikzpicture}
        \node[anchor=north west] (image) at (0,0) 
        {\includegraphics[width=0.6\columnwidth]{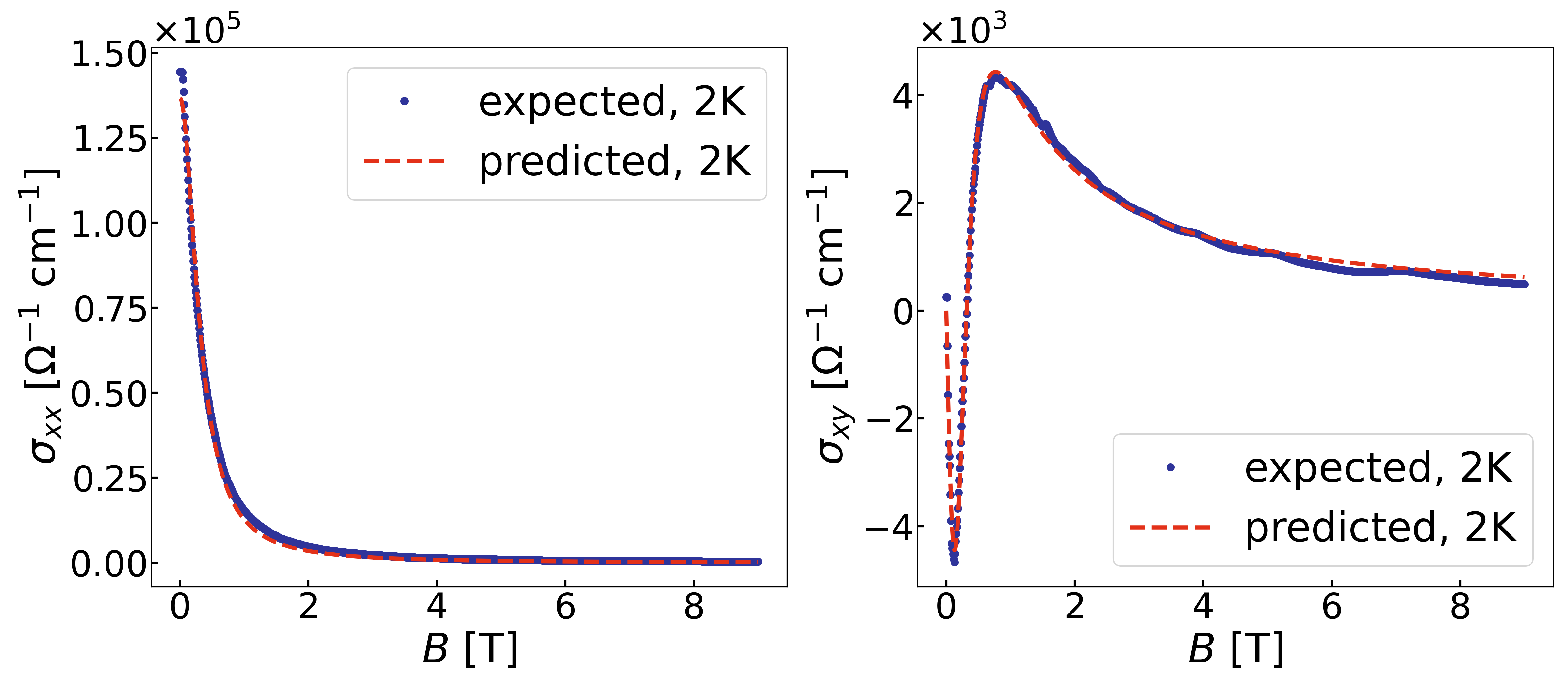}};
        \node[anchor=north west, scale = 1.2] (c) at (0,0) {\textbf{c}};
        \node[anchor=north west, scale = 1.2] (d) at (0.3\columnwidth,0) {\textbf{d}};
    \end{tikzpicture}
    \begin{tikzpicture}
        \node[anchor=north west] (image) at (0,0) 
        {\includegraphics[width=0.6\columnwidth]{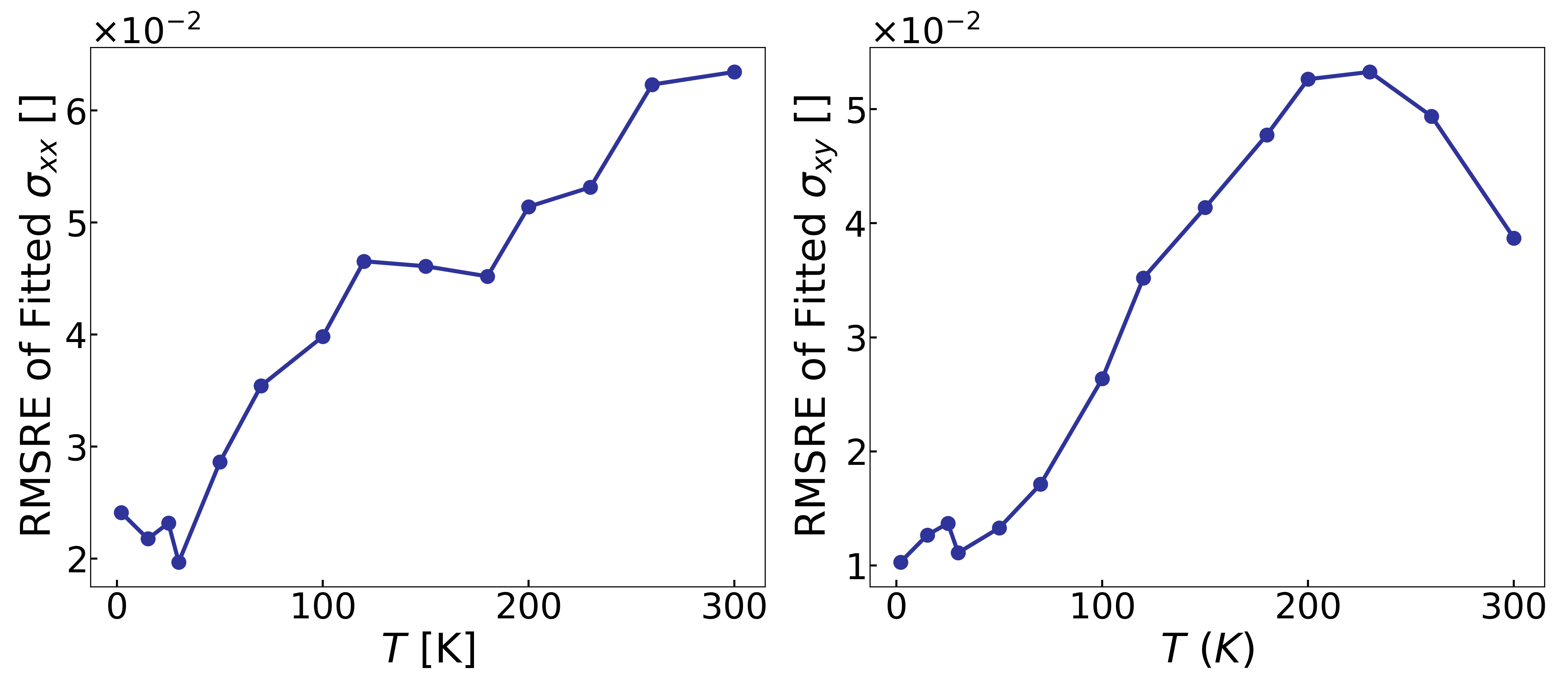}};
        \node[anchor=north west, scale = 1.2] (e) at (0,0) {\textbf{e}};
        \node[anchor=north west, scale = 1.2] (f) at (0.3\columnwidth,0) {\textbf{f}};
    \end{tikzpicture}
    \caption{Temperature variation of \textbf{a} longitudinal conductivity, $\sigma_{xx}$, and \textbf{b} transverse conductivity, $\sigma_{xy}$, for S$_{3m}$. The fitting results of the curves in \textbf{a}, and \textbf{b} at 2 K with two-band model are shown in \textbf{c}, and \textbf{d}, respectively. The fitting RMSRE (Root Mean Squared Relative Error) values at different temperatures for \textbf{e} $\sigma_{xx}$, and \textbf{f} $\sigma_{xy}$.}
    \label{figS5}
\end{figure}

\begin{figure}[ht!]
    \centering
    \begin{tikzpicture}
        \node[anchor=north west] (image) at (0,0) 
        {\includegraphics[width=0.6\columnwidth]{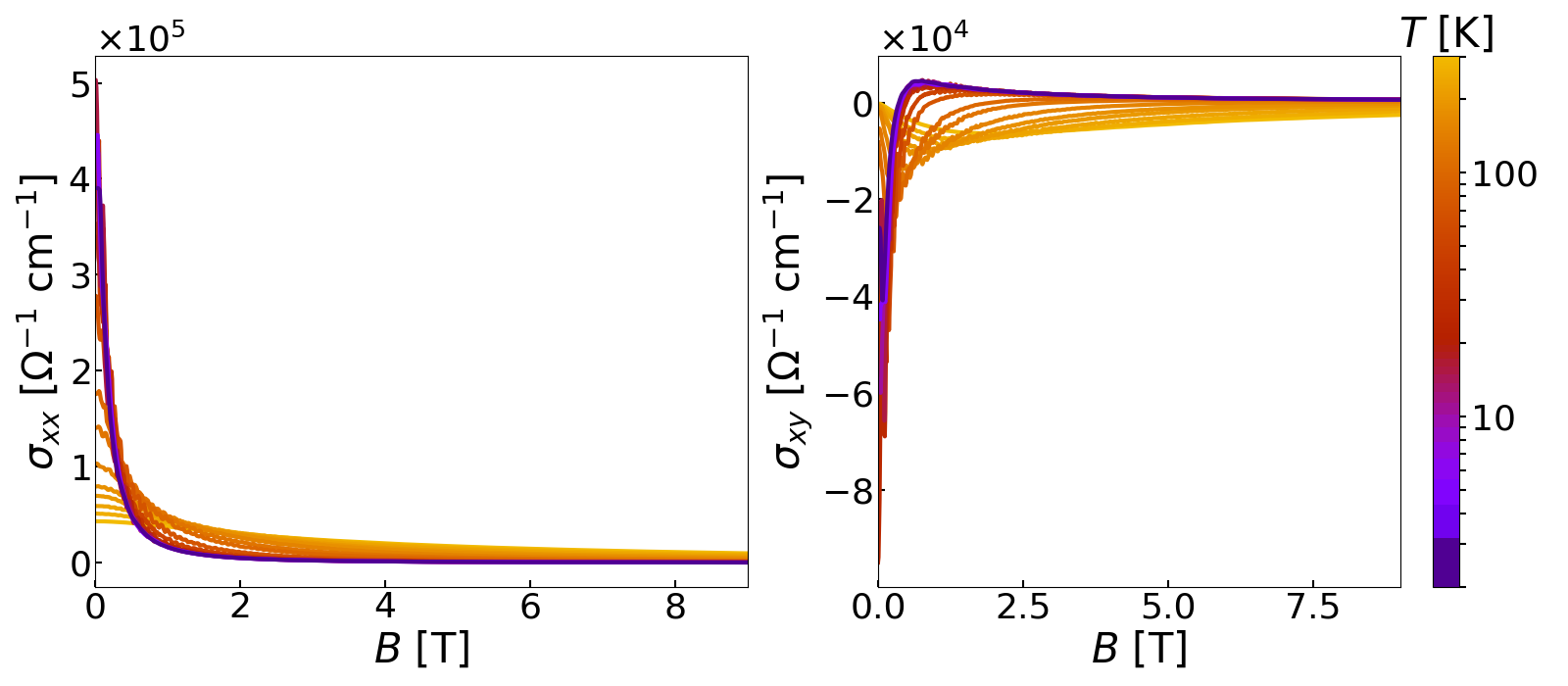}};
        \node[anchor=north west, scale = 1.2] (a) at (0,0) {\textbf{a}};
        \node[anchor=north west, scale = 1.2] (b) at (0.3\columnwidth,0) {\textbf{b}};

        \node[anchor=north west, scale = 1.2] at (0.225\columnwidth,-1) {S$_{20m}$};
    \end{tikzpicture}
    \begin{tikzpicture}
        \node[anchor=north west] (image) at (0,0) 
        {\includegraphics[width=0.6\columnwidth]{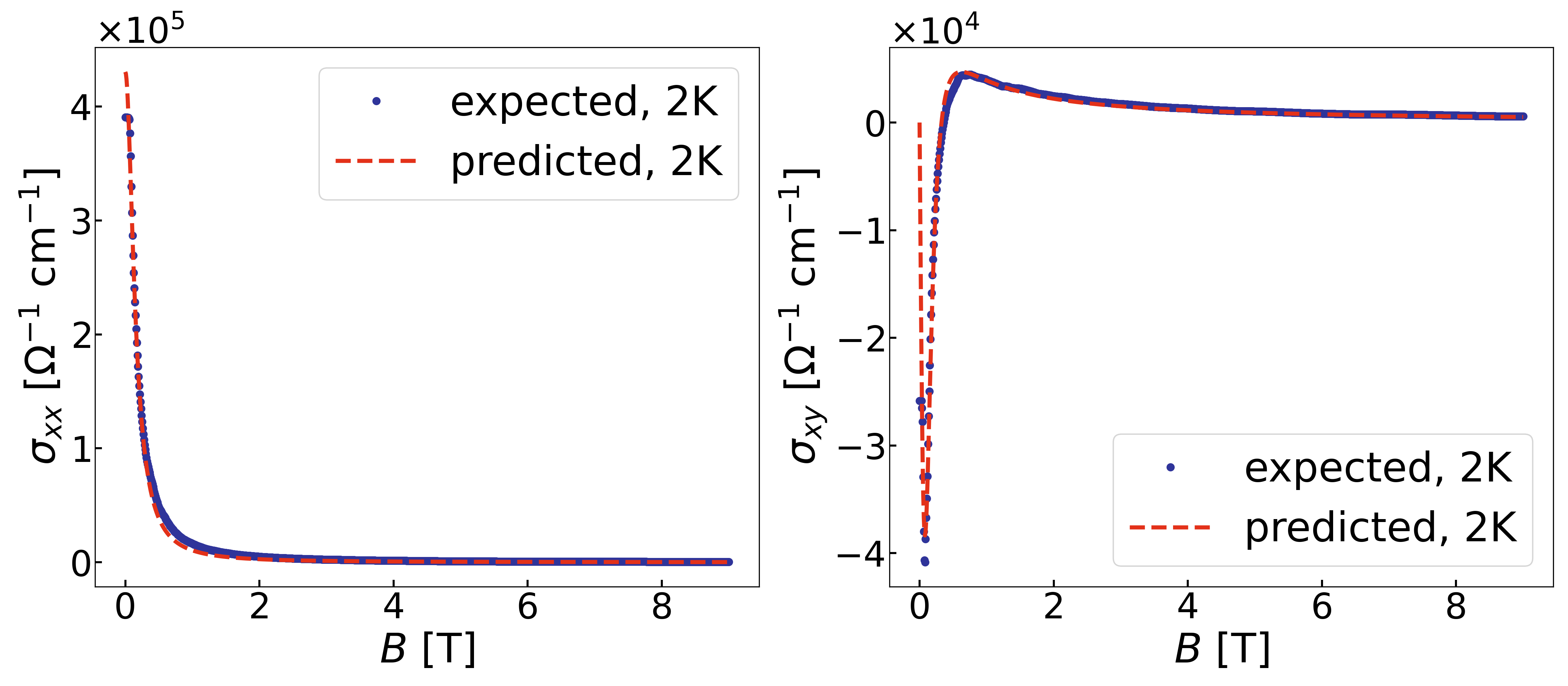}};
        \node[anchor=north west, scale = 1.2] (c) at (0,0) {\textbf{c}};
        \node[anchor=north west, scale = 1.2] (d) at (0.3\columnwidth,0) {\textbf{d}};
    \end{tikzpicture}
    \begin{tikzpicture}
        \node[anchor=north west] (image) at (0,0) 
        {\includegraphics[width=0.6\columnwidth]{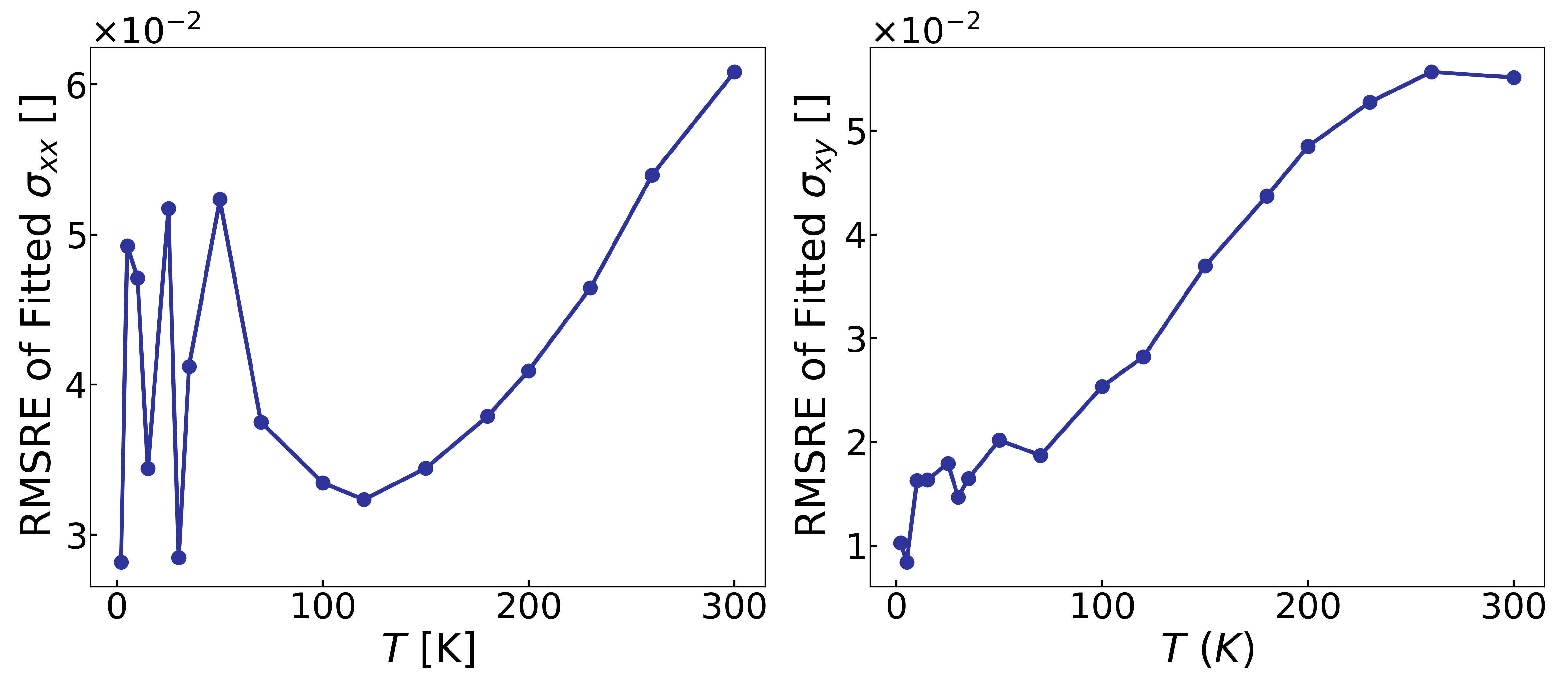}};
        \node[anchor=north west, scale = 1.2] (e) at (0,0) {\textbf{e}};
        \node[anchor=north west, scale = 1.2] (f) at (0.3\columnwidth,0) {\textbf{f}};
    \end{tikzpicture}
    \caption{Temperature variation of \textbf{a} longitudinal conductivity, $\sigma_{xx}$, and \textbf{b} transverse conductivity, $\sigma_{xy}$, for S$_{20m}$. The fitting results of the curves in \textbf{a}, and \textbf{b} at 2 K with two-band model are shown in \textbf{c}, and \textbf{d}, respectively. The fitting RMSRE (Root Mean Squared Relative Error) values at different temperatures for \textbf{e} $\sigma_{xx}$, and \textbf{f} $\sigma_{xy}$.}
    \label{figS6}
\end{figure}

To investigate the impact of irradiation on carrier concentration and mobility, we calculated the longitudinal conductivity, $\sigma_{xx}$, and transverse conductivity, $\sigma_{xy}$, using the following equations:
\begin{equation}
    \sigma_{xx} = \frac{\rho_{xx}} {{\rho_{xx}^2 + \rho_{xy}^2}}, \,
    \sigma_{xy} = -\frac{\rho_{xy}} {{\rho_{xx}^2 + \rho_{xy}^2}}.
\end{equation}
The field dependence of $\sigma_{xx}$ and $\sigma_{xy}$ at different temperatures are illustrated in \figref{figS5}a and \figref{figS5}b for S$_{3m}$, respectively, and \figref{figS6}a and \figref{figS6}b for S$_{20m}$, respectively. To extract the carrier concentration and mobility, we simultaneously fit the $\sigma_{xx}$ and $\sigma_{xy}$ data using two-band model defined by:
\begin{equation}
    \sigma_{xx} = \frac{n_{e}\mu_{e}e}{1+(\mu_{e}B)^2} +\frac{n_{h}\mu_{h}e}{1+(\mu_{h}B)^2}, \,   
    \sigma_{xy} = \left( \frac{n_{h}\mu_{h}^2}{1+(\mu_{h}B)^2} - \frac{n_{e}\mu_{e}^2}{1+(\mu_{e}B)^2} \right)eB
\end{equation}

\figref{figS5}c-d, and \figref{figS6}c-d show example fitting results of the curves for 2 K temperature for S$_{3m}$, and S$_{20m}$, respectively. The root mean squared relative error (RMSRE) linked to the fitting procedure at different temperatures using the two-band model is displayed in \figref{figS5}e-f, and \figref{figS6}e-f for S$_{3m}$, and S$_{20m}$, respectively. The main text includes the temperature variation of electron and hole concentration, along with their corresponding mobility, for each doping level and compares them with pristine TaP (\figref{fig4}). Our findings reveal that H$^{-}$-ion implantation significantly impacts the Fermi surface by considerably elevating the electron-hole compensation temperature despite the mobility being reduced by an order of magnitude.

\subsection{Analysis of Quantum Oscillation}
\begin{figure}[ht!]
    \centering
    \begin{tikzpicture}
        \node[anchor=north west] (image) at (0,0) 
        {\includegraphics[width=0.3\columnwidth]{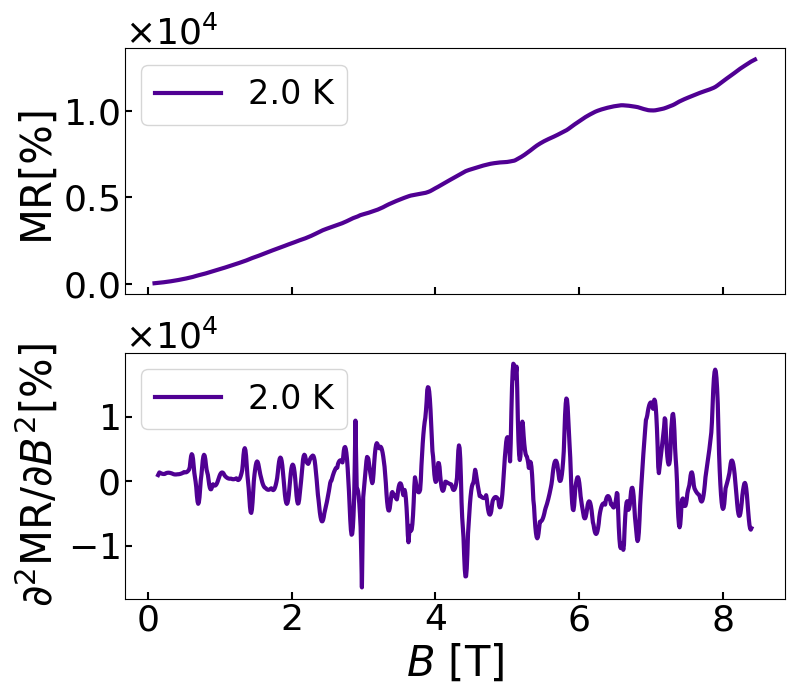}};
        \node[anchor=north west] (image) at (0.3\columnwidth,0) 
        {\includegraphics[width=0.3\columnwidth]{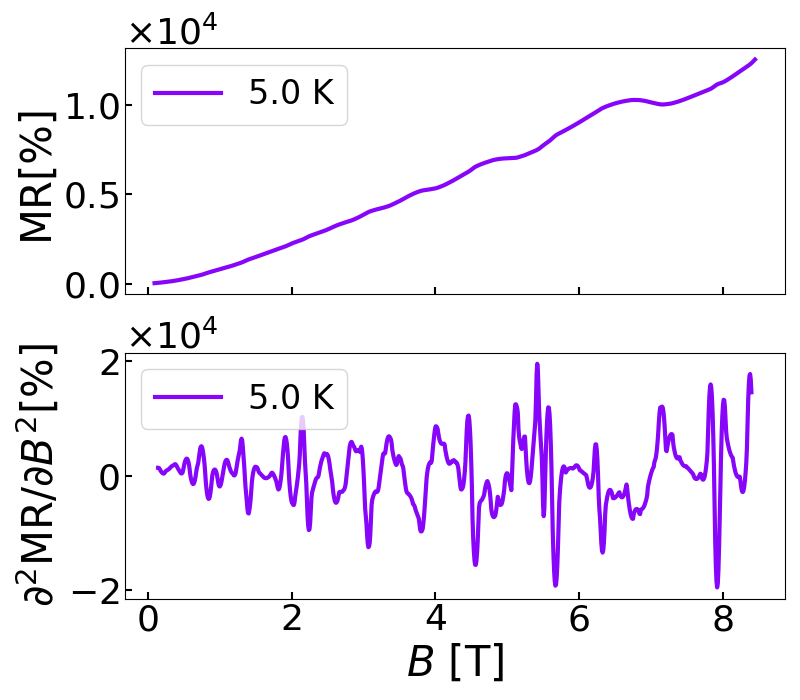}};
        \node[anchor=north west] (image) at (0.6\columnwidth,0) 
        {\includegraphics[width=0.3\columnwidth]{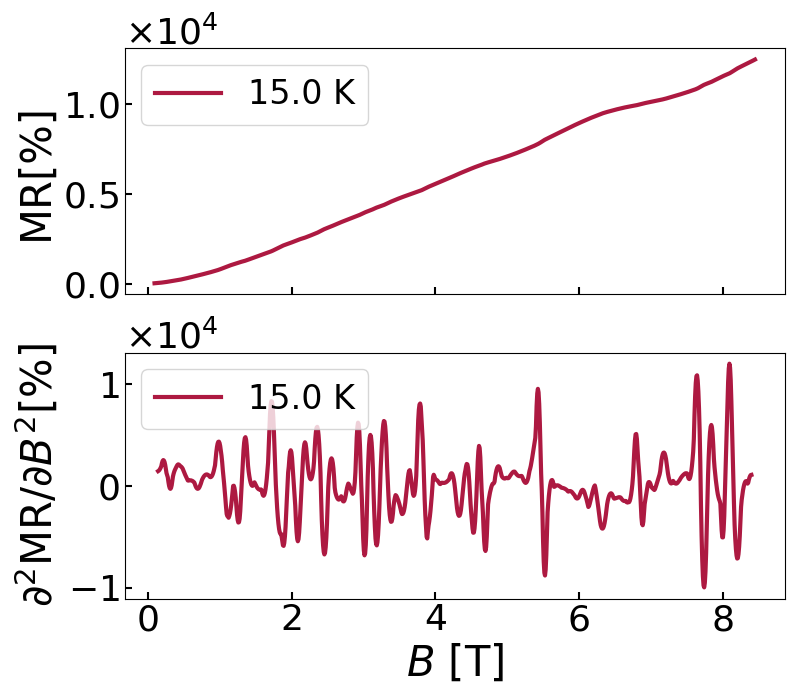}};
        \node[anchor=north west, scale = 1.2] (a) at (0,0) {\textbf{a}};
        \node[anchor=north west, scale = 1.2] (b) at (0.3\columnwidth,0) {\textbf{b}};
        \node[anchor=north west, scale = 1.2] (c) at (0.6\columnwidth,0) {\textbf{c}};

        \node[anchor=north west, scale = 1.2] at (0.225\columnwidth,-1.25) {S$_{3m}$};
    \end{tikzpicture}
    \begin{tikzpicture}
        \node[anchor=north west] (image) at (0,0) 
        {\includegraphics[width=0.3\columnwidth]{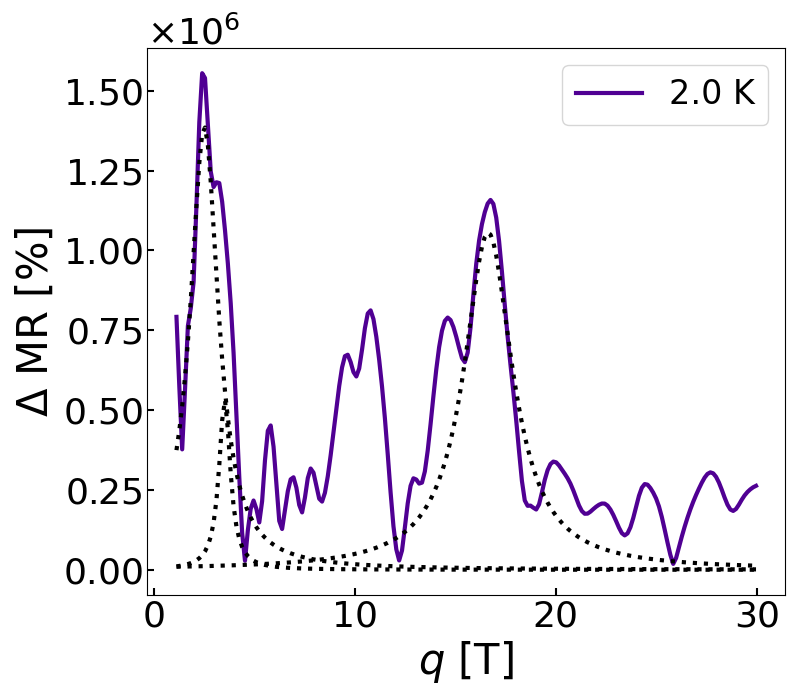}};
        \node[anchor=north west] (image) at (0.3\columnwidth,0) 
        {\includegraphics[width=0.3\columnwidth]{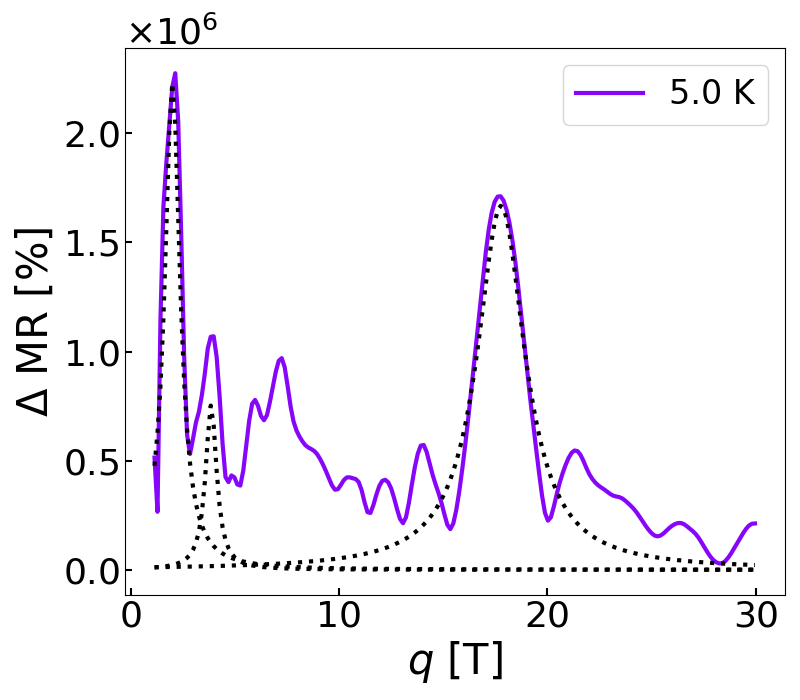}};
        \node[anchor=north west] (image) at (0.6\columnwidth,0) 
        {\includegraphics[width=0.3\columnwidth]{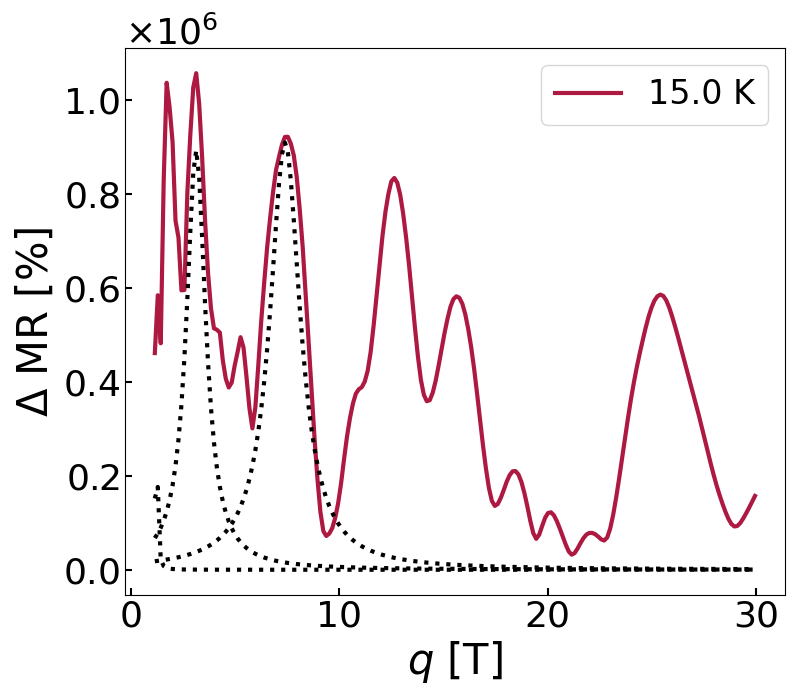}};
        \node[anchor=north west, scale = 1.2] (d) at (0,0) {\textbf{d}};
        \node[anchor=north west, scale = 1.2] (e) at (0.3\columnwidth,0) {\textbf{e}};
        \node[anchor=north west, scale = 1.2] (f) at (0.6\columnwidth,0) {\textbf{f}};
    \end{tikzpicture}
    \begin{tikzpicture}
        \node[anchor=north west] (image) at (0,0) 
        {\includegraphics[width=0.3\columnwidth]{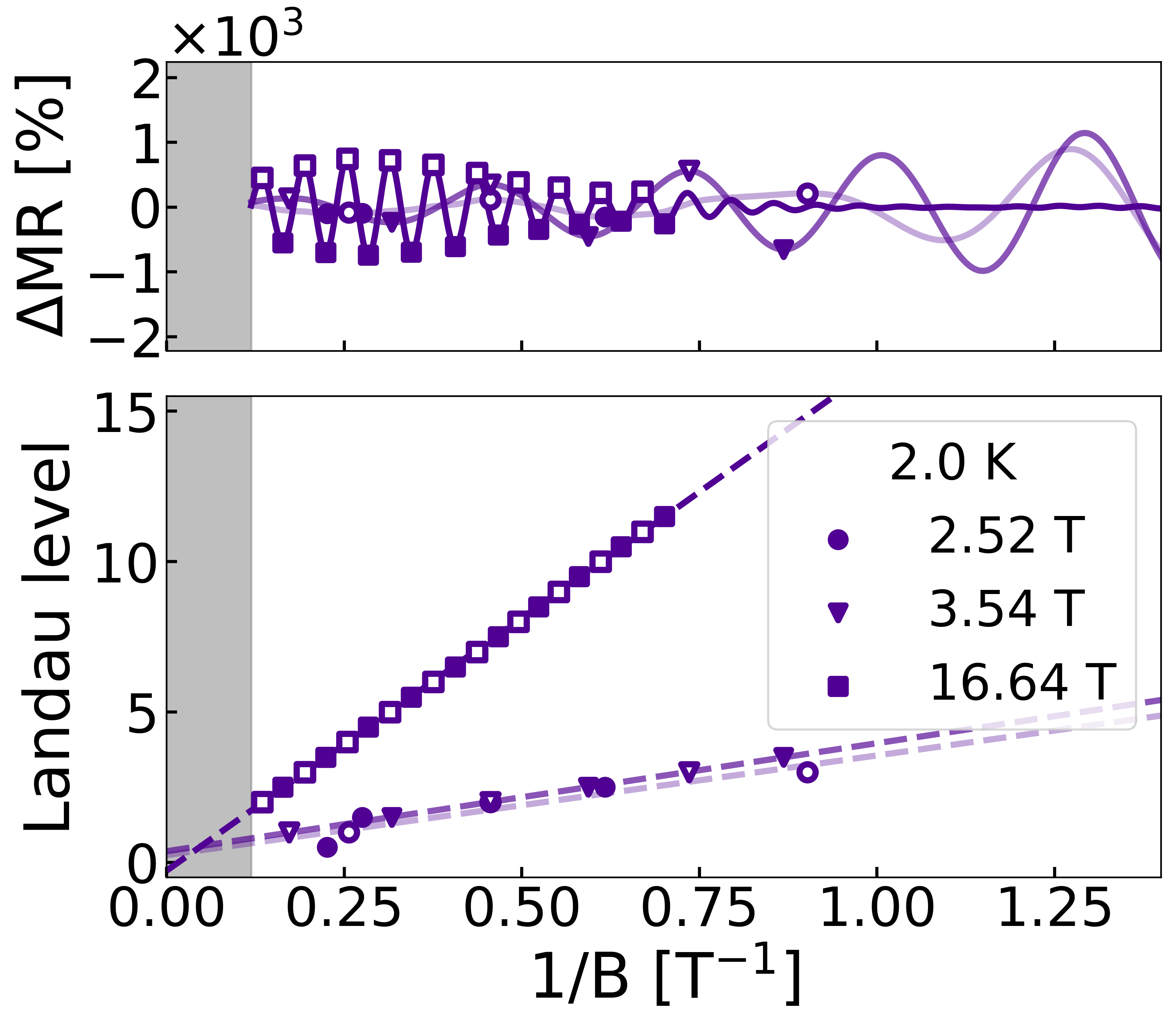}};
        \node[anchor=north west] (image) at (0.3\columnwidth,0) 
        {\includegraphics[width=0.3\columnwidth]{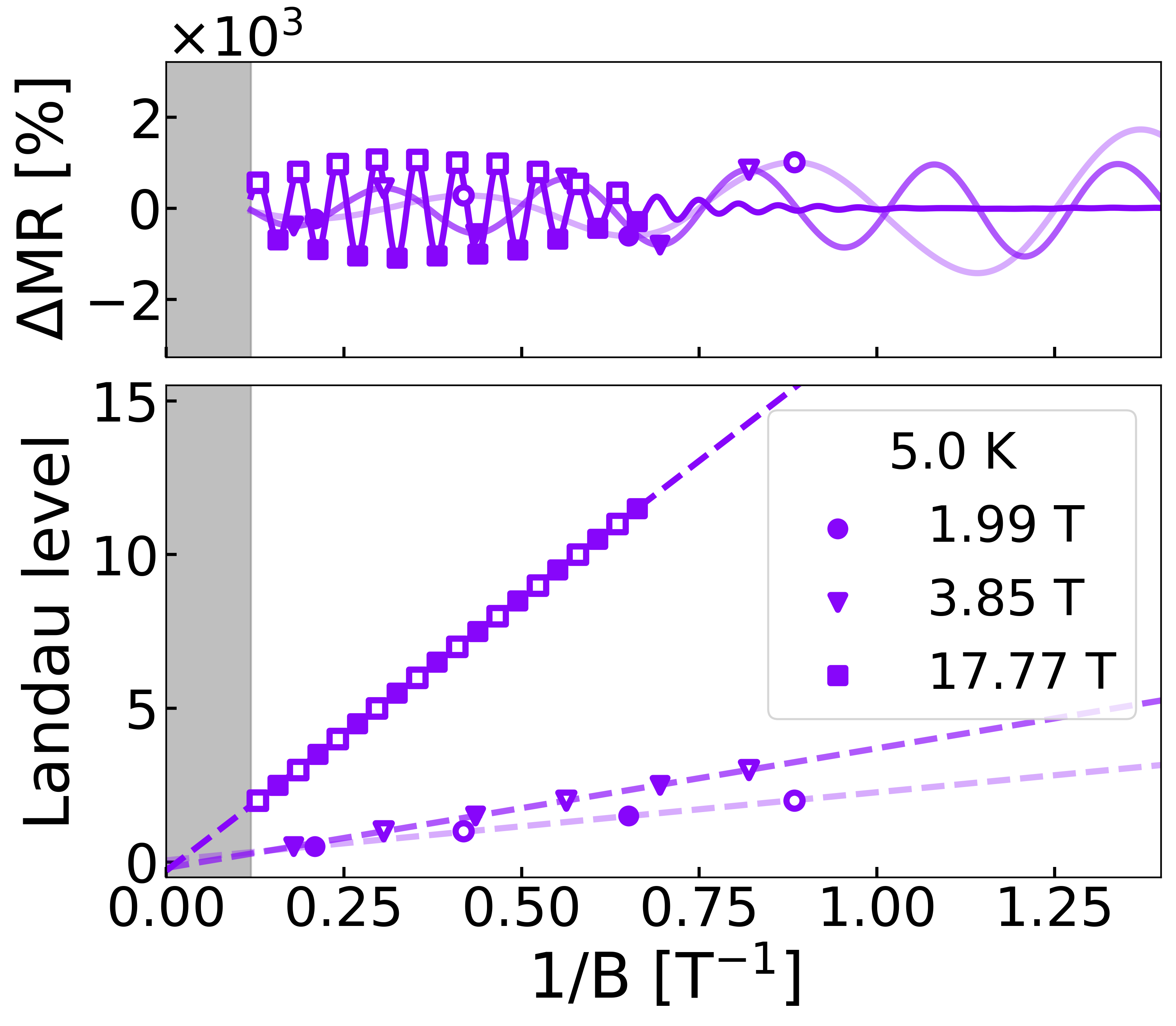}};
        \node[anchor=north west] (image) at (0.6\columnwidth,0) 
        {\includegraphics[width=0.3\columnwidth]{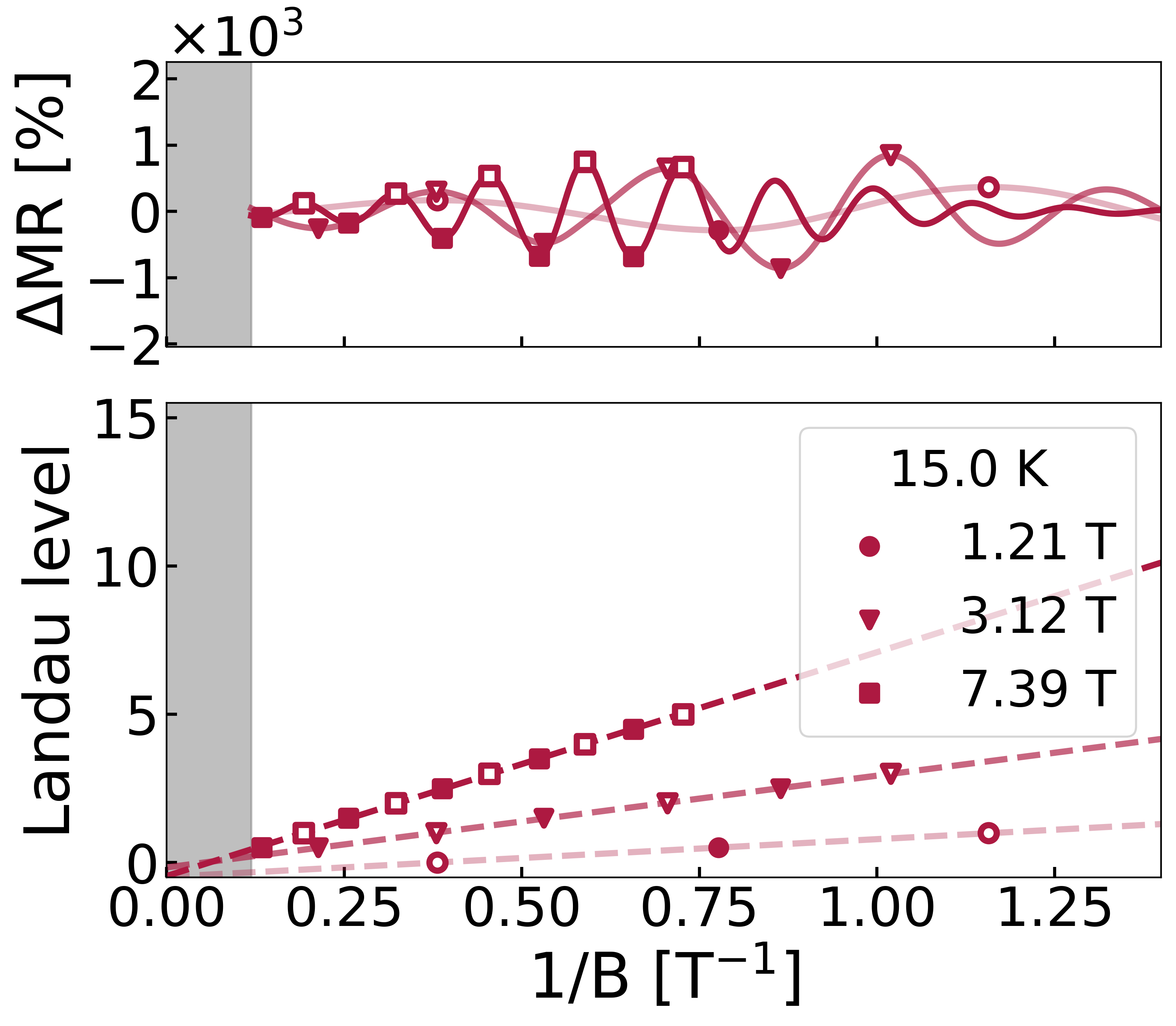}};
        \node[anchor=north west, scale = 1.2] (g) at (0,0) {\textbf{g}};
        \node[anchor=north west, scale = 1.2] (h) at (0.3\columnwidth,0) {\textbf{h}};
        \node[anchor=north west, scale = 1.2] (i) at (0.6\columnwidth,0) {\textbf{i}};
    \end{tikzpicture}
    \caption{MR analysis of S$_{3m}$ with background-free curvature method. MR plots and their corresponding second derivatives which work as $\Delta$MR at \textbf{a} 2 K, \textbf{b} 5 K, and \textbf{c} 15 K. \textbf{d}-\textbf{f} Fast Fourier transform (FFT) of \textbf{a}-\textbf{c}, respectively. \textbf{g}-\textbf{i} display the Shubnikov-de Haas (SdH) oscillation and Landau level fan plot of \textbf{d}-\textbf{f}, respectively.}
    \label{figS7}
\end{figure}

\begin{figure}[ht!]
    \centering
    \begin{tikzpicture}
        \node[anchor=north west] (image) at (0,0) 
        {\includegraphics[width=0.3\columnwidth]{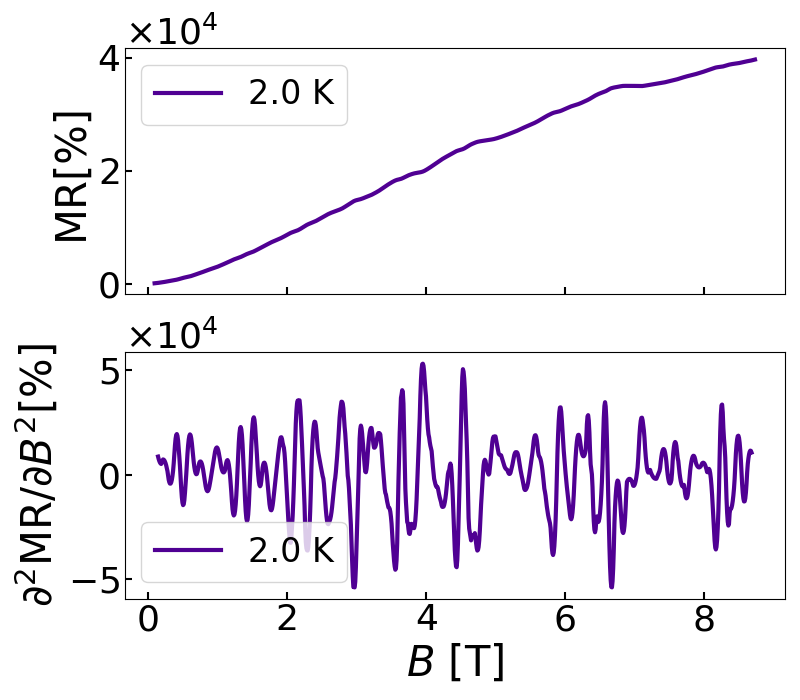}};
        \node[anchor=north west] (image) at (0.3\columnwidth,0) 
        {\includegraphics[width=0.3\columnwidth]{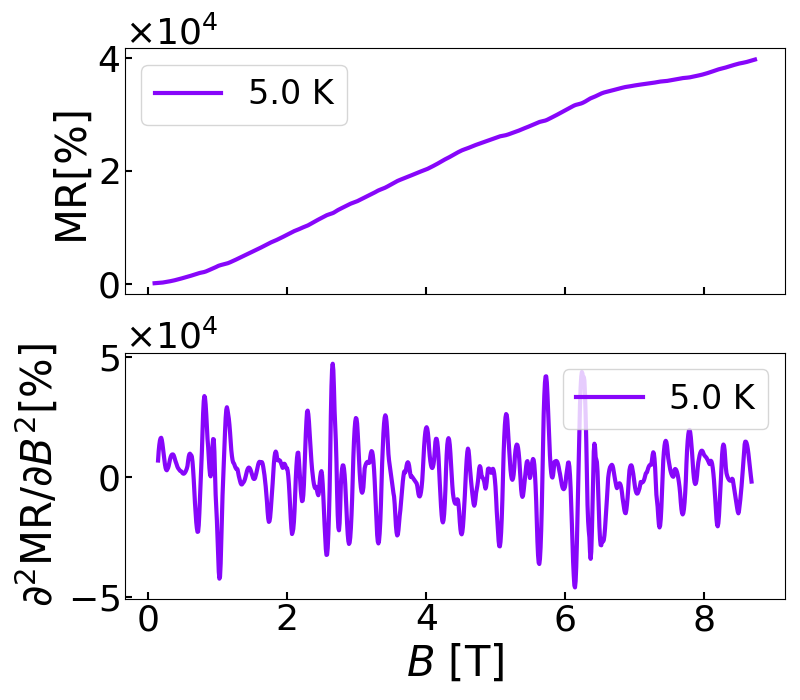}};
        \node[anchor=north west] (image) at (0.6\columnwidth,0) 
        {\includegraphics[width=0.3\columnwidth]{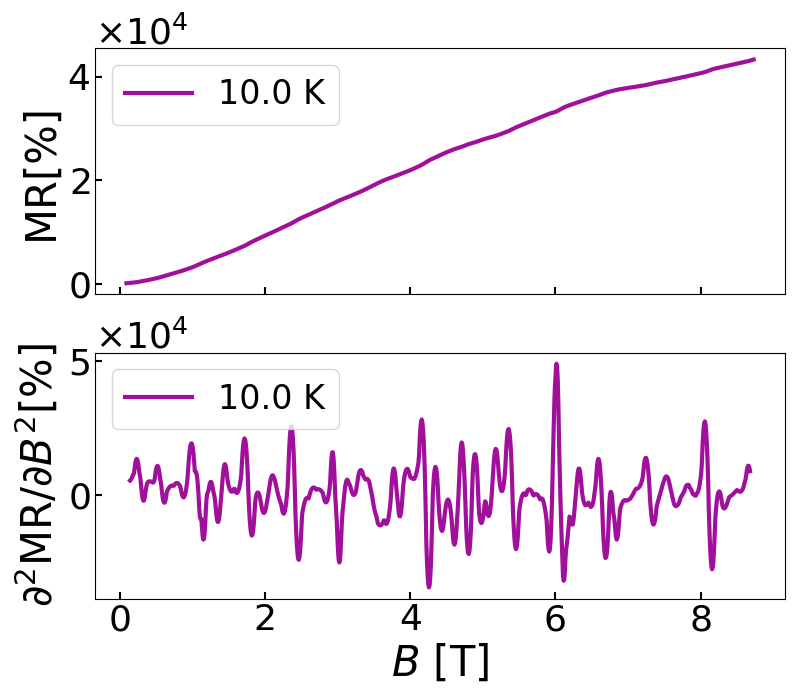}};
        \node[anchor=north west, scale = 1.2] (a) at (0,0) {\textbf{a}};
        \node[anchor=north west, scale = 1.2] (b) at (0.3\columnwidth,0) {\textbf{b}};
        \node[anchor=north west, scale = 1.2] (c) at (0.6\columnwidth,0) {\textbf{c}};

        \node[anchor=north west, scale = 1.2] at (0.225\columnwidth,-1.25) {S$_{20m}$};
    \end{tikzpicture}
    \begin{tikzpicture}
        \node[anchor=north west] (image) at (0,0) 
        {\includegraphics[width=0.3\columnwidth]{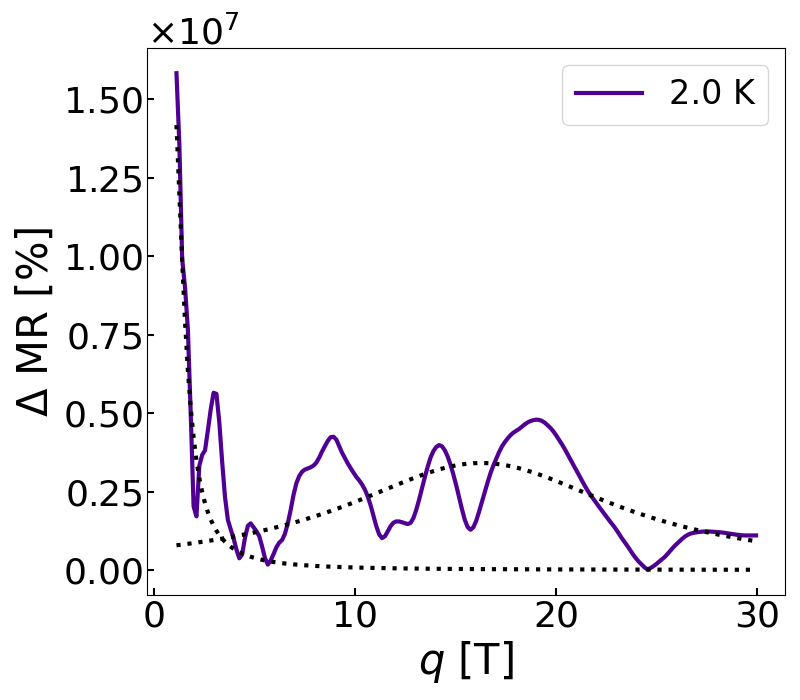}};
        \node[anchor=north west] (image) at (0.3\columnwidth,0) 
        {\includegraphics[width=0.3\columnwidth]{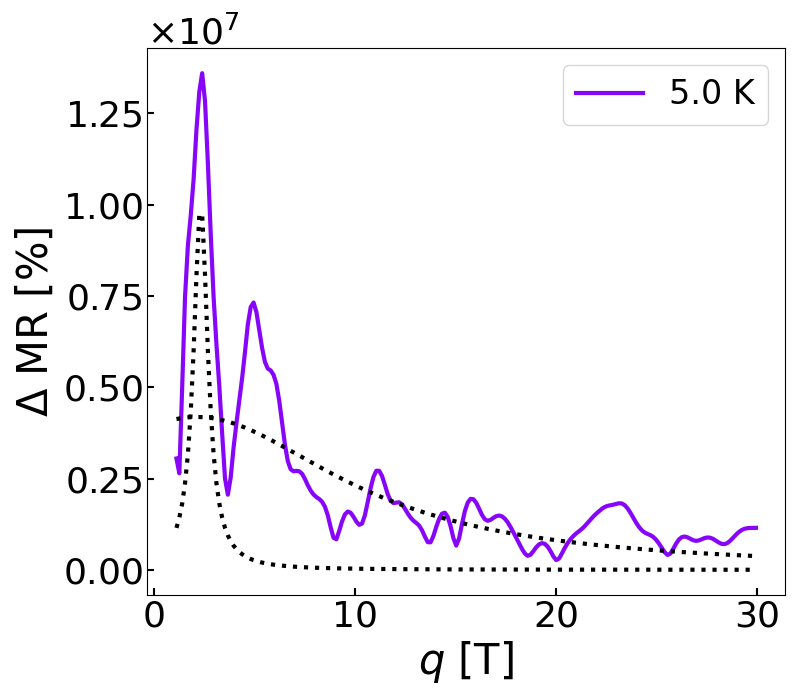}};
        \node[anchor=north west] (image) at (0.6\columnwidth,0) 
        {\includegraphics[width=0.3\columnwidth]{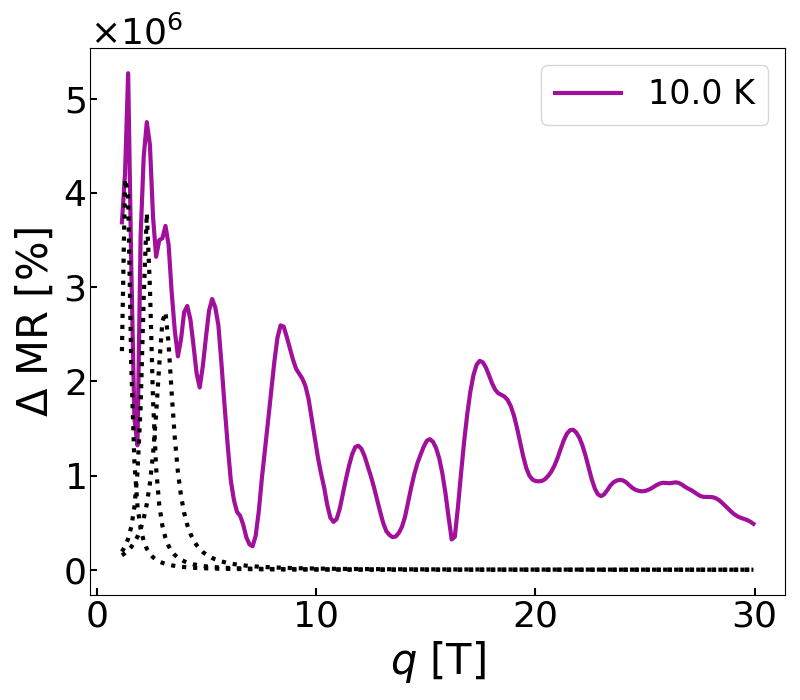}};
        \node[anchor=north west, scale = 1.2] (d) at (0,0) {\textbf{d}};
        \node[anchor=north west, scale = 1.2] (e) at (0.3\columnwidth,0) {\textbf{e}};
        \node[anchor=north west, scale = 1.2] (f) at (0.6\columnwidth,0) {\textbf{f}};
    \end{tikzpicture}
    \begin{tikzpicture}
        \node[anchor=north west] (image) at (0,0) 
        {\includegraphics[width=0.3\columnwidth]{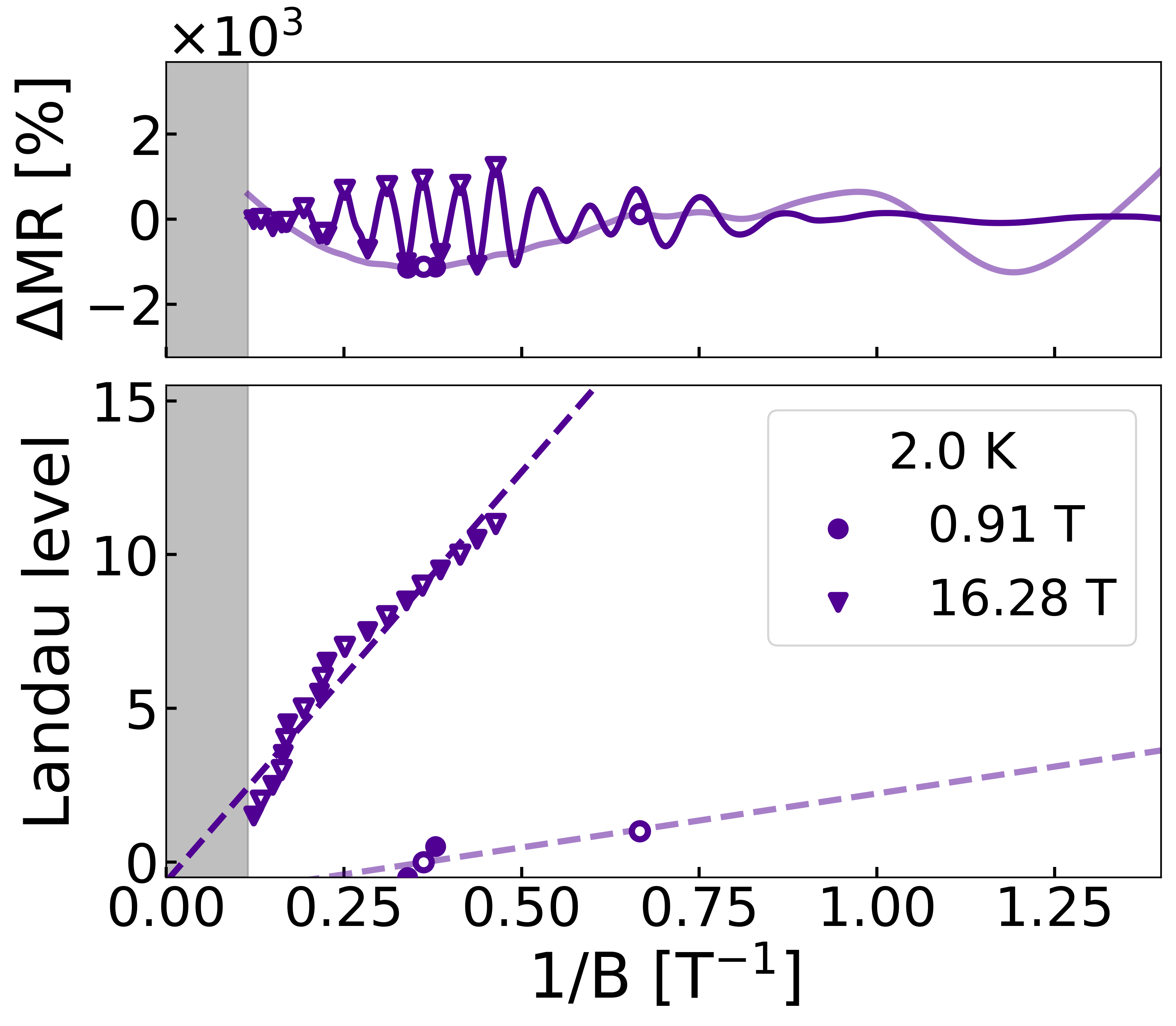}};
        \node[anchor=north west] (image) at (0.3\columnwidth,0) 
        {\includegraphics[width=0.3\columnwidth]{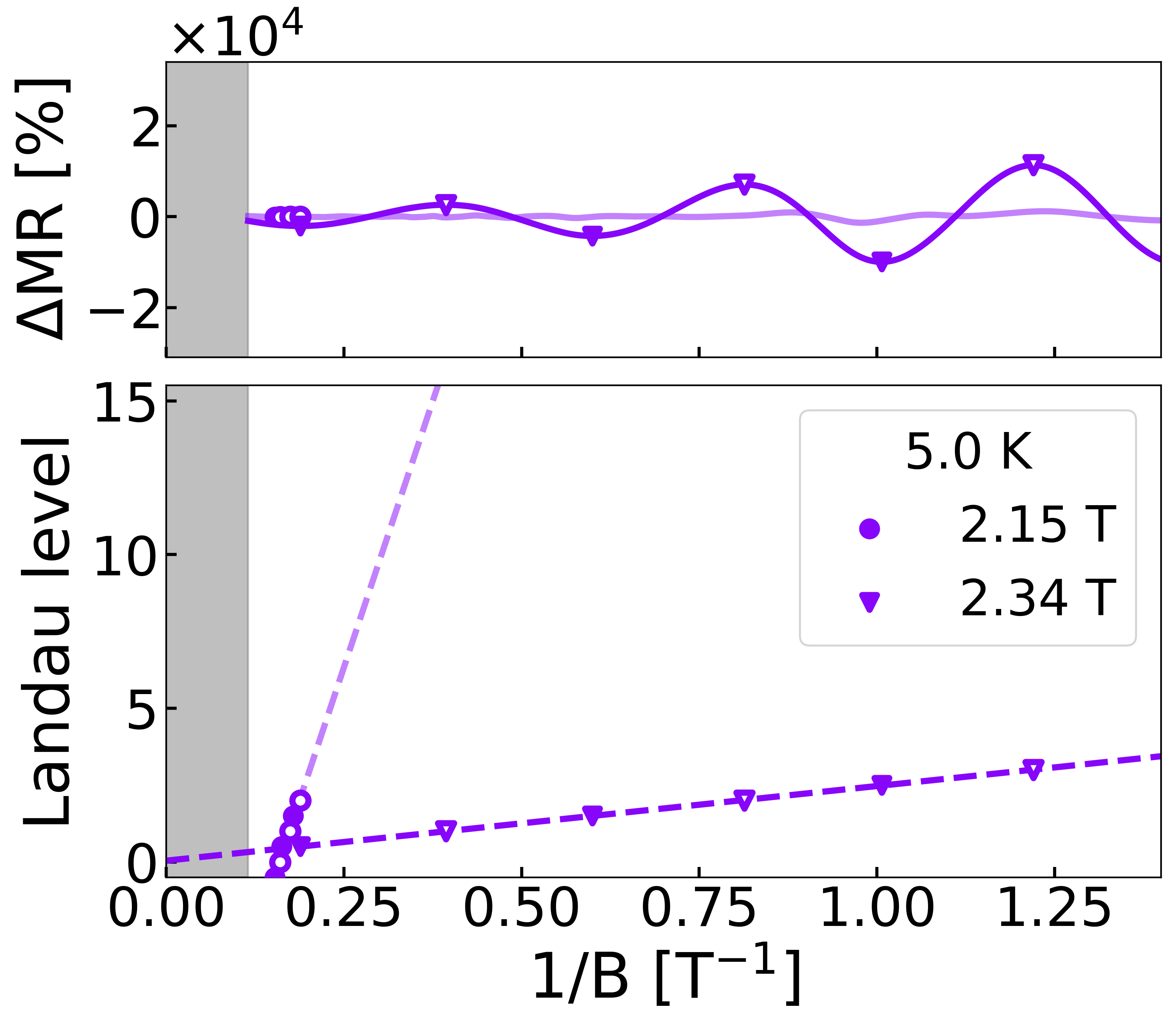}};
        \node[anchor=north west] (image) at (0.6\columnwidth,0) 
        {\includegraphics[width=0.3\columnwidth]{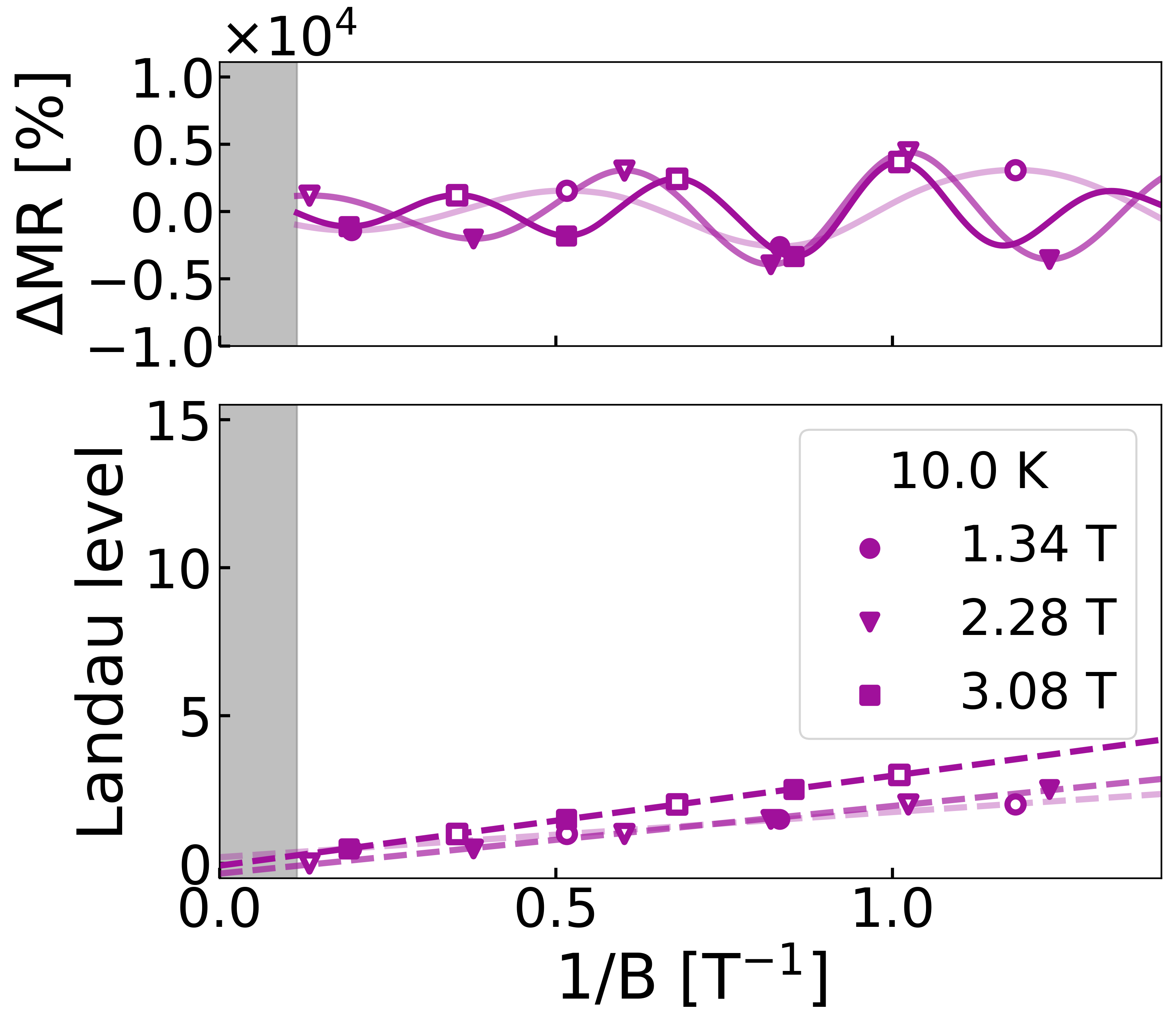}};
        \node[anchor=north west, scale = 1.2] (g) at (0,0) {\textbf{g}};
        \node[anchor=north west, scale = 1.2] (h) at (0.3\columnwidth,0) {\textbf{h}};
        \node[anchor=north west, scale = 1.2] (i) at (0.6\columnwidth,0) {\textbf{i}};
    \end{tikzpicture}
    \caption{MR analysis of S$_{20m}$ with background-free curvature method. MR plots and their corresponding second derivatives, which work as $\Delta$MR at \textbf{a} 2 K, \textbf{b} 5 K, and \textbf{c} 10 K. \textbf{d}-\textbf{f} Fast Fourier transform (FFT) of \textbf{a}-\textbf{c}, respectively. \textbf{g}-\textbf{i} display the Shubnikov-de Haas (SdH) oscillation and Landau level fan plot of \textbf{d}-\textbf{f}, respectively.}
    \label{figS8}
\end{figure}

\begin{figure}[ht!]
    \centering
    \begin{tikzpicture}
        \node[anchor=north west] (image) at (0,0) 
        {\includegraphics[width=0.3\columnwidth]{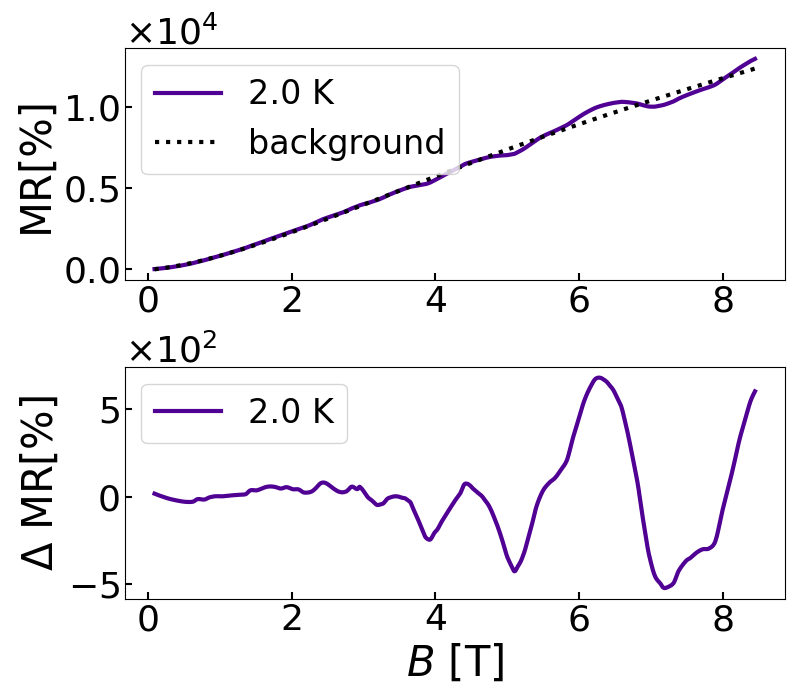}};
        \node[anchor=north west] (image) at (0.3\columnwidth,0) 
        {\includegraphics[width=0.3\columnwidth]{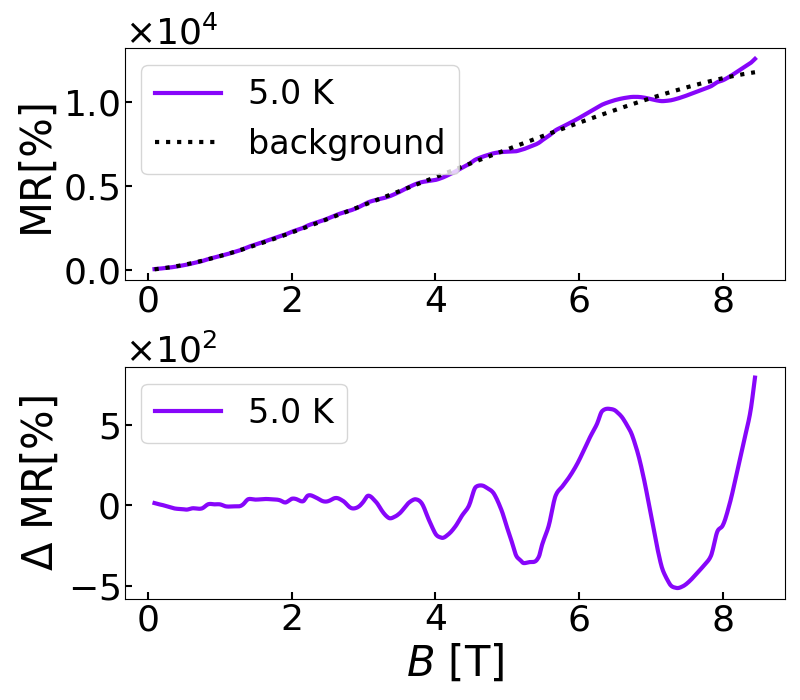}};
        \node[anchor=north west] (image) at (0.6\columnwidth,0) 
        {\includegraphics[width=0.3\columnwidth]{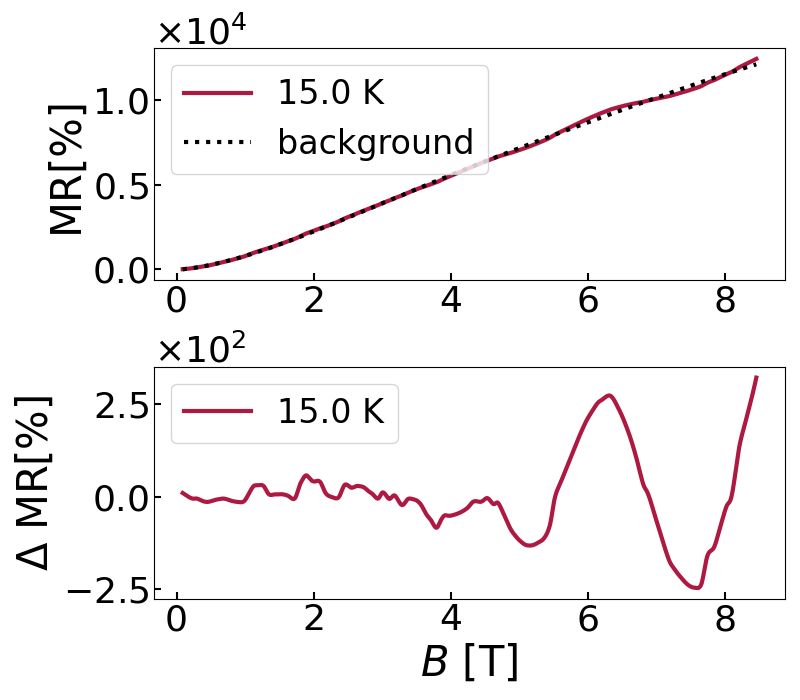}};
        \node[anchor=north west, scale = 1.2] (a) at (0,0) {\textbf{a}};
        \node[anchor=north west, scale = 1.2] (b) at (0.3\columnwidth,0) {\textbf{b}};
        \node[anchor=north west, scale = 1.2] (c) at (0.6\columnwidth,0) {\textbf{c}};

        \node[anchor=north west, scale = 1.2] at (0.225\columnwidth,-1.25) {S$_{3m}$};
    \end{tikzpicture}
    \begin{tikzpicture}
        \node[anchor=north west] (image) at (0,0) 
        {\includegraphics[width=0.3\columnwidth]{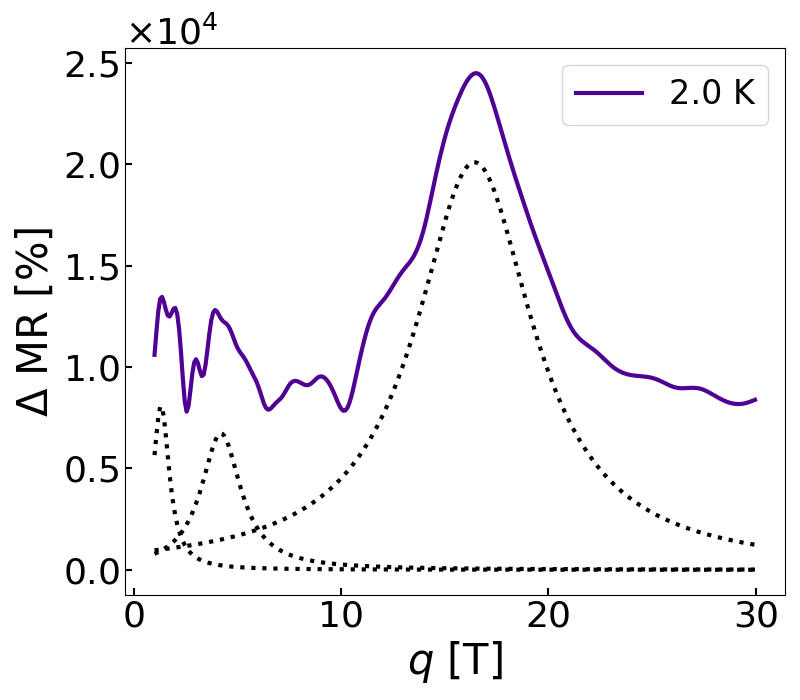}};
        \node[anchor=north west] (image) at (0.3\columnwidth,0) 
        {\includegraphics[width=0.3\columnwidth]{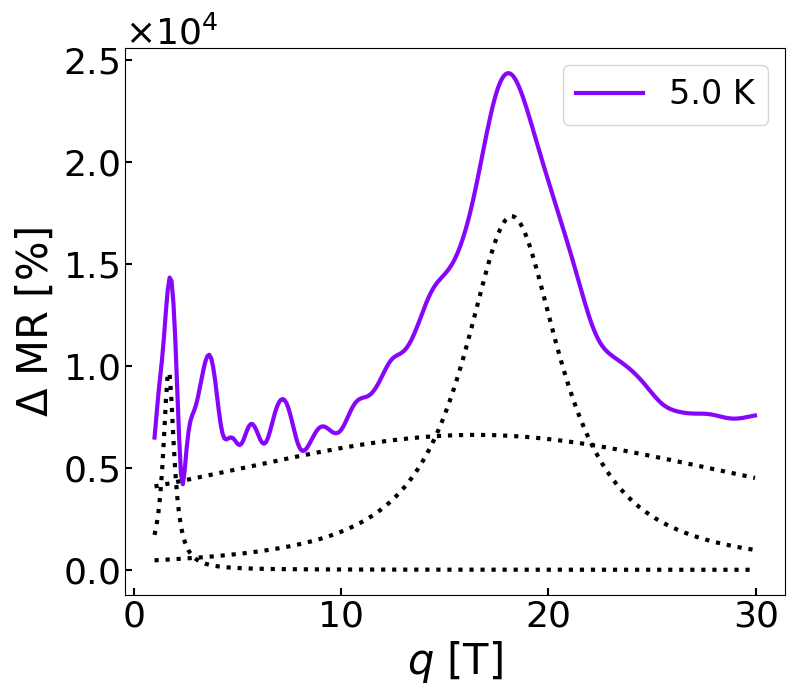}};
        \node[anchor=north west] (image) at (0.6\columnwidth,0) 
        {\includegraphics[width=0.3\columnwidth]{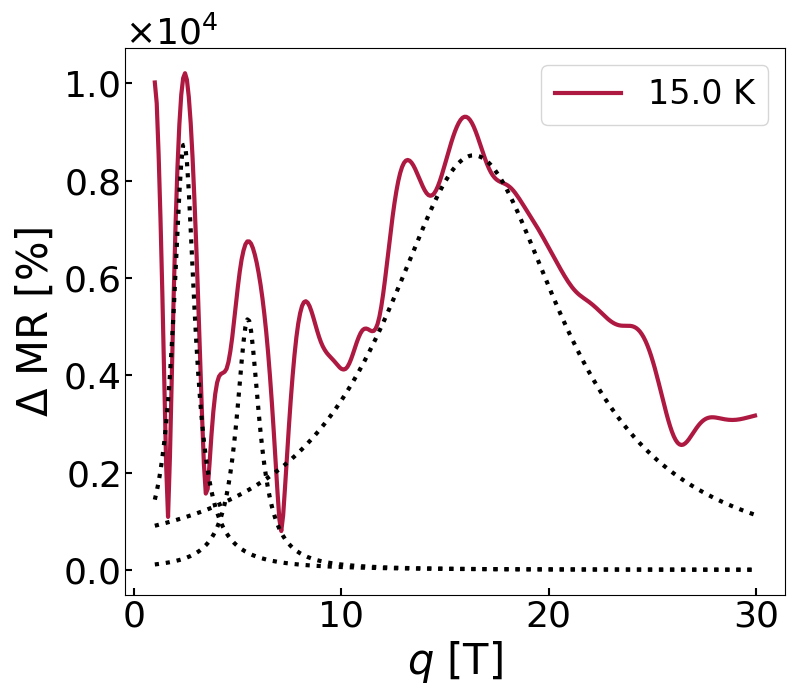}};
        \node[anchor=north west, scale = 1.2] (d) at (0,0) {\textbf{d}};
        \node[anchor=north west, scale = 1.2] (e) at (0.3\columnwidth,0) {\textbf{e}};
        \node[anchor=north west, scale = 1.2] (f) at (0.6\columnwidth,0) {\textbf{f}};
    \end{tikzpicture}
    \begin{tikzpicture}
        \node[anchor=north west] (image) at (0,0) 
        {\includegraphics[width=0.3\columnwidth]{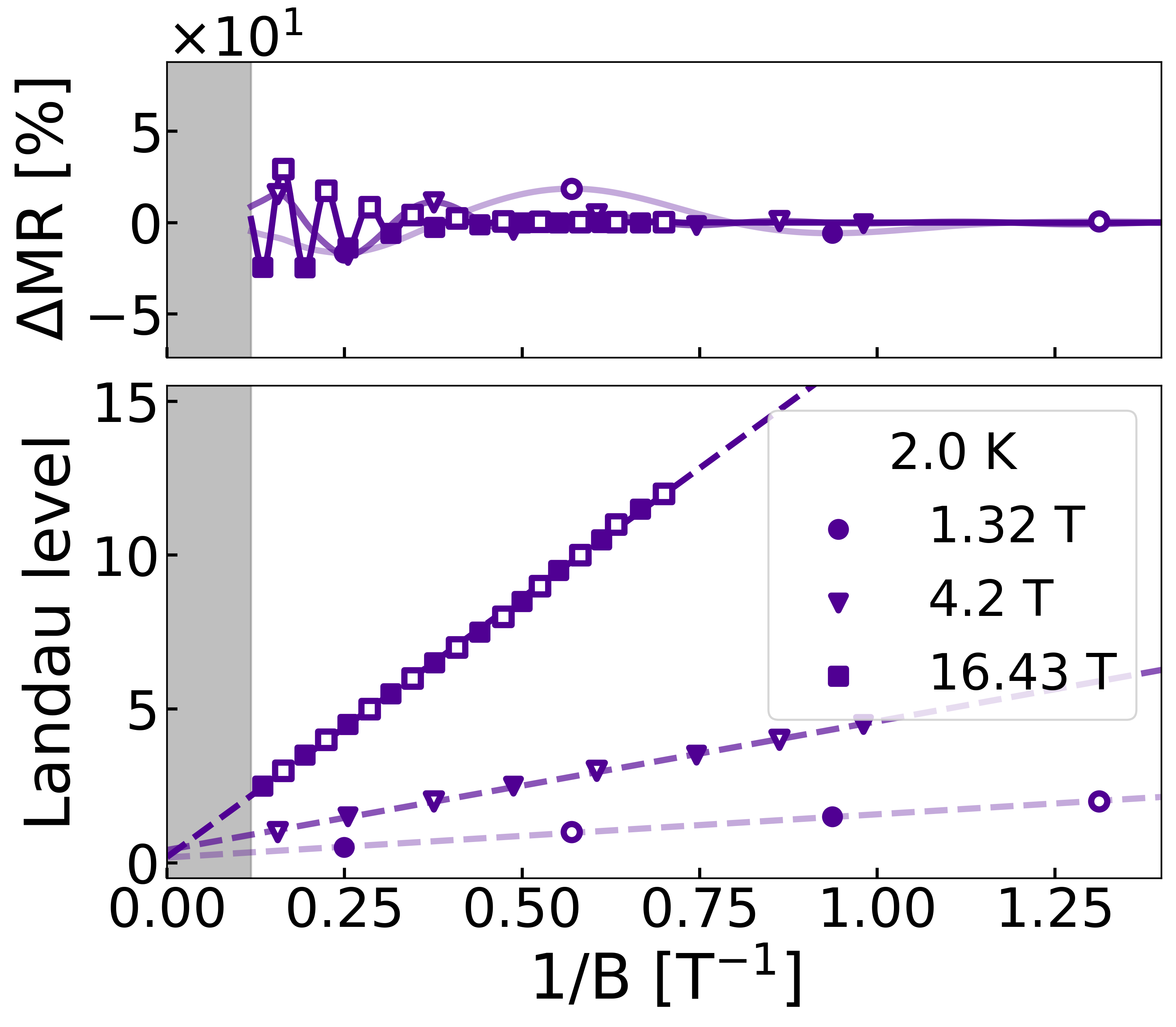}};
        \node[anchor=north west] (image) at (0.3\columnwidth,0) 
        {\includegraphics[width=0.3\columnwidth]{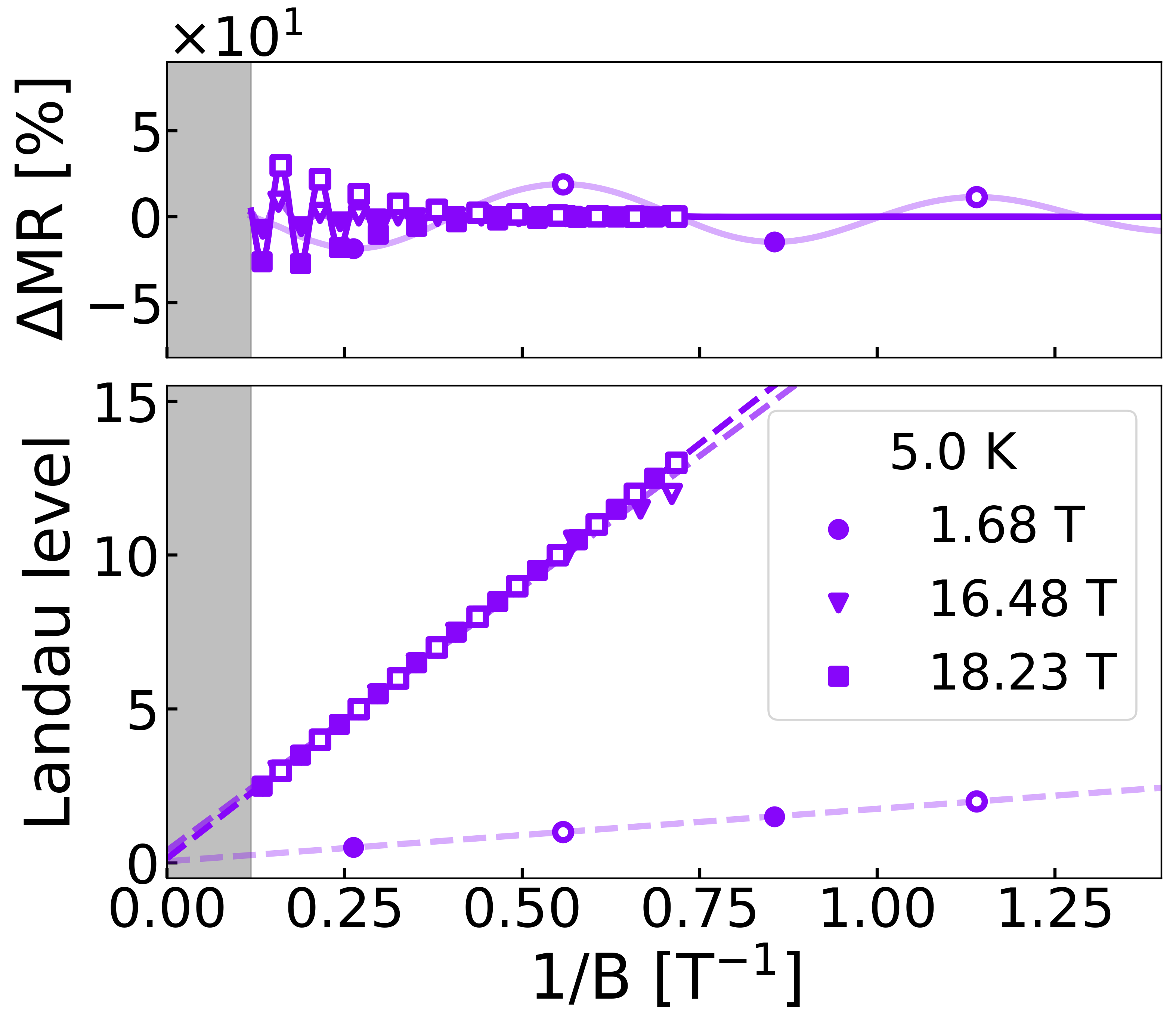}};
        \node[anchor=north west] (image) at (0.6\columnwidth,0) 
        {\includegraphics[width=0.3\columnwidth]{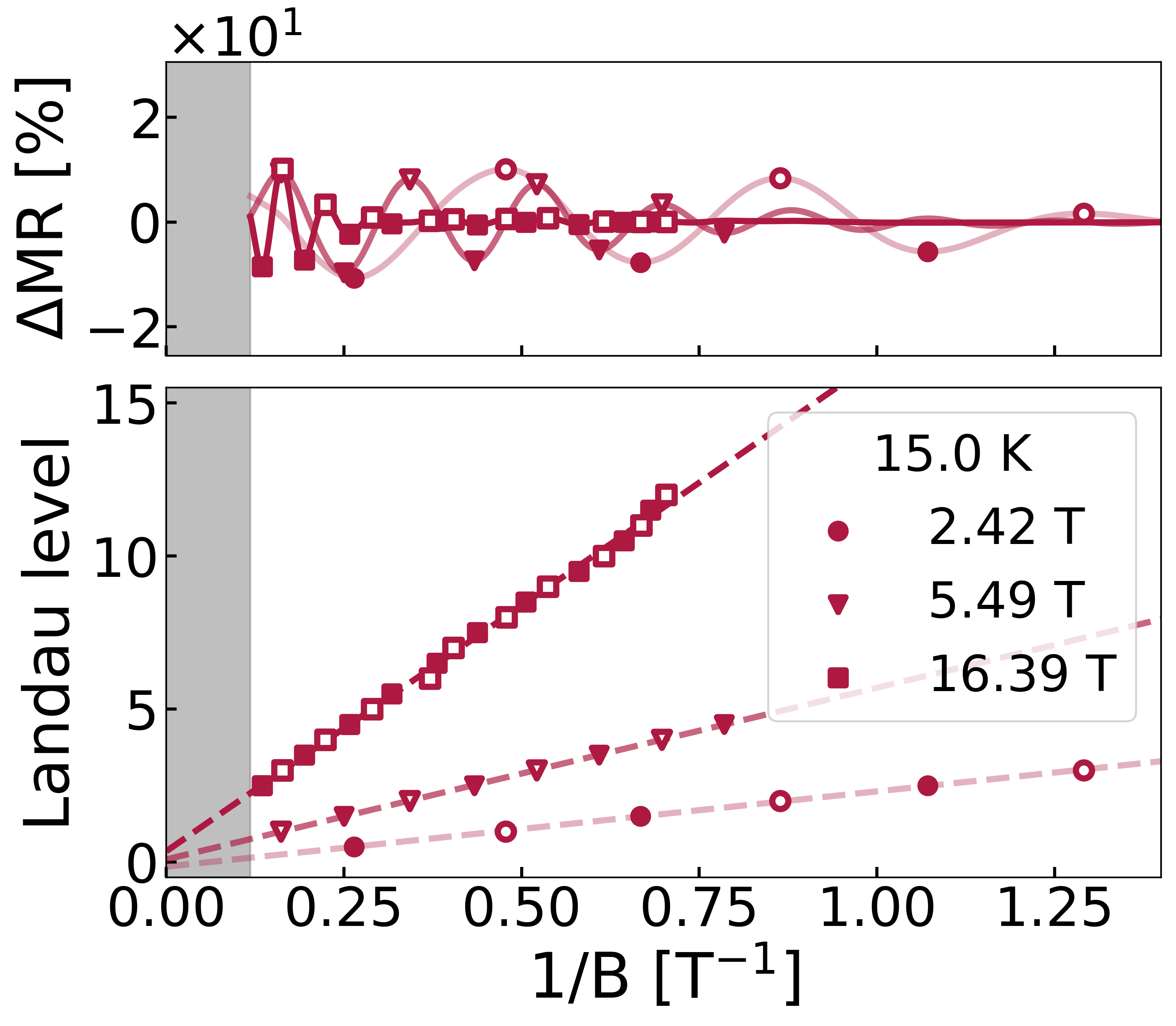}};
        \node[anchor=north west, scale = 1.2] (g) at (0,0) {\textbf{g}};
        \node[anchor=north west, scale = 1.2] (h) at (0.3\columnwidth,0) {\textbf{h}};
        \node[anchor=north west, scale = 1.2] (i) at (0.6\columnwidth,0) {\textbf{i}};
    \end{tikzpicture}
    \caption{MR analysis of S$_{3m}$ with polynomial background method. MR plots and $\Delta$MR at \textbf{a} 2 K, \textbf{b} 5 K, and \textbf{c} 15 K. \textbf{d}-\textbf{f} Fast Fourier transform (FFT) of \textbf{a}-\textbf{c}, respectively. \textbf{g}-\textbf{i} display the Shubnikov-de Haas (SdH) oscillation and Landau level fan plot of \textbf{d}-\textbf{f}, respectively.}
    \label{figS9}
\end{figure}

\begin{figure}[ht!]
    \centering
    \begin{tikzpicture}
        \node[anchor=north west] (image) at (0,0) 
        {\includegraphics[width=0.3\columnwidth]{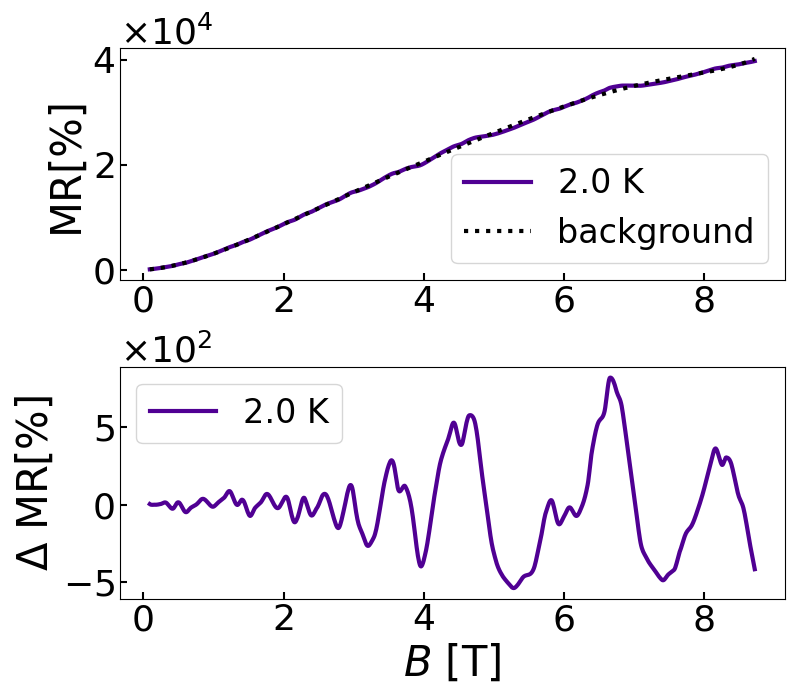}};
        \node[anchor=north west] (image) at (0.3\columnwidth,0) 
        {\includegraphics[width=0.3\columnwidth]{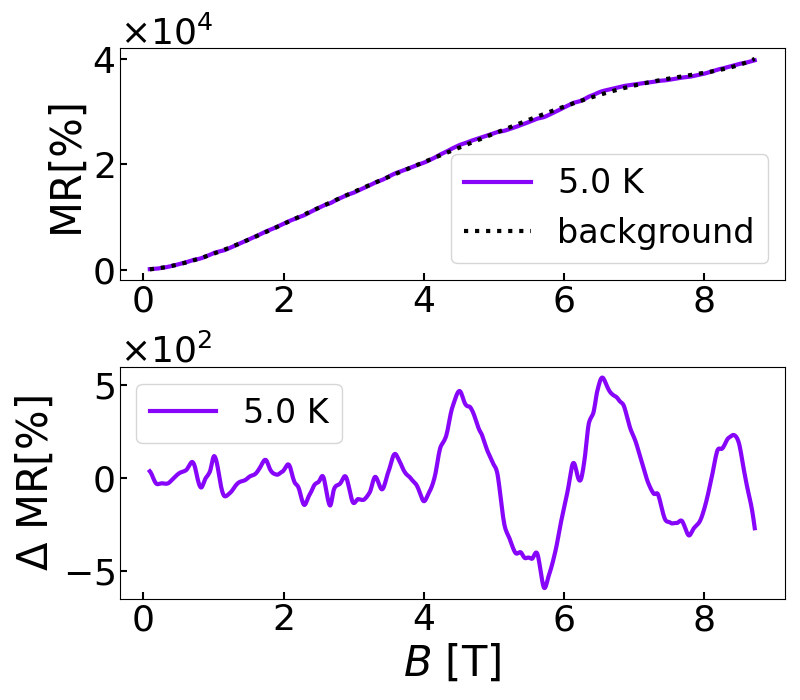}};
        \node[anchor=north west] (image) at (0.6\columnwidth,0) 
        {\includegraphics[width=0.3\columnwidth]{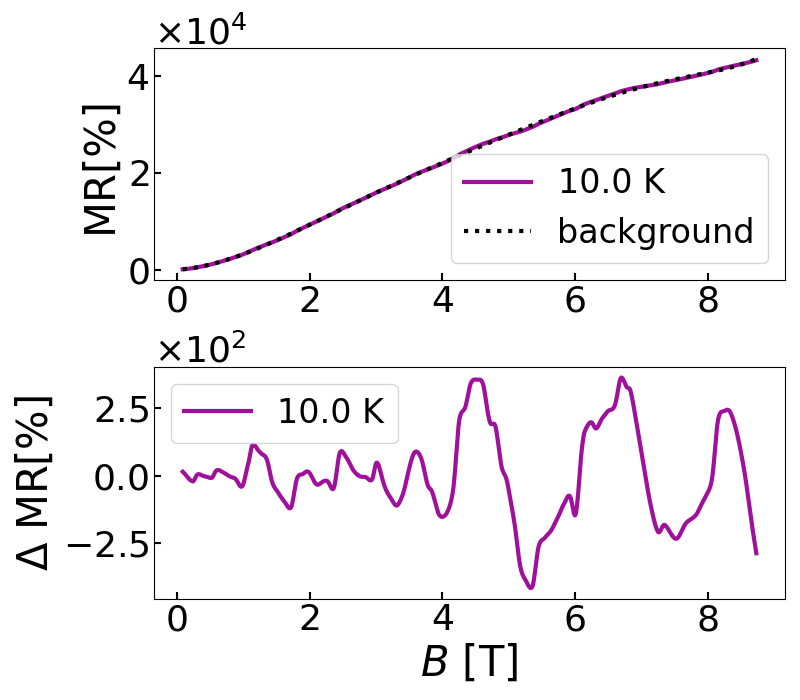}};
        \node[anchor=north west, scale = 1.2] (a) at (0,0) {\textbf{a}};
        \node[anchor=north west, scale = 1.2] (b) at (0.3\columnwidth,0) {\textbf{b}};
        \node[anchor=north west, scale = 1.2] (c) at (0.6\columnwidth,0) {\textbf{c}};

        \node[anchor=north west, scale = 1.2] at (0.075\columnwidth,-0.75) {S$_{20m}$};
    \end{tikzpicture}
    \begin{tikzpicture}
        \node[anchor=north west] (image) at (0,0) 
        {\includegraphics[width=0.3\columnwidth]{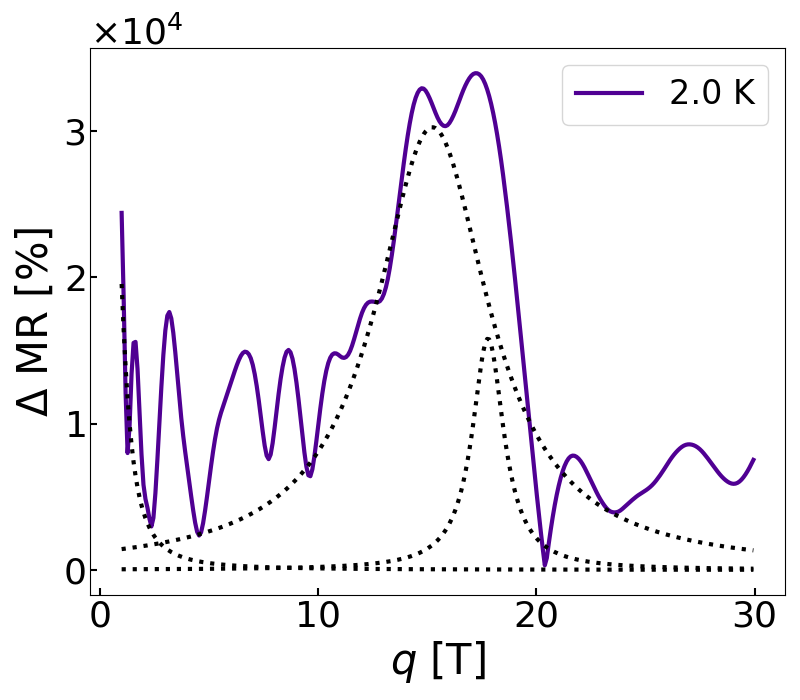}};
        \node[anchor=north west] (image) at (0.3\columnwidth,0) 
        {\includegraphics[width=0.3\columnwidth]{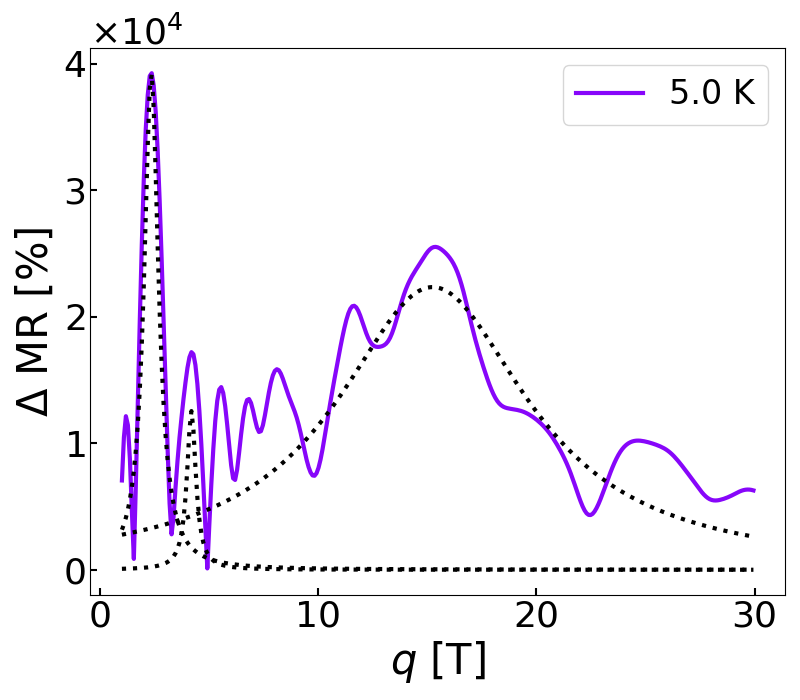}};
        \node[anchor=north west] (image) at (0.6\columnwidth,0) 
        {\includegraphics[width=0.3\columnwidth]{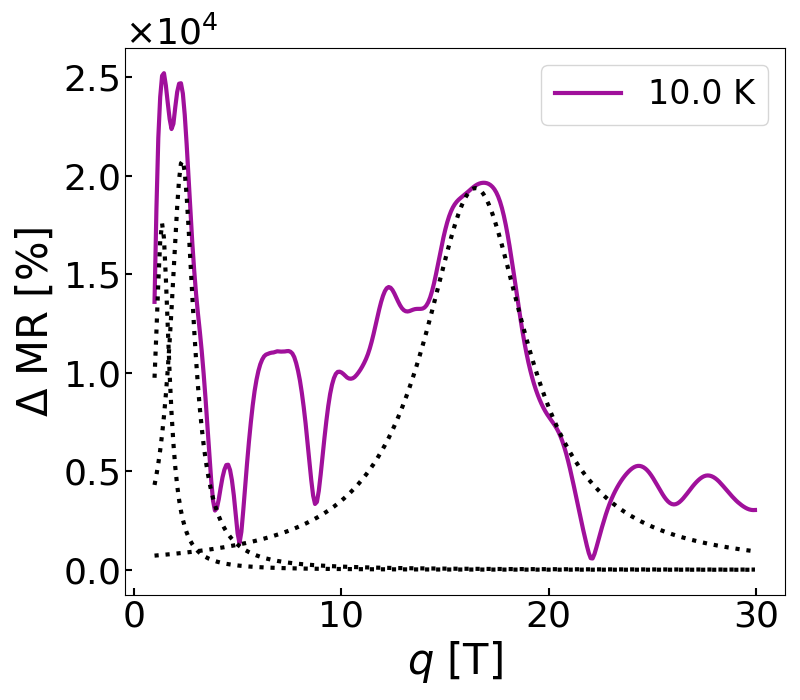}};
        \node[anchor=north west, scale = 1.2] (d) at (0,0) {\textbf{d}};
        \node[anchor=north west, scale = 1.2] (e) at (0.3\columnwidth,0) {\textbf{e}};
        \node[anchor=north west, scale = 1.2] (f) at (0.6\columnwidth,0) {\textbf{f}};
    \end{tikzpicture}
    \begin{tikzpicture}
        \node[anchor=north west] (image) at (0,0) 
        {\includegraphics[width=0.3\columnwidth]{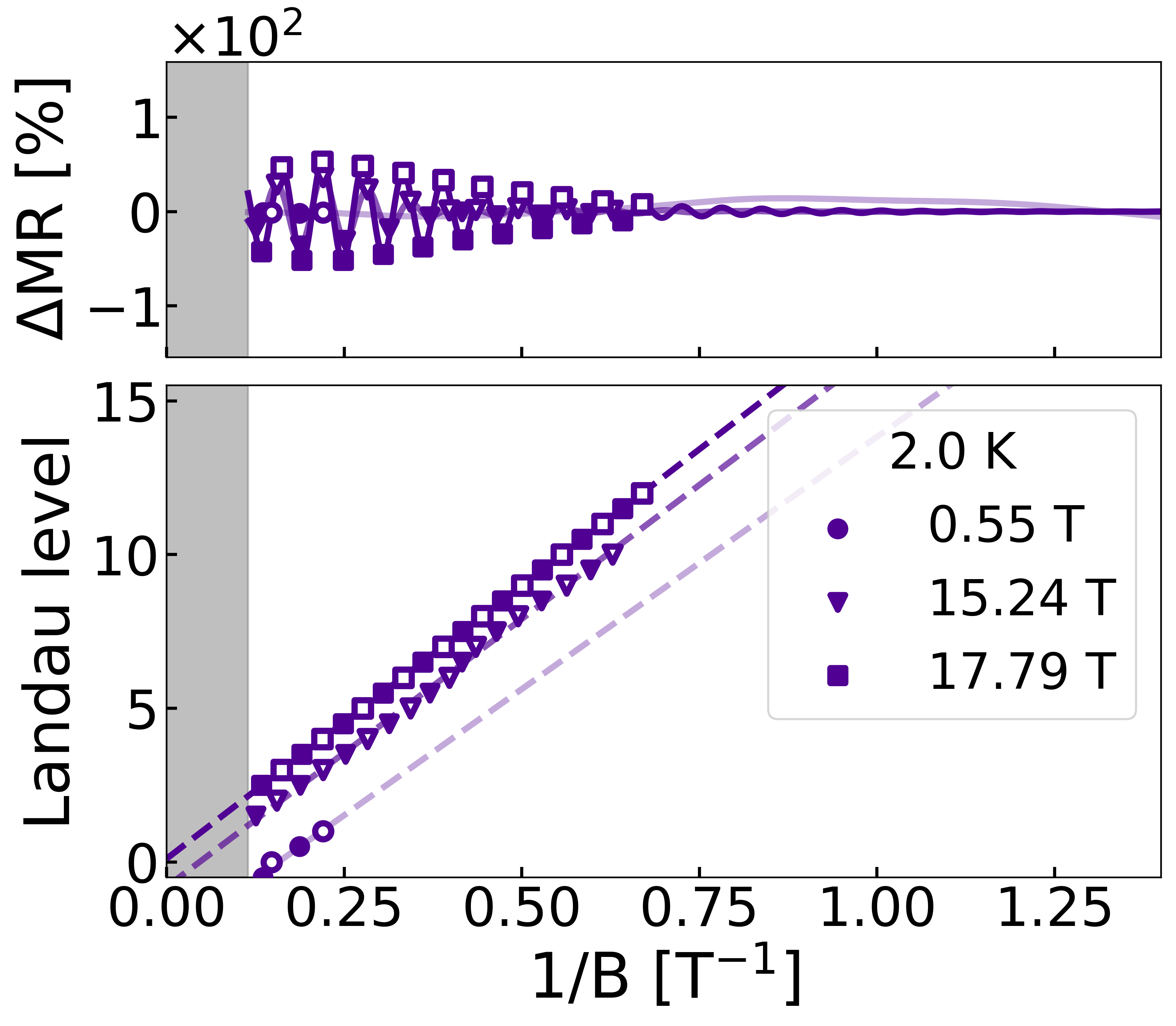}};
        \node[anchor=north west] (image) at (0.3\columnwidth,0) 
        {\includegraphics[width=0.3\columnwidth]{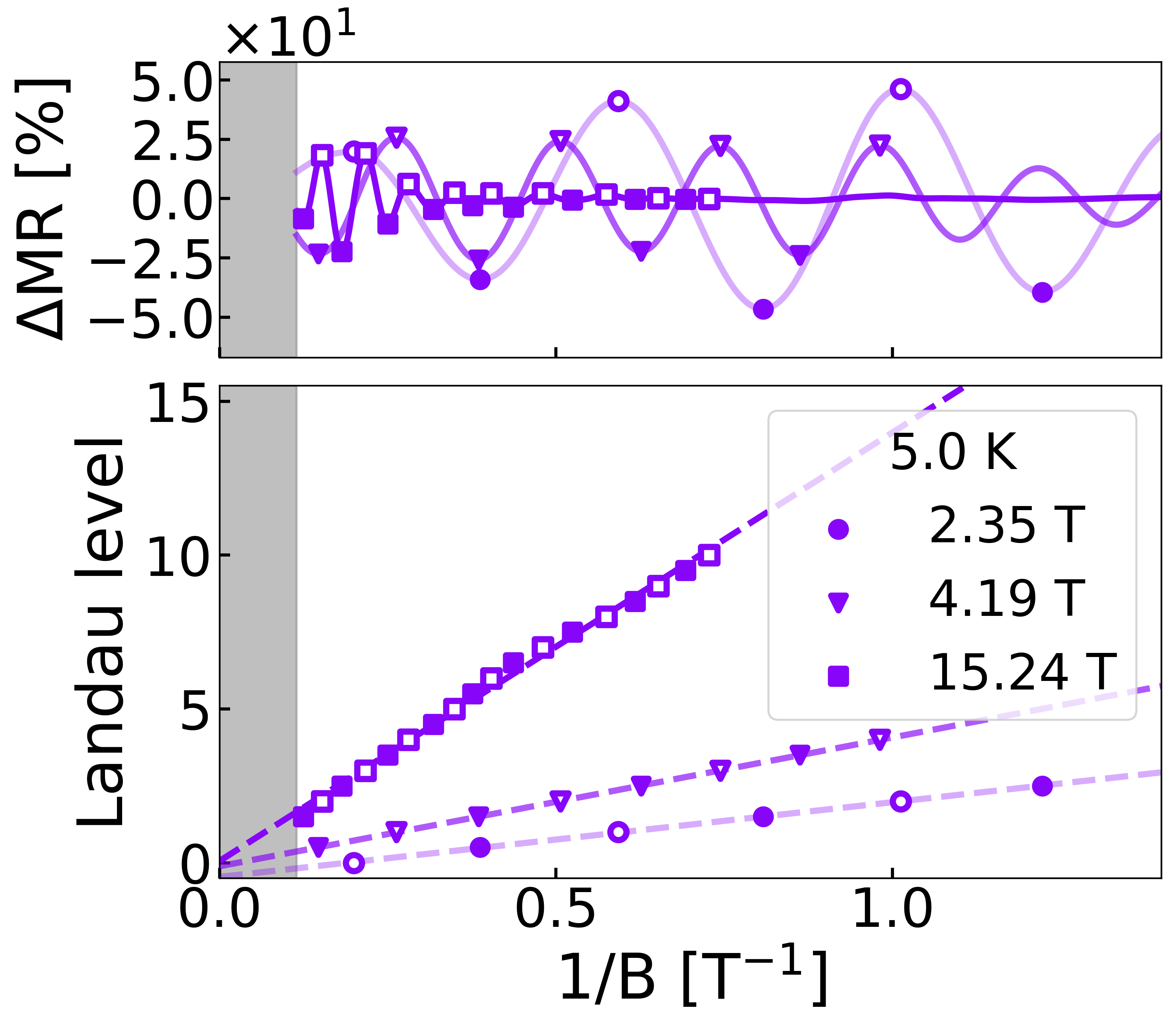}};
        \node[anchor=north west] (image) at (0.6\columnwidth,0) 
        {\includegraphics[width=0.3\columnwidth]{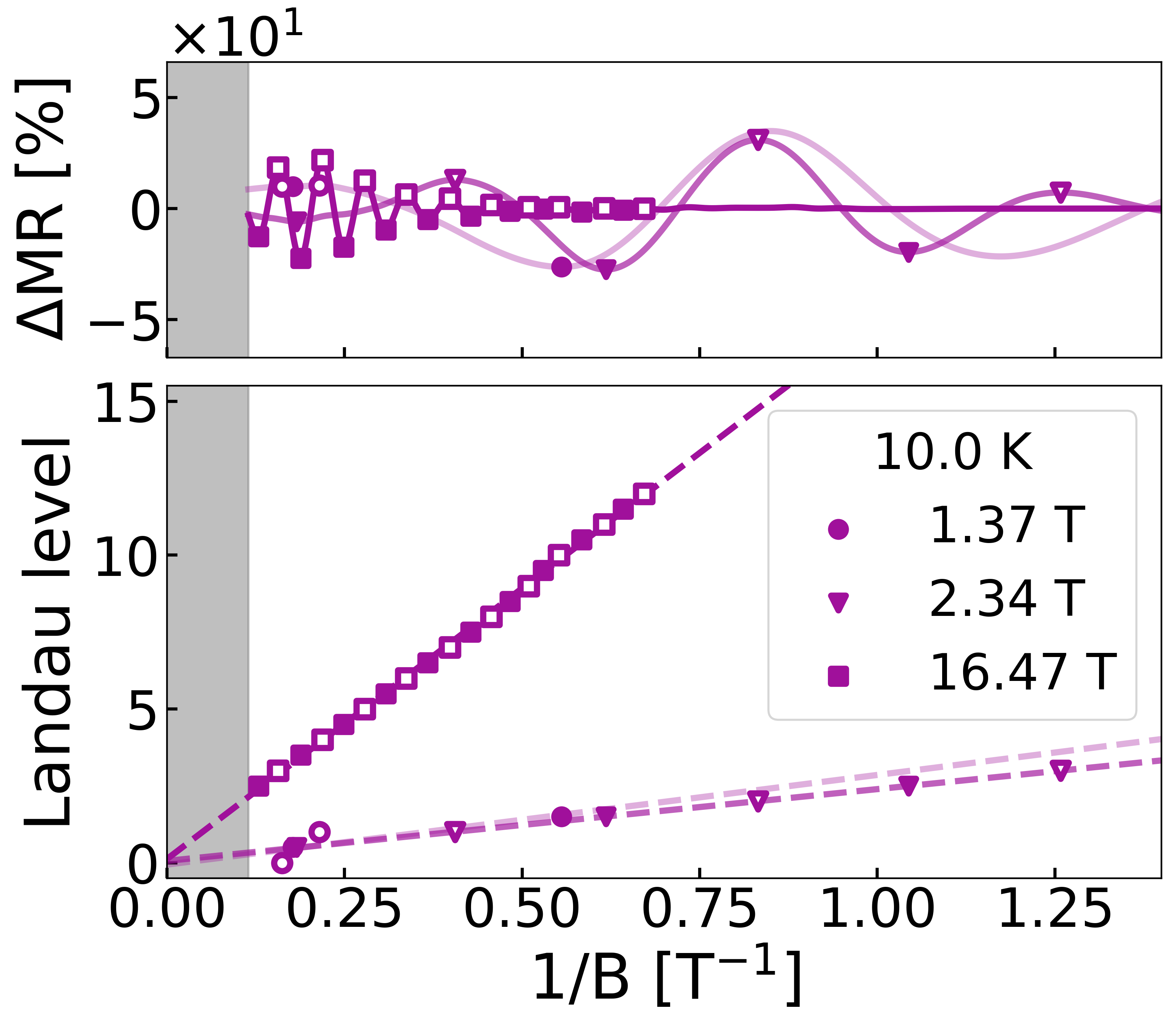}};
        \node[anchor=north west, scale = 1.2] (g) at (0,0) {\textbf{g}};
        \node[anchor=north west, scale = 1.2] (h) at (0.3\columnwidth,0) {\textbf{h}};
        \node[anchor=north west, scale = 1.2] (i) at (0.6\columnwidth,0) {\textbf{i}};
    \end{tikzpicture}
    \caption{MR analysis of S$_{20m}$ with polynomial background method. MR plots and $\Delta$MR at \textbf{a} 2 K, \textbf{b} 5 K, and \textbf{c} 10 K. \textbf{d}-\textbf{f} Fast Fourier transform (FFT) of \textbf{a}-\textbf{c}, respectively. \textbf{g}-\textbf{i} display the Shubnikov-de Haas (SdH) oscillation and Landau level fan plot of \textbf{d}-\textbf{f}, respectively.}
    \label{figS10}
\end{figure}

\begin{figure}[ht!]
    \centering
    \begin{tikzpicture}
        \node[anchor=north west] (image) at (0,0) 
        {\includegraphics[width=0.3\columnwidth]{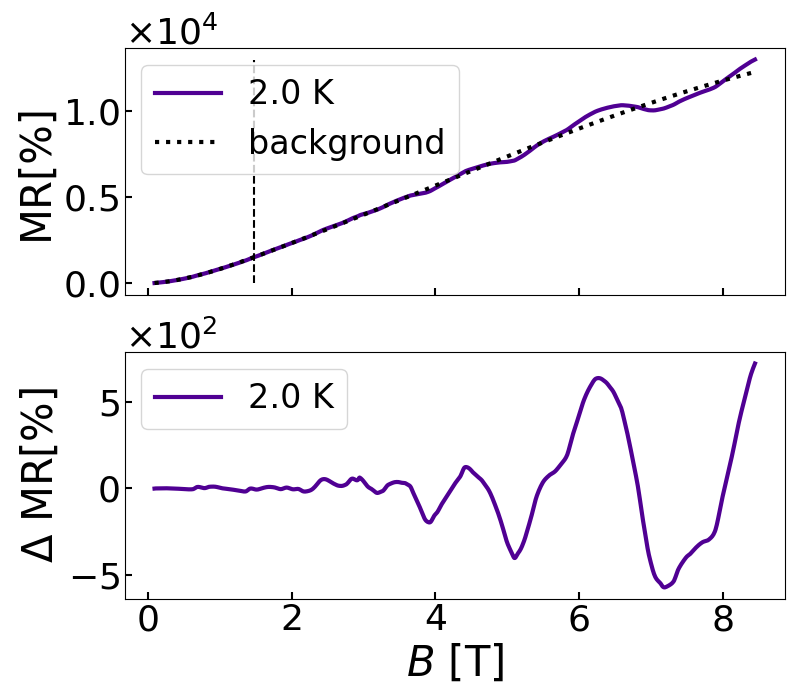}};
        \node[anchor=north west] (image) at (0.3\columnwidth,0) 
        {\includegraphics[width=0.3\columnwidth]{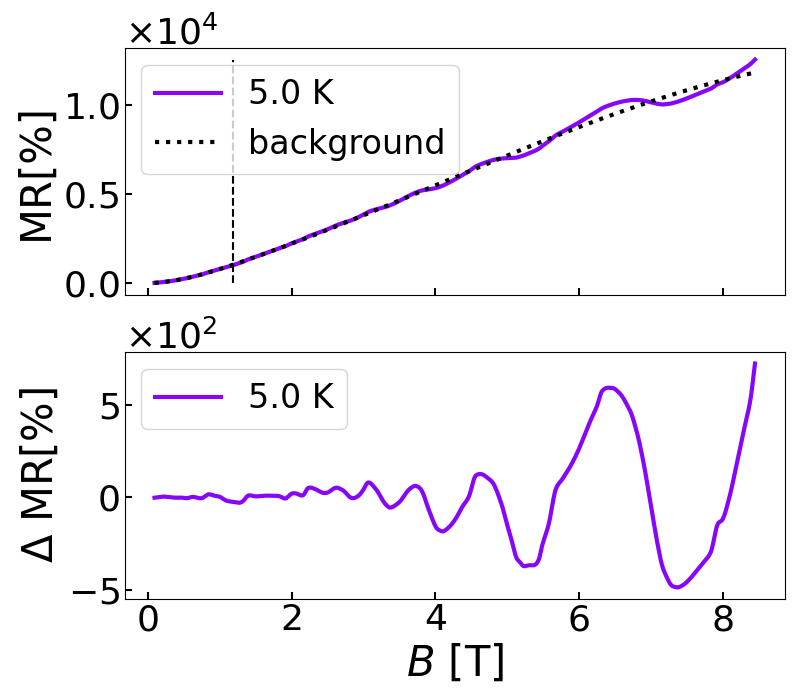}};
        \node[anchor=north west] (image) at (0.6\columnwidth,0) 
        {\includegraphics[width=0.3\columnwidth]{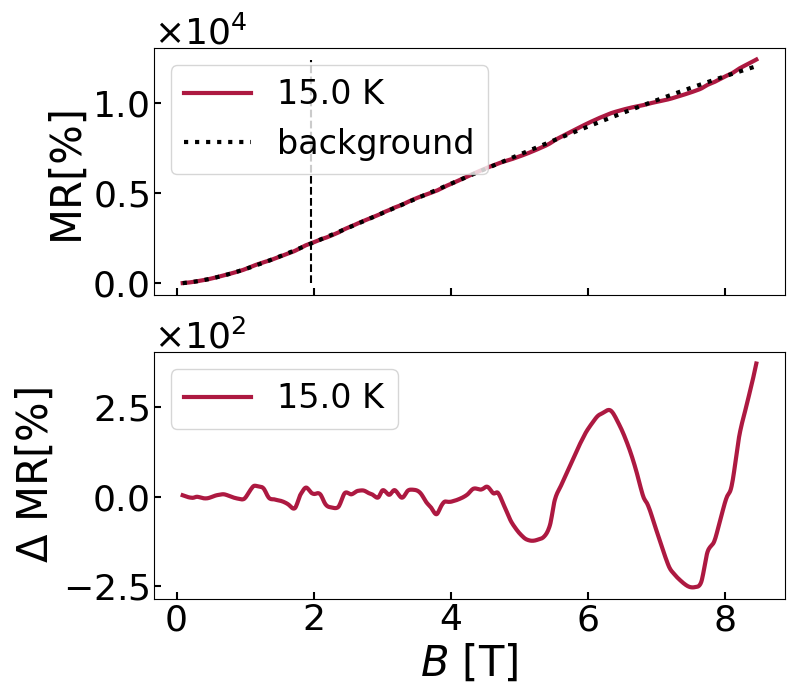}};
        \node[anchor=north west, scale = 1.2] (a) at (0,0) {\textbf{a}};
        \node[anchor=north west, scale = 1.2] (b) at (0.3\columnwidth,0) {\textbf{b}};
        \node[anchor=north west, scale = 1.2] (c) at (0.6\columnwidth,0) {\textbf{c}};

        \node[anchor=north west, scale = 1.2] at (0.225\columnwidth,-1.25) {S$_{3m}$};
    \end{tikzpicture}
    \begin{tikzpicture}
        \node[anchor=north west] (image) at (0,0) 
        {\includegraphics[width=0.3\columnwidth]{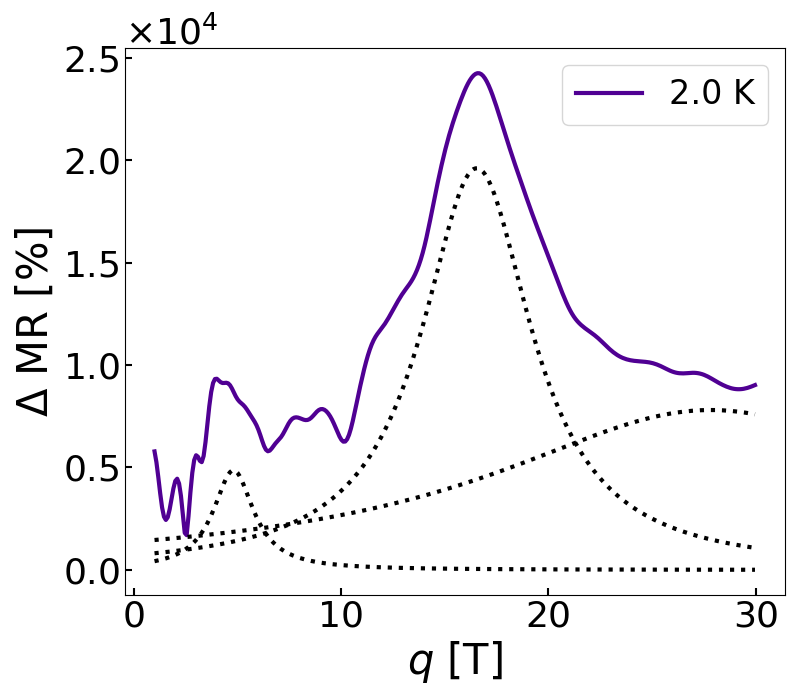}};
        \node[anchor=north west] (image) at (0.3\columnwidth,0) 
        {\includegraphics[width=0.3\columnwidth]{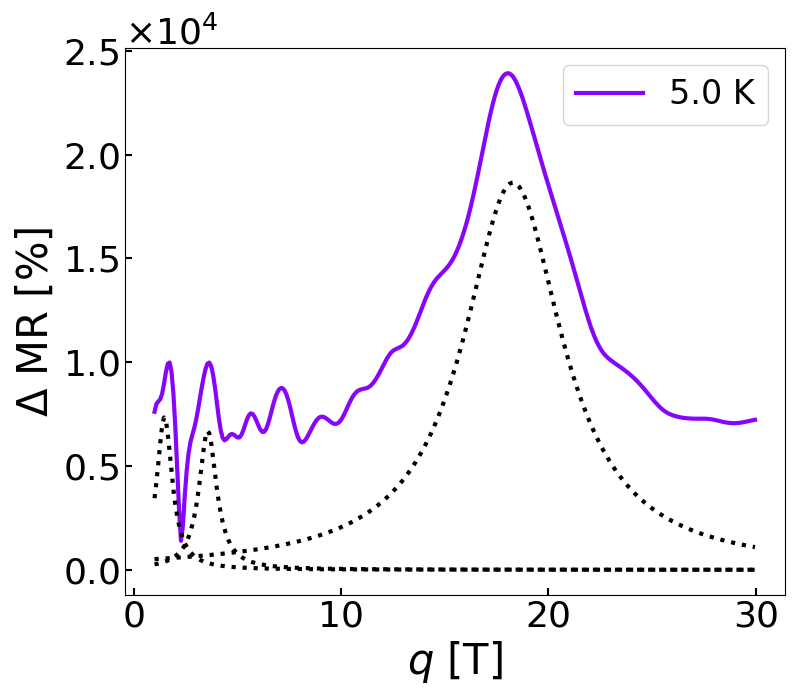}};
        \node[anchor=north west] (image) at (0.6\columnwidth,0) 
        {\includegraphics[width=0.3\columnwidth]{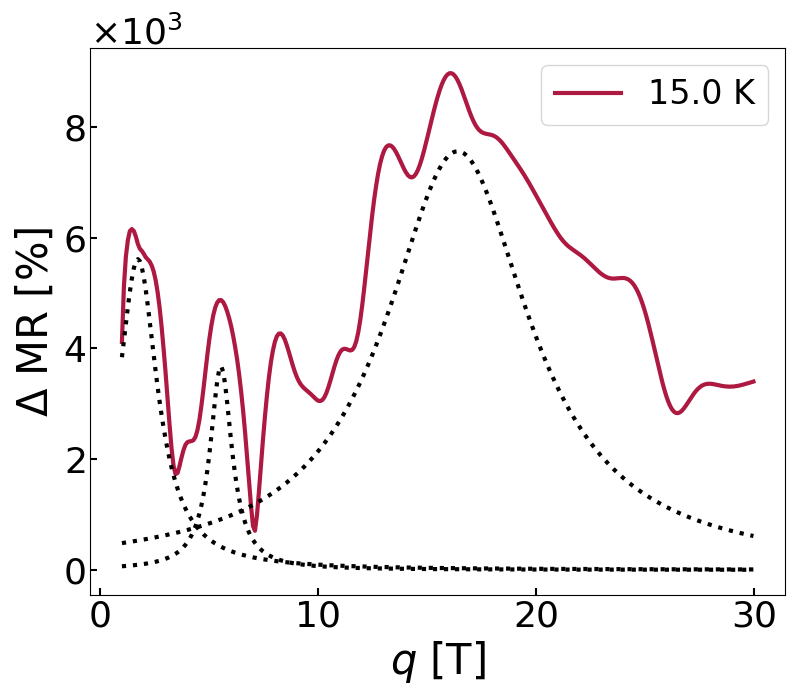}};
        \node[anchor=north west, scale = 1.2] (d) at (0,0) {\textbf{d}};
        \node[anchor=north west, scale = 1.2] (e) at (0.3\columnwidth,0) {\textbf{e}};
        \node[anchor=north west, scale = 1.2] (f) at (0.6\columnwidth,0) {\textbf{f}};
    \end{tikzpicture}
    \begin{tikzpicture}
        \node[anchor=north west] (image) at (0,0) 
        {\includegraphics[width=0.3\columnwidth]{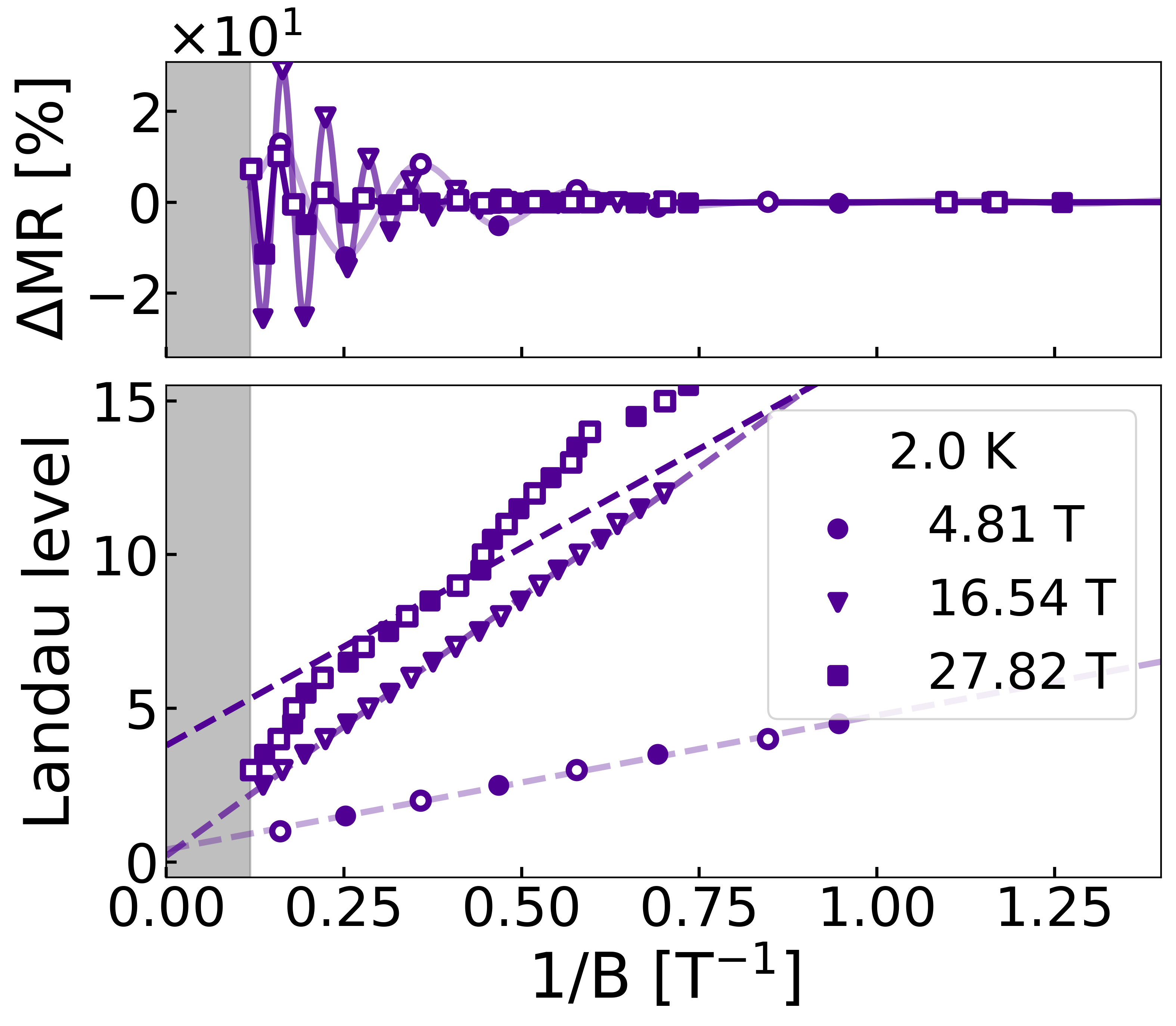}};
        \node[anchor=north west] (image) at (0.3\columnwidth,0) 
        {\includegraphics[width=0.3\columnwidth]{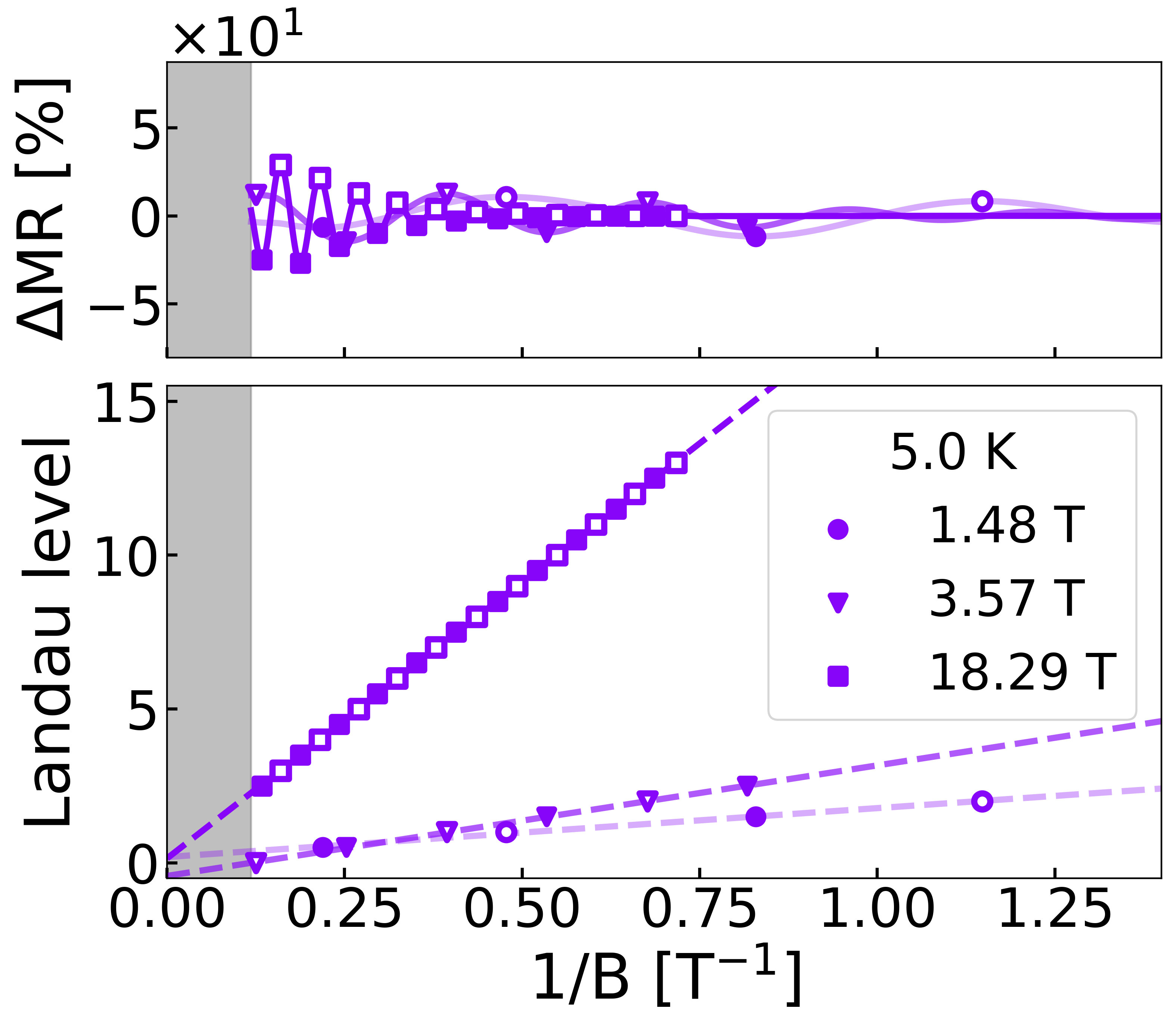}};
        \node[anchor=north west] (image) at (0.6\columnwidth,0) 
        {\includegraphics[width=0.3\columnwidth]{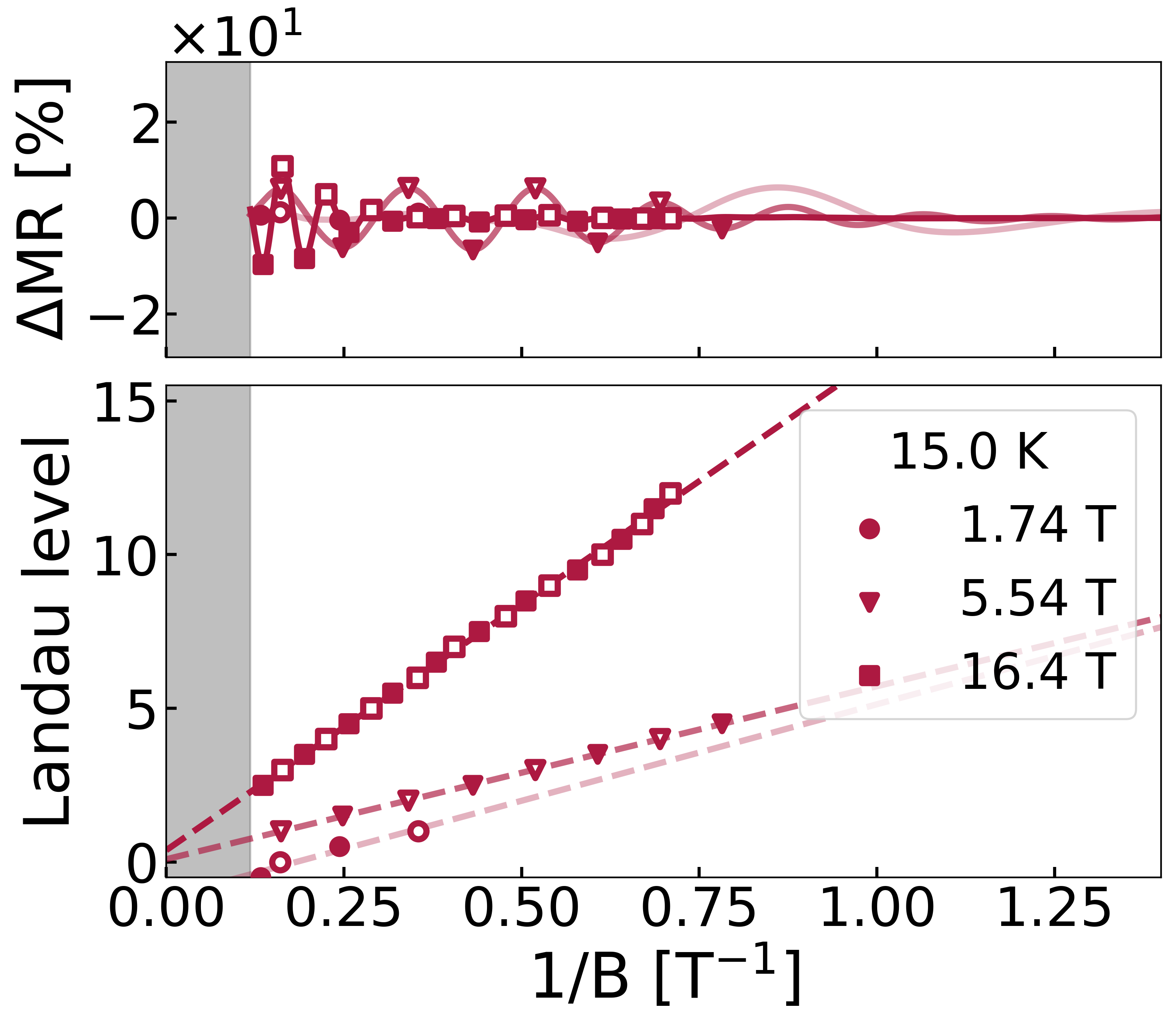}};
        \node[anchor=north west, scale = 1.2] (g) at (0,0) {\textbf{g}};
        \node[anchor=north west, scale = 1.2] (h) at (0.3\columnwidth,0) {\textbf{h}};
        \node[anchor=north west, scale = 1.2] (i) at (0.6\columnwidth,0) {\textbf{i}};
    \end{tikzpicture}
    \caption{MR analysis of S$_{3m}$ with piece-wise polynomial background method. MR plots and $\Delta$MR at \textbf{a} 2 K, \textbf{b} 5 K, and \textbf{c} 15 K. \textbf{d}-\textbf{f} Fast Fourier transform (FFT) of \textbf{a}-\textbf{c}, respectively. \textbf{g}-\textbf{i} display the Shubnikov-de Haas (SdH) oscillation and Landau level fan plot of \textbf{d}-\textbf{f}, respectively.}
    \label{figS11}
\end{figure}

\begin{figure}[ht!]
    \centering
    \begin{tikzpicture}
        \node[anchor=north west] (image) at (0,0) 
        {\includegraphics[width=0.3\columnwidth]{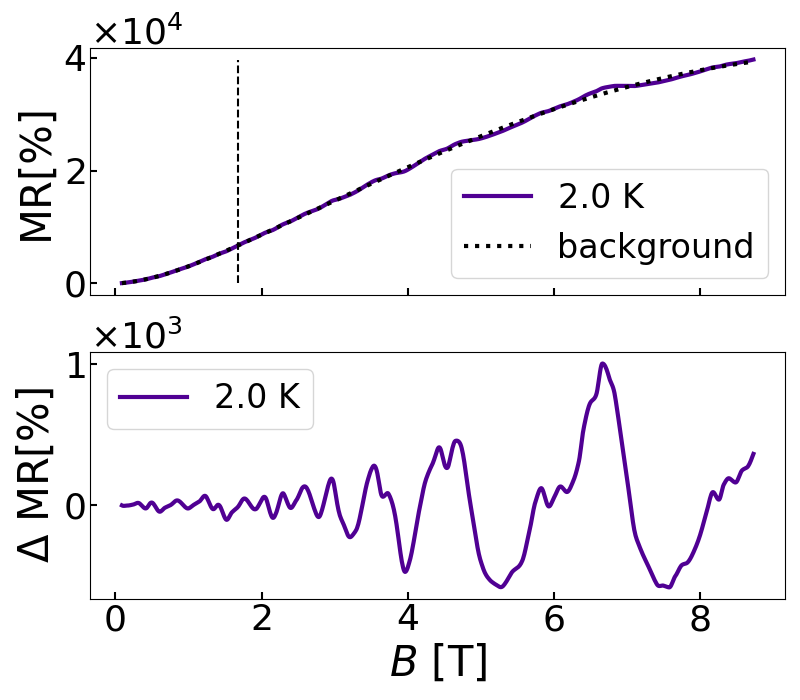}};
        \node[anchor=north west] (image) at (0.3\columnwidth,0) 
        {\includegraphics[width=0.3\columnwidth]{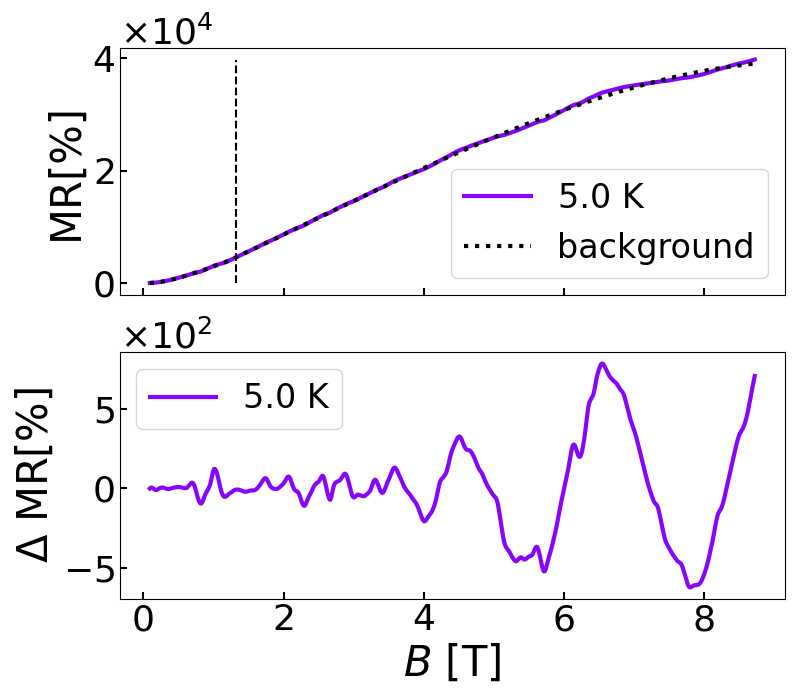}};
        \node[anchor=north west] (image) at (0.6\columnwidth,0) 
        {\includegraphics[width=0.3\columnwidth]{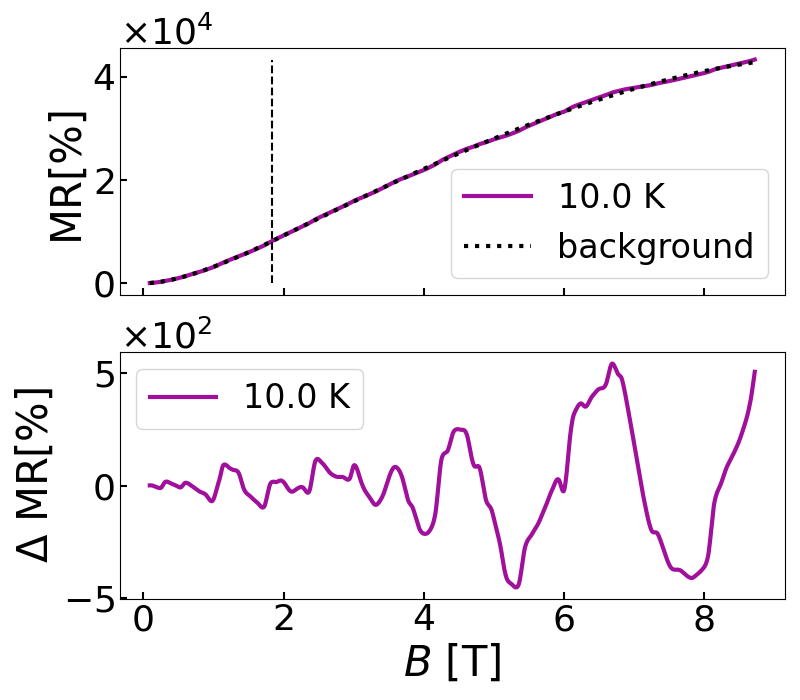}};
        \node[anchor=north west, scale = 1.2] (a) at (0,0) {\textbf{a}};
        \node[anchor=north west, scale = 1.2] (b) at (0.3\columnwidth,0) {\textbf{b}};
        \node[anchor=north west, scale = 1.2] (c) at (0.6\columnwidth,0) {\textbf{c}};

        \node[anchor=north west, scale = 1.2] at (0.075\columnwidth,-0.75) {S$_{20m}$};
    \end{tikzpicture}
    \begin{tikzpicture}
        \node[anchor=north west] (image) at (0,0) 
        {\includegraphics[width=0.3\columnwidth]{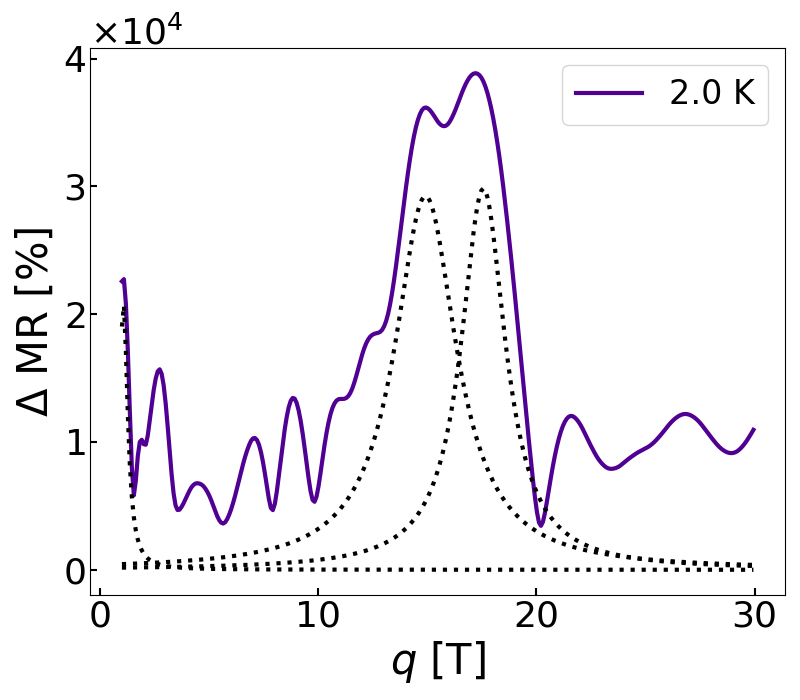}};
        \node[anchor=north west] (image) at (0.3\columnwidth,0) 
        {\includegraphics[width=0.3\columnwidth]{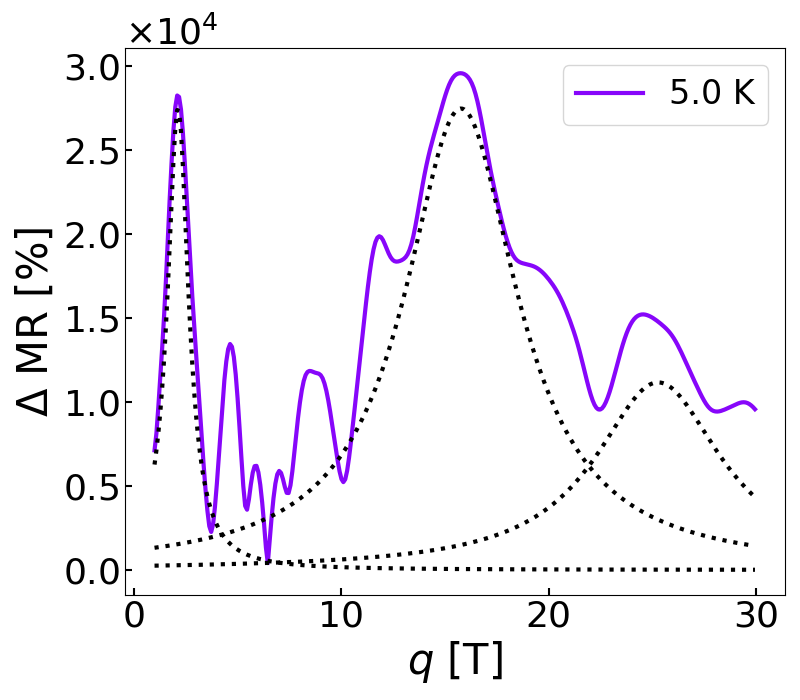}};
        \node[anchor=north west] (image) at (0.6\columnwidth,0) 
        {\includegraphics[width=0.3\columnwidth]{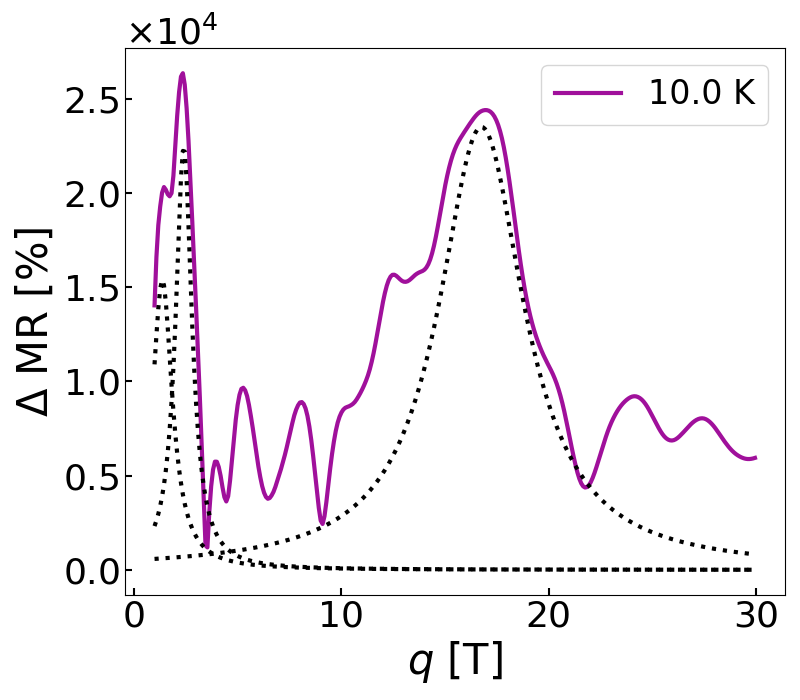}};
        \node[anchor=north west, scale = 1.2] (d) at (0,0) {\textbf{d}};
        \node[anchor=north west, scale = 1.2] (e) at (0.3\columnwidth,0) {\textbf{e}};
        \node[anchor=north west, scale = 1.2] (f) at (0.6\columnwidth,0) {\textbf{f}};
    \end{tikzpicture}
    \begin{tikzpicture}
        \node[anchor=north west] (image) at (0,0) 
        {\includegraphics[width=0.3\columnwidth]{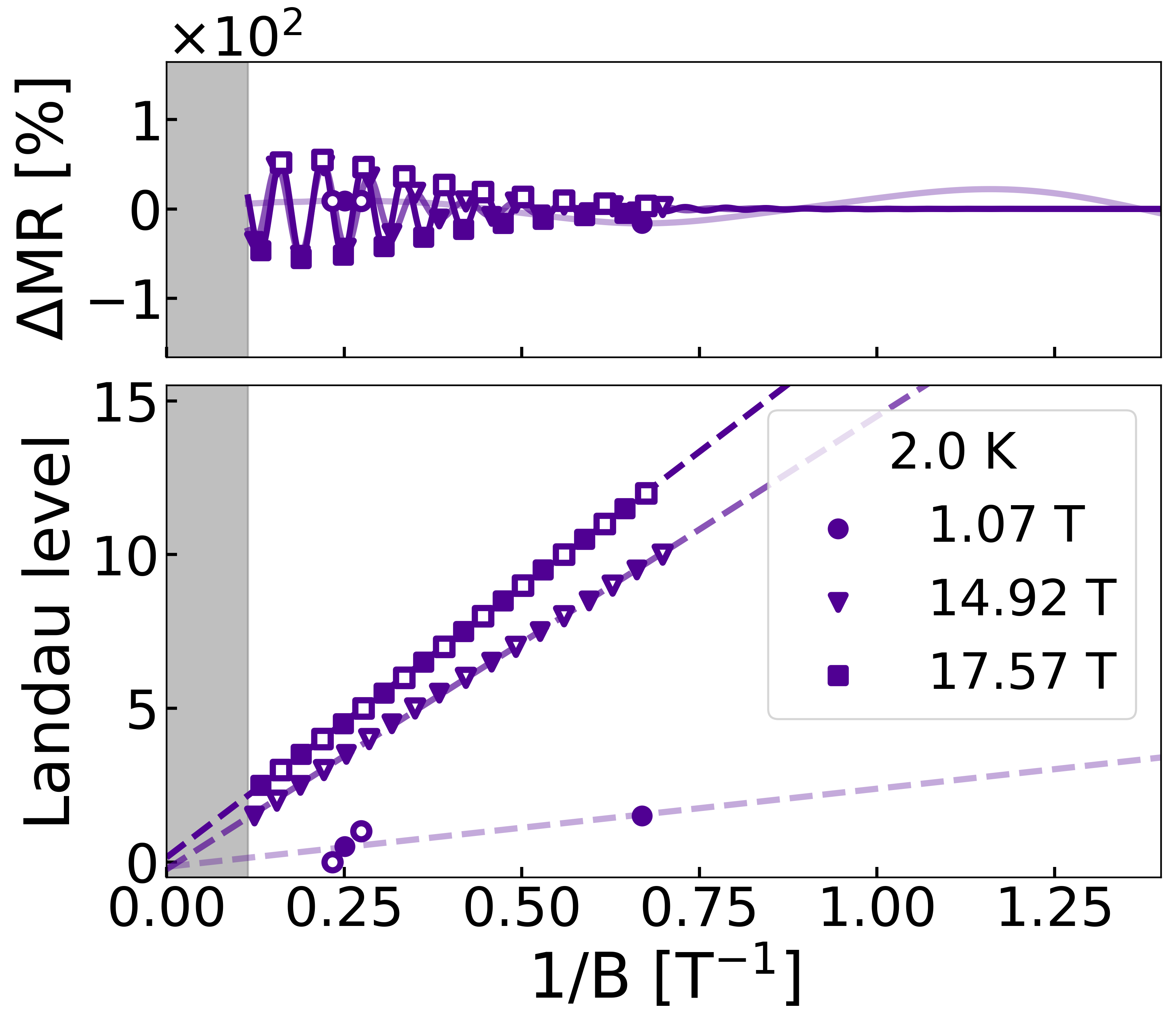}};
        \node[anchor=north west] (image) at (0.3\columnwidth,0) 
        {\includegraphics[width=0.3\columnwidth]{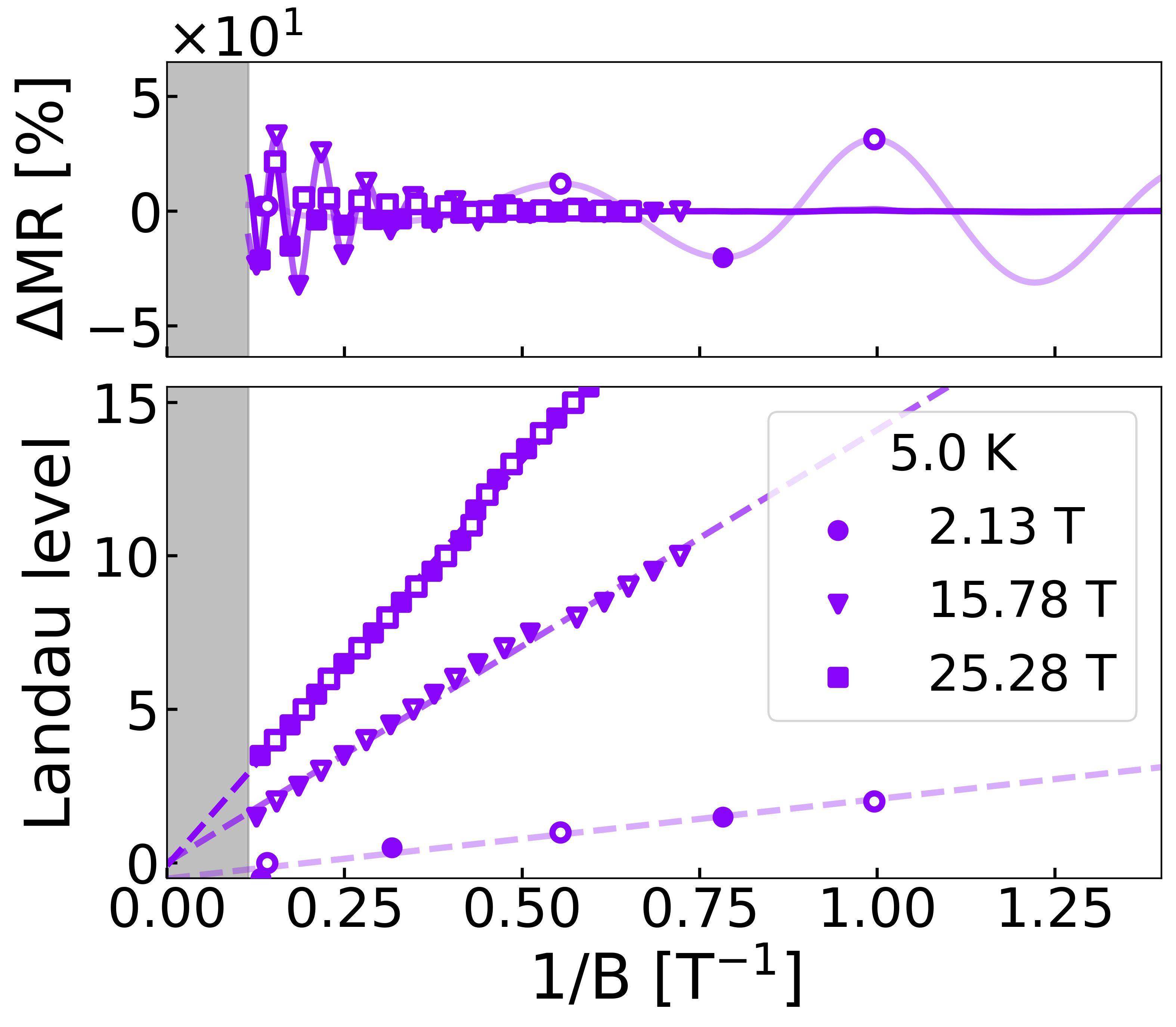}};
        \node[anchor=north west] (image) at (0.6\columnwidth,0) 
        {\includegraphics[width=0.3\columnwidth]{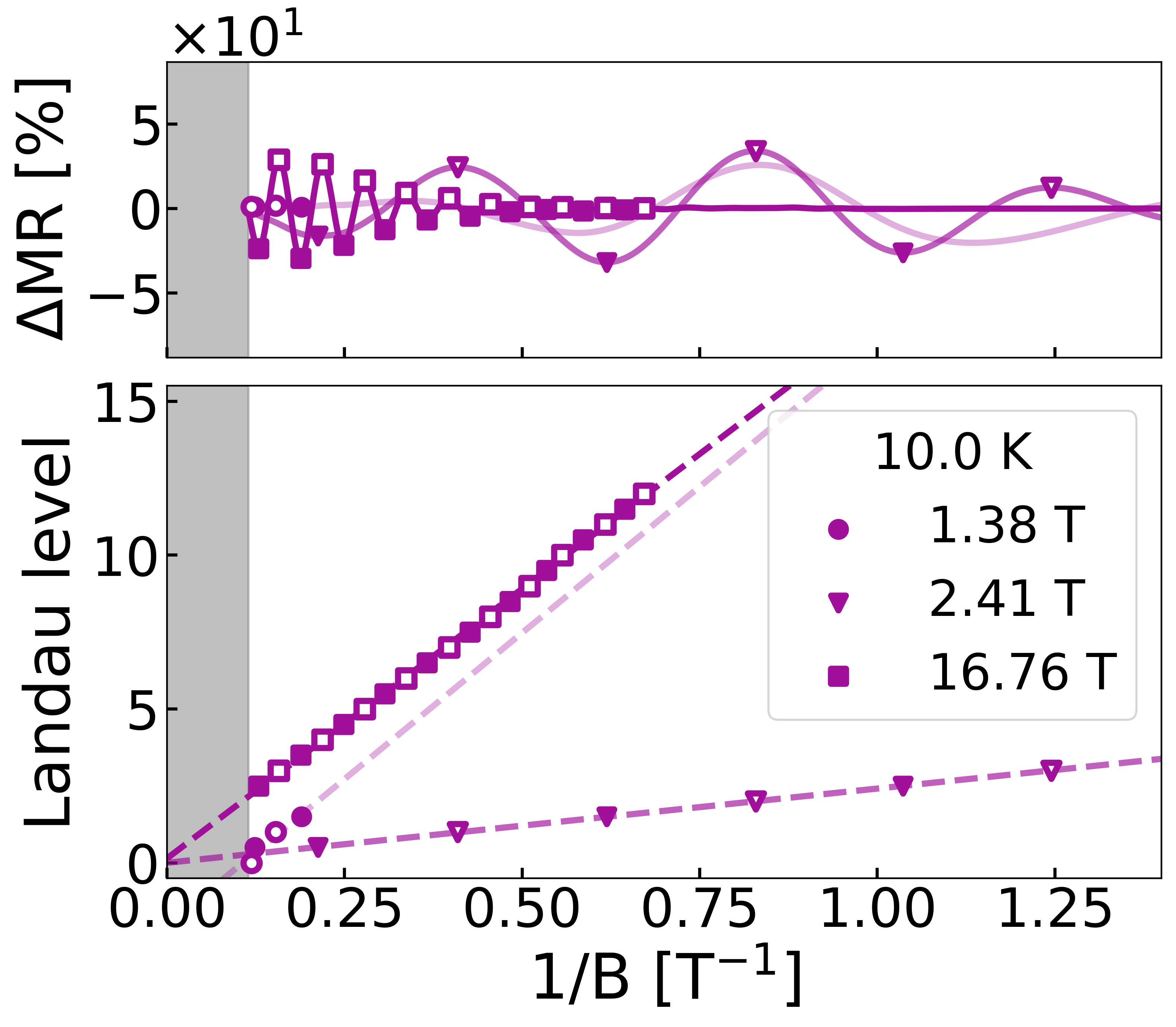}};
        \node[anchor=north west, scale = 1.2] (g) at (0,0) {\textbf{g}};
        \node[anchor=north west, scale = 1.2] (h) at (0.3\columnwidth,0) {\textbf{h}};
        \node[anchor=north west, scale = 1.2] (i) at (0.6\columnwidth,0) {\textbf{i}};
    \end{tikzpicture}
    \caption{MR analysis of S$_{20m}$ with piece-wise polynomial background method. MR plots and $\Delta$MR at \textbf{a} 2 K, \textbf{b} 5 K, and \textbf{c} 10 K. \textbf{d}-\textbf{f} Fast Fourier transform (FFT) of \textbf{a}-\textbf{c}, respectively. \textbf{g}-\textbf{i} display the Shubnikov-de Haas (SdH) oscillation and Landau level fan plot of \textbf{d}-\textbf{f}, respectively.}
    \label{figS12}
\end{figure}

Magnetoresistance (MR) data was extracted from $\rho_{xx}$ data, using the following relation:
\begin{equation}
\text{MR} = \frac{\rho_{xx}(B) - \rho_{xx}(0)}{\rho_{xx}(0)} \times 100\%
\end{equation}
The quantum oscillation data, $\Delta$MR, was obtained by subtracting background signals from the MR data. To ensure robustness, we employed three independent methods to extract and analyze $\Delta$MR: (a) background-free curvature (\figref{figS7} and \figref{figS8}), (b) polynomial background (partially in \figref{fig3}, and fully in \figref{figS9} and \figref{figS10}), and (c) piece-wise polynomial background (\figref{figS11} and \figref{figS12}). In (a), a second-order derivative with respect to the magnetic field $B$ is applied to the MR data, automatically eliminating all constant, linear, and quadratic terms, which mostly contribute to background instead of oscillation, without the need for actual background selection.

Notably, the fast Fourier transform (FFT) analysis unveiled distinct oscillation frequencies. After applying a standard signal filtering process through FFT against $1/B$, we isolated the oscillation components from the data and determined the corresponding Landau levels (LLs). The LL index fan corresponding to the different frequencies for each fitting model has been plotted. All three methods consistently support the existence of a low-frequency carrier pocket, F$_{\alpha}$ 1.87 T - 3.94 T for S$_{3m}$ and 0.35 T - 1.14 T for S$_{20m}$. In contrast, the pristine TaP exhibits the same behavior in the range F$_{\alpha}$ 2.3 T - 4 T and F$_{\beta}$ 18 T - 19 T \cite{TaP}. Our comprehensive analysis confirms the presence of new oscillation frequencies, suggesting that the carrier pockets of the Weyl node can indeed reach the desired 0 Landau level, and additional Weyl nodes appear with H$^{-}$ ion irradiation.

\subsection{Extended Kohler’s rule of magnetoresistance}
\begin{figure}[ht!]
    \centering
    \begin{tikzpicture}
        \node[anchor=north west] (image) at (0,0) 
        {\includegraphics[width=0.3\columnwidth]{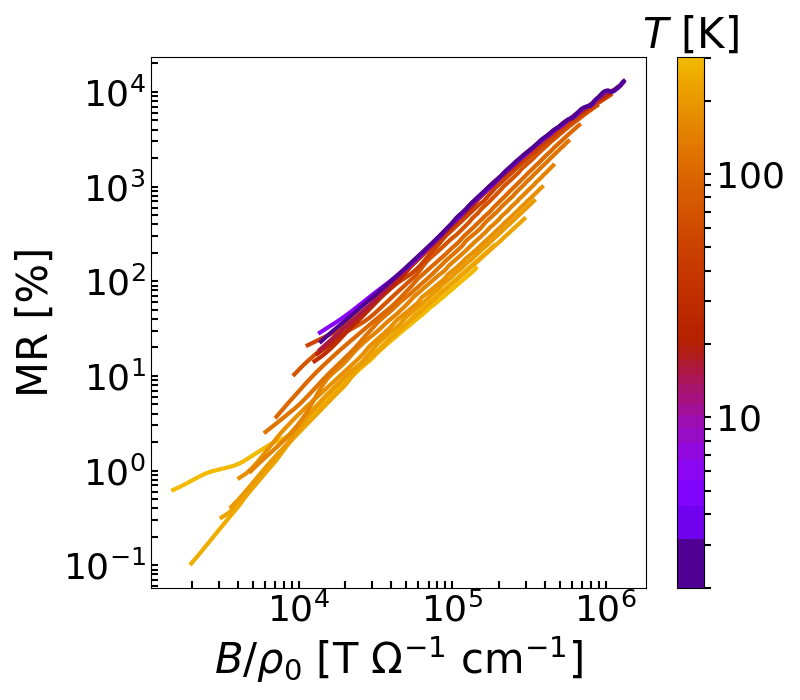}};
        \node[anchor=north west] (image) at (0.3\columnwidth,0) 
        {\includegraphics[width=0.3\columnwidth]{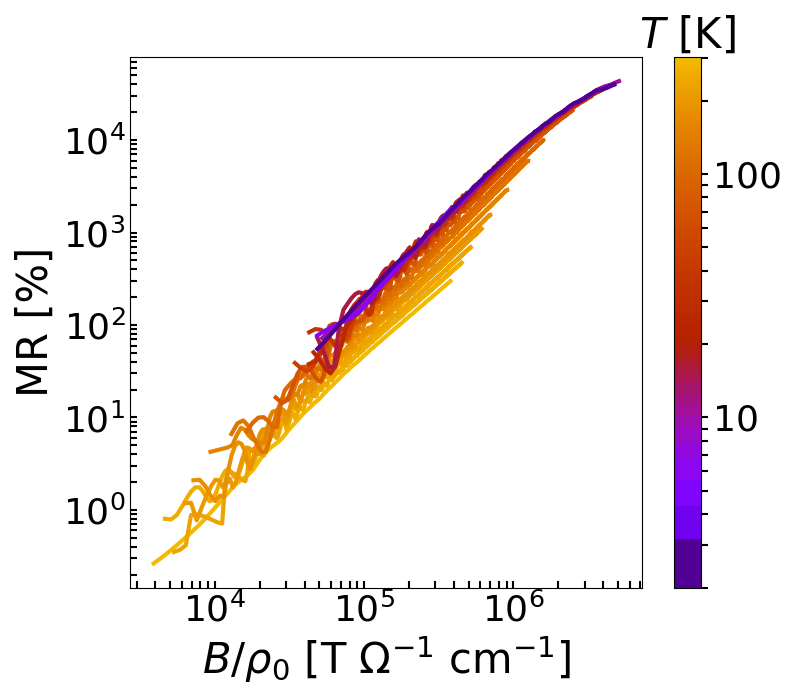}};
        \node[anchor=north west, scale = 1.2] (a) at (0,0) {\textbf{a}};
        \node[anchor=north west, scale = 1.2] (b) at (0.3\columnwidth,0) {\textbf{b}};

        \node[anchor=north west, scale = 1.2] at (0.075\columnwidth,-0.75) {S$_{3m}$};
        \node[anchor=north west, scale = 1.2] at (0.375\columnwidth,-0.75) {S$_{20m}$};
    \end{tikzpicture}
    \begin{tikzpicture}
        \node[anchor=north west] (image) at (0,0) 
        {\includegraphics[width=0.3\columnwidth]{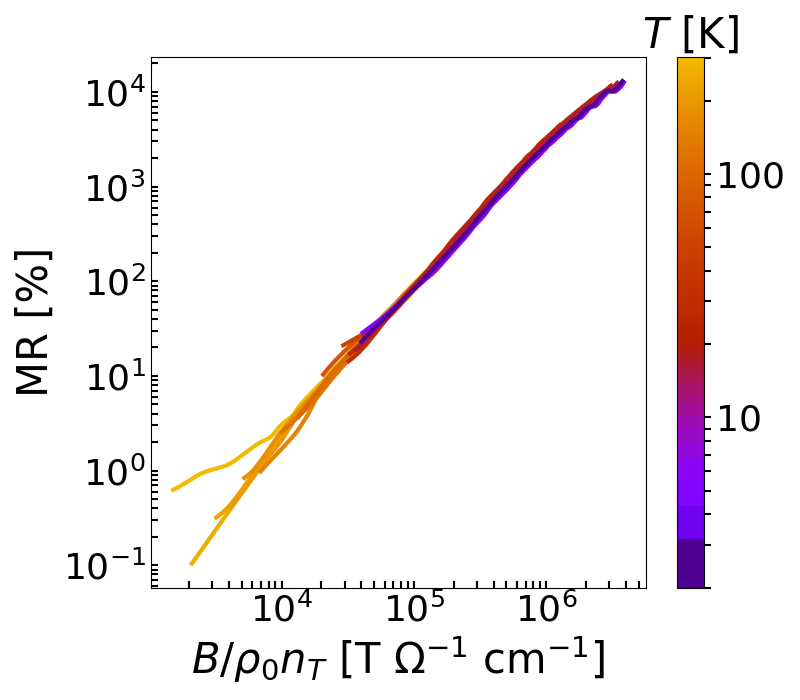}};
        \node[anchor=north west] (image) at (0.3\columnwidth,0) 
        {\includegraphics[width=0.3\columnwidth]{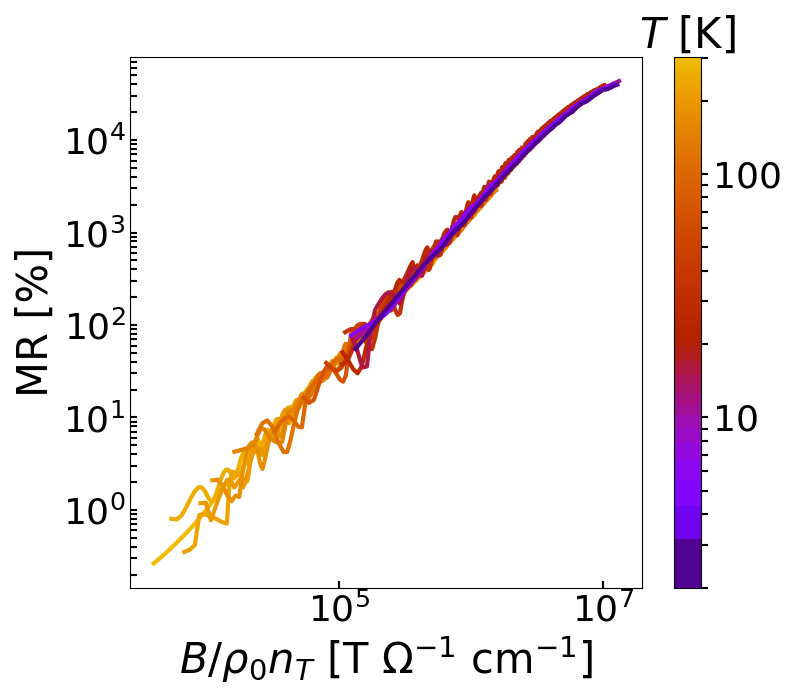}};
        \node[anchor=north west, scale = 1.2] (c) at (0,0) {\textbf{c}};
        \node[anchor=north west, scale = 1.2] (d) at (0.3\columnwidth,0) {\textbf{d}};

        \node[anchor=north west, scale = 1.2] at (0.075\columnwidth,-0.75) {S$_{3m}$};
        \node[anchor=north west, scale = 1.2] at (0.375\columnwidth,-0.75) {S$_{20m}$};
    \end{tikzpicture}
    \caption{\textbf{a}-\textbf{b} Kohler’s rule plot of the data for S$_{3m}$, and S$_{20m}$, respectively. \textbf{c}-\textbf{d} Extended Kohler’s rule plot of the data in \textbf{a}-\textbf{b}, respectively.}
    \label{figS13}
\end{figure}

\begin{figure}[ht!]
    \centering
    \begin{tikzpicture}
        \node[anchor=north west] (image) at (0,0) 
        {\includegraphics[width=0.3\columnwidth]{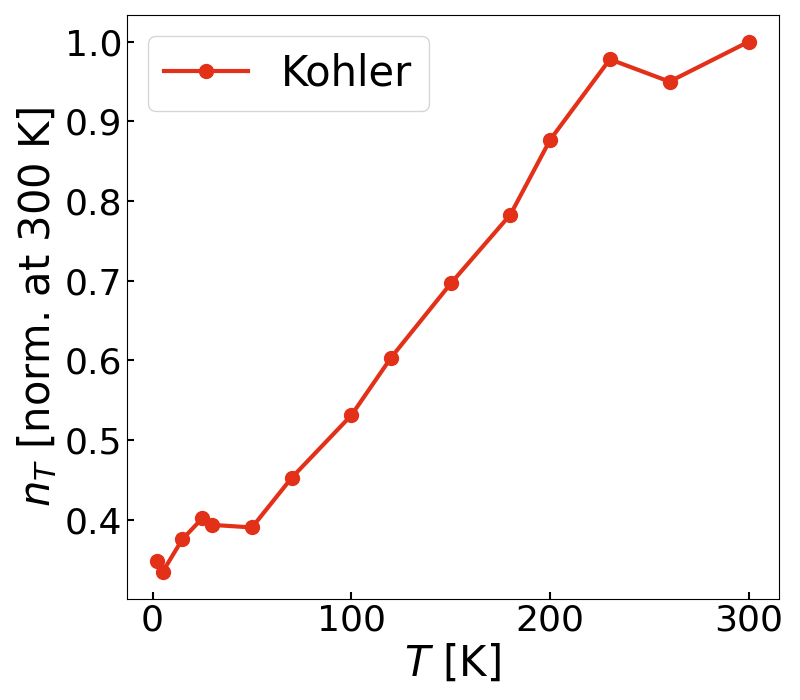}};
        \node[anchor=north west] (image) at (0.3\columnwidth,0) 
        {\includegraphics[width=0.3\columnwidth]{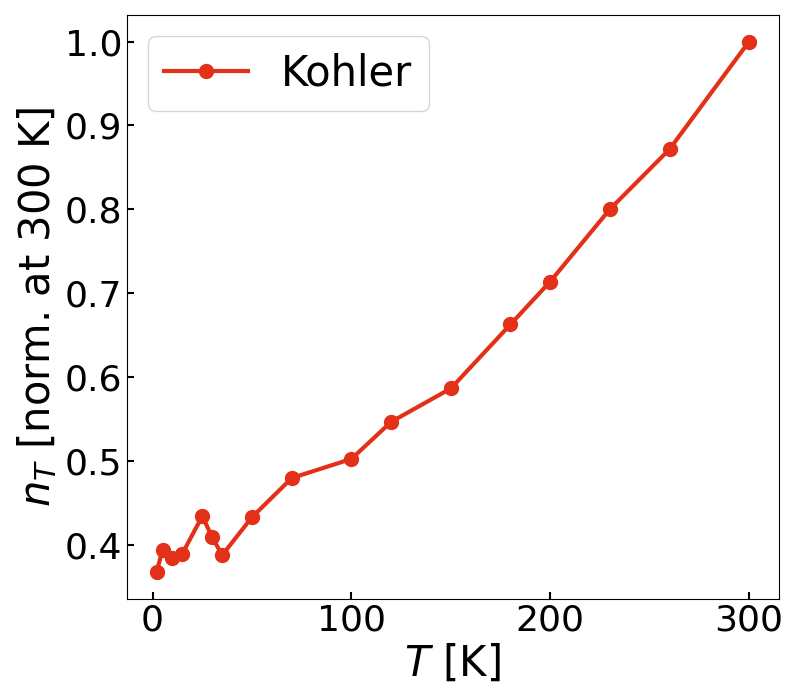}};
        \node[anchor=north west, scale = 1.2] (a) at (0,0) {\textbf{a}};
        \node[anchor=north west, scale = 1.2] (b) at (0.3\columnwidth,0) {\textbf{b}};

        \node[anchor=north west, scale = 1.2] at (0.225\columnwidth,-3.25) {S$_{3m}$};
        \node[anchor=north west, scale = 1.2] at (0.525\columnwidth,-3.25) {S$_{20m}$};
    \end{tikzpicture}
    \begin{tikzpicture}
        \node[anchor=north west] (image) at (0,0) 
        {\includegraphics[width=0.3\columnwidth]{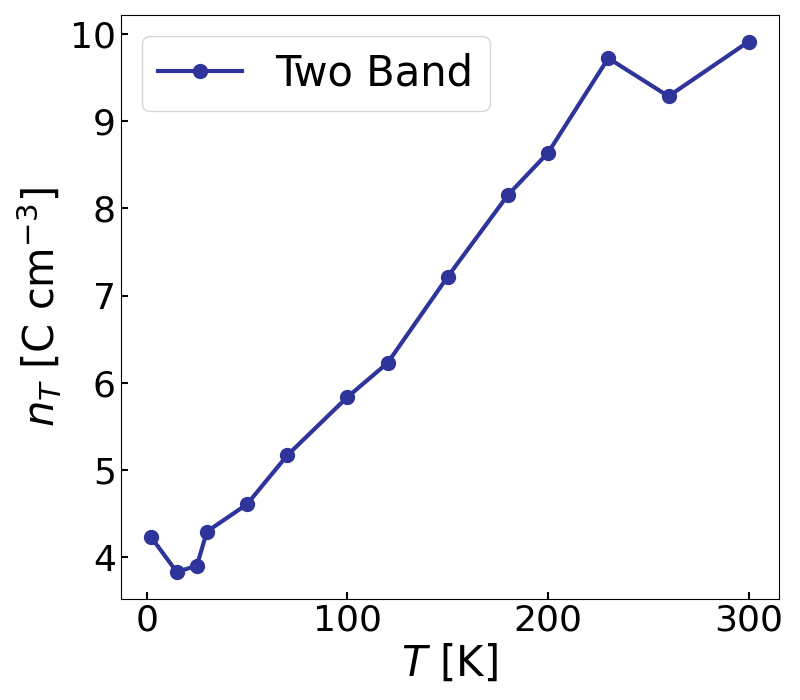}};
        \node[anchor=north west] (image) at (0.3\columnwidth,0) 
        {\includegraphics[width=0.3\columnwidth]{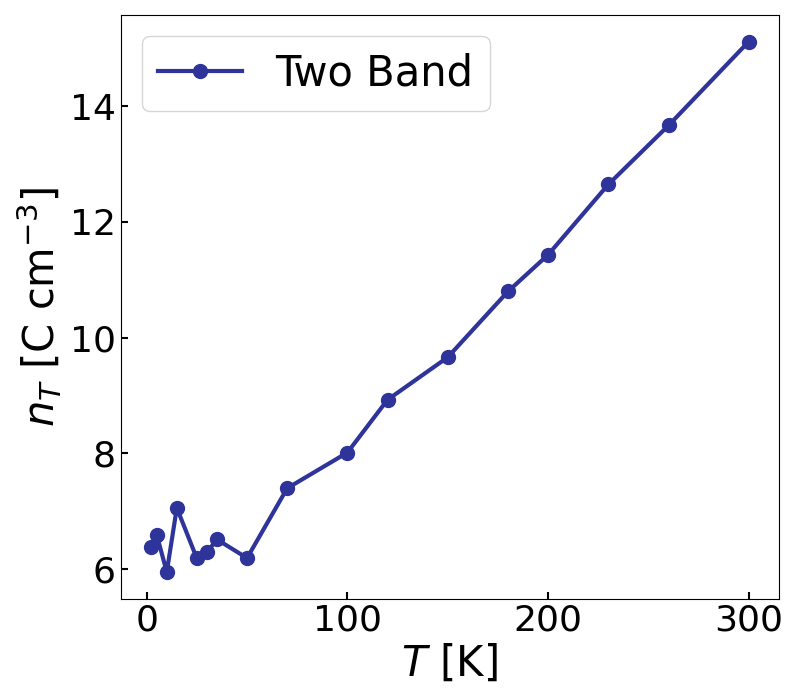}};
        \node[anchor=north west, scale = 1.2] (c) at (0,0) {\textbf{c}};
        \node[anchor=north west, scale = 1.2] (d) at (0.3\columnwidth,0) {\textbf{d}};

        \node[anchor=north west, scale = 1.2] at (0.225\columnwidth,-3.25) {S$_{3m}$};
        \node[anchor=north west, scale = 1.2] at (0.525\columnwidth,-3.25) {S$_{20m}$};
    \end{tikzpicture}
    \begin{tikzpicture}
        \node[anchor=north west] (image) at (0,0) 
        {\includegraphics[width=0.3\columnwidth]{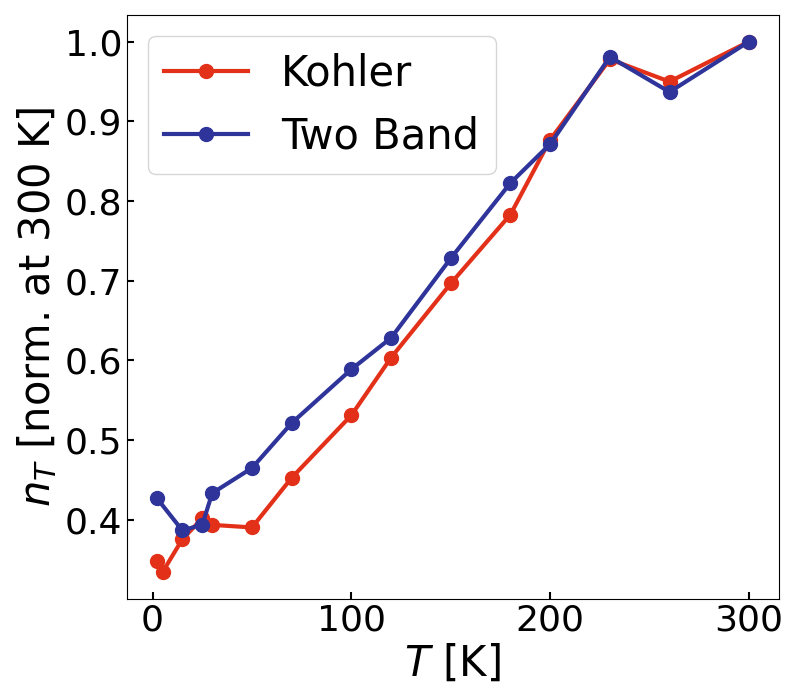}};
        \node[anchor=north west] (image) at (0.3\columnwidth,0) 
        {\includegraphics[width=0.3\columnwidth]{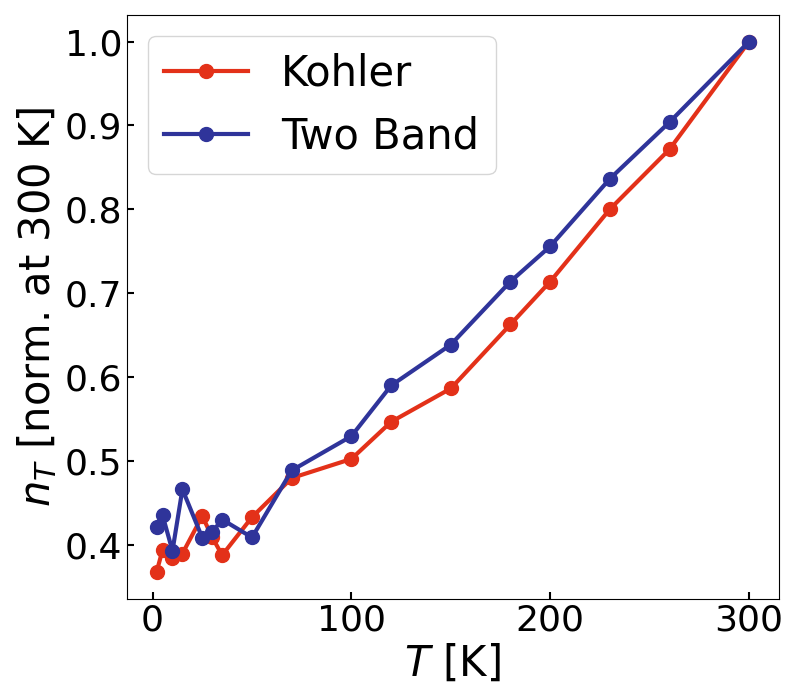}};
        \node[anchor=north west, scale = 1.2] (e) at (0,0) {\textbf{e}};
        \node[anchor=north west, scale = 1.2] (f) at (0.3\columnwidth,0) {\textbf{f}};

        \node[anchor=north west, scale = 1.2] at (0.225\columnwidth,-3.25) {S$_{3m}$};
        \node[anchor=north west, scale = 1.2] at (0.525\columnwidth,-3.25) {S$_{20m}$};
    \end{tikzpicture}
    \caption{The temperature dependence of the adjusted parameter $n_{T}$ extracted from Extended Kohler’s rule for \textbf{a} S$_{3m}$, and \textbf{b} S$_{20m}$. \textbf{c}-\textbf{d} The same as \textbf{a}-\textbf{b}, but $n_{T}$ are calculated with fitted parameters from the two-band model using the relation given in equation \eqref{nT}. \textbf{e}-\textbf{f} depict the normalized $n_{T}$ values, enabling a direct comparison between simple extended Kohler’s rule and fitted two-band model.}
    \label{figS14}
\end{figure}

The MR curves do not coincide onto a single curve when plotted against $B/\rho_{xx}(0)$, indicating that the $\text{MR} = f(B/\rho_{xx}(0))$ rule proposed by Kohler is not applicable to S$_{3m}$ and S$_{20m}$, as observed in our study. However, an intriguing observation is that all the curves in \figref{figS13}a-b are nearly parallel with each other. This suggests the possibility of a single temperature-dependent multiplier affecting MR (y-axis) or $B/\rho_{xx}(0)$ (x-axis), causing them to coincide or merge into one curve. To understand this behavior further, MR curves are normalized by a scaling procedure using the 300 K curve, and adjust the parameter $n_{T}$ individually for each curve \cite{Kohler1}. The outcome is clearly visible in \figref{figS13}c-d, where all the curves seamlessly coincide onto the 300 K curve when the data is scaled using $B/\rho_{xx}(0)n_{T}$. The corresponding values of $n_{T}$ for the different curves at various temperatures can be found in \figref{figS14}. This scaling procedure sheds light on the temperature-dependent behavior of the MR and provides valuable insights into the underlying physical mechanisms governing the observed phenomena. We have further normalized $n_{T}$ and compared it with the data extracted by using a two-band model from the relation \cite{Kohler1}:
\begin{equation}
n_{T} = e \left[\frac{\left(\sum_{i} n_{i} \mu_{i}\right)^{3/2}}{\left(\sum_{i} n_{i} \mu_{i}^{3}\right)^{1/2}}\right].
\label{nT}
\end{equation}

\end{document}